\documentclass[twocolumn,prb,floatfix,nobibnotes,nofootinbib]{revtex4-1}
\usepackage{hyperref}
\usepackage{amsmath}
\usepackage{bbold}
\usepackage{amssymb}
\usepackage{graphicx}
\usepackage{bm}
\usepackage{xcolor}
\newcommand{\editRRR}[1] {\textcolor{black}{#1}}
\newcommand{\editRR}[1] {\textcolor{black}{#1}}
\newcommand{\editEE}[1] {\textcolor{black}{#1}}

\newcommand{\edit}[1] {\textcolor{black}{#1}}
\newcommand{\editP}[1] {\textcolor{black}{#1}}
\newcommand{\editE}[1] {\textcolor{black}{#1}}
\newcommand{\editJJ}[1] {\textcolor{black}{#1}}
\newcommand{\editJ}[1] {\textcolor{black}{#1}}

\providecommand{\dd}{\textrm{d}}

\begin{document}

\title{Superconducting islands with semiconductor-nanowire-based topological Josephson junctions}

\author{J. \' Avila$^1$, E. Prada$^2$, P. San-Jose$^1$, R. Aguado$^1$}
\affiliation{\\
$^1$Instituto de Ciencia de Materiales de Madrid (ICMM), Consejo Superior de Investigaciones Cient\'{i}ficas (CSIC), Sor Juana In\'{e}s de la Cruz 3, 28049 Madrid, Spain. Research Platform on Quantum Technologies (CSIC).\\$^2$Departamento de F\'isica de la Materia Condensada, Condensed Matter Physics Center (IFIMAC) and Instituto Nicol\'as Cabrera, Universidad Aut\'onoma de Madrid, E-28049 Madrid, Spain}

\date{\today}
\begin{abstract}
We theoretically study superconducting islands based on semiconductor-nanowire Josephson junctions and take into account the presence of subgap quasiparticle excitations in the spectrum of the junction. Our method extends the standard model Hamiltonian for a superconducting charge qubit and replaces the Josephson potential by the Bogoliubov--de Gennes Hamiltonian of the nanowire junction, projected onto the relevant low-energy subgap subspace. This allows to fully incorporate the coherent dynamics of subgap levels in the junction. The combined effect of spin-orbit coupling and Zeeman energy in the nanowires forming the junction triggers a topological transition, where the subgap levels evolve from finite-energy Andreev bound states into near-zero energy Majorana bound states. The interplay between the microscopic energy scales governing the nanowire junction (the Josephson energy, the Majorana coupling and the Majorana energy splitting), with the charging energy of the superconducting island, gives rise to a great variety of physical regimes. Based on this interplay of different energy scales, we fully characterize the microwave response of the junction, from the Cooper pair box to the transmon regimes, and show how the presence of Majoranas can be detected through distinct spectroscopic features. \editRR{In split-junction geometries, the plasma mode couples to the phase-dispersing subgap levels resulting from Majorana hybridization via a Jaynes--Cummings-like interaction. As a consequence of this interaction, higher order plasma excitations in the junction inherit Majorana properties, including the $4\pi$ effect.}
\end{abstract}\maketitle


\section{Introduction}
\editP{Josephson junctions (JJ) involving mesoscopic superconducting islands} 
are one of the most versatile platforms for quantum state engineering and solid-state qubit implementations \cite{PhysRevA.69.062320,Devoret1169,Wendin_2017}. Their physics is governed by the competition between two energy scales: the charging energy \editP{$E_C$ of the island and the Josephson coupling $E_J$ across the junction}.
This competition is described by the Hamiltonian \cite{Bouchiat_1998}
\editP{
\begin{eqnarray}
\label{transmonH}
H&=&4E_C(\hat N-n_g)^2 + V_J(\hat \varphi),\\
V_J(\hat \varphi) &=& -E_J\cos\hat \varphi \nonumber,
\end{eqnarray}
where $V_J(\hat \varphi)$ is the Josephson potential, $\hat N$ is the number of Cooper pairs in the island}, conjugate to the \editP{junction superconducting phase difference $\hat\varphi$, and $n_g=Q_g/2e=V_g/(2eC_g)$ is a gate-induced charge offset in the island in units of a Cooper pair. The latter is controlled by a gate at potential $V_g$ with gate-island capacitance $C_g$}. Equation \eqref{transmonH} can be simply interpreted as the energy stored in a LC oscillator where the standard (linear) inductance $L$ is replaced by the (nonlinear) Josephson inductance $L_J^{-1}(\varphi)=(2e^2/\hbar)^2d^2V_{J}(\varphi)/d\varphi^2=(2e^2/\hbar)^2E_J\cos(\varphi)$.

In the limit $E_J\ll E_C$, charge quantization is strong, which manifests as Coulomb Blockade oscillations in units of $2e$. \editP{At points with half-integer $n_g = m+1/2$, $N$ and $N+1$ states become nearly degenerate, defining a charge qubit. In this} so-called Cooper pair box (CPB) regime, the charge dispersion of the qubit frequency (i. e., its variation as a function of the gate-induced offset charge) is large, \editP{since charge eigenenergies depend strongly on gate $V_g$, making the qubit} very susceptible to charge noise. 
In the opposite $E_J\gg E_C$ so-called transmon regime~\cite{koch2007charge}, quantum fluctuations suppress charge quantization and charge dispersion is exponentially-suppressed. \editP{As a result, the qubit susceptibility to noise is strongly suppressed and quantum coherence is correspondingly enhanced}. This comes, however, at the cost of reduced anharmonicity (the transmon spectrum is almost harmonic with a frequency given by the Josephson plasma frequency $\omega_{pl}=\sqrt{8E_JE_C}/\hbar$), 
which reduces the operation time due to leakage out of the  qubit subspace.

The above discussion assumes a a sinusoidal current-phase relation which gives a Josephson relation of the form $V_{J}(\varphi)=-E_J\cos(\varphi)$. This is an \editE{excellent} description of a superconductor-insulator-superconductor (SIS) tunnel junction, which forms the basis of almost all state-of-the-art superconducting qubits. More recently, alternative technologies are sought in order to replace the weak link in the JJ and reach further operational functionalities. Such alternatives include semiconducting nanowires (NWs) --also known as gatemons \cite{PhysRevLett.115.127001,PhysRevLett.115.127002,PhysRevB.97.060508,PhysRevLett.116.150505,PhysRevLett.120.100502,PhysRevB.99.085434}--, two-dimensional gases \cite{Casparis2018} and van der Waals heterostructures \cite{Kroll2018,Schmidt2018,Wang2019,Tahan19}. Arguably, their main goal is to have compatibility with large magnetic fields and tunability by means of gate voltages, both \editP{of which are} key requirements to reach a topological superconductor state, as predicted in many platforms \cite{Leijnse:SSAT12,Alicea:RPP12,Beenakker:ARCMP13,Sato:JPSJ16,Aguado:RNC17,Sato:ROPIP17,Lutchyn:NRM18}. This opens the possibility of using standard circuit QED techniques for microwave (MW) readout of topological qubits based on Majorana bound states in such platforms~\cite{Hassler_2011,PhysRevB.88.235401,PhysRevLett.111.107007,PhysRevB.88.144507,ginossar2014microwave,PhysRevB.92.075143,PhysRevB.92.245432,PhysRevB.92.134508,PhysRevB.94.085409,PhysRevLett.118.126803,PhysRevB.97.041415,10.21468/SciPostPhys.7.4.050}.  

The physics of most of the \editE{alternative} \editP{weak link junctions cited above differ} considerably from standard SIS tunnel junctions. In particular, the Josephson effect in \editP{NW junctions} is typically dominated by a small number of highly transmitting channels, see e. g. Refs. \onlinecite{PhysRevB.97.060508,Goffman_2017,PhysRevX.9.011010}. This implies that the current-phase relation is no longer sinusoidal and thus $V_{J}(\varphi)\neq -E_J\cos(\varphi)$ \editP{in Eq. \eqref{transmonH}}. A proper description of superconducting islands \editE{presenting} such non-sinusoidal Josephson potentials thus needs a correct treatment of the microscopic mechanisms governing the subgap spectrum (Andreev levels) of the weak link, which in turn dictates the final form of $V_{J}(\varphi)$.

\begin{figure}
\centering \includegraphics[width=\columnwidth]{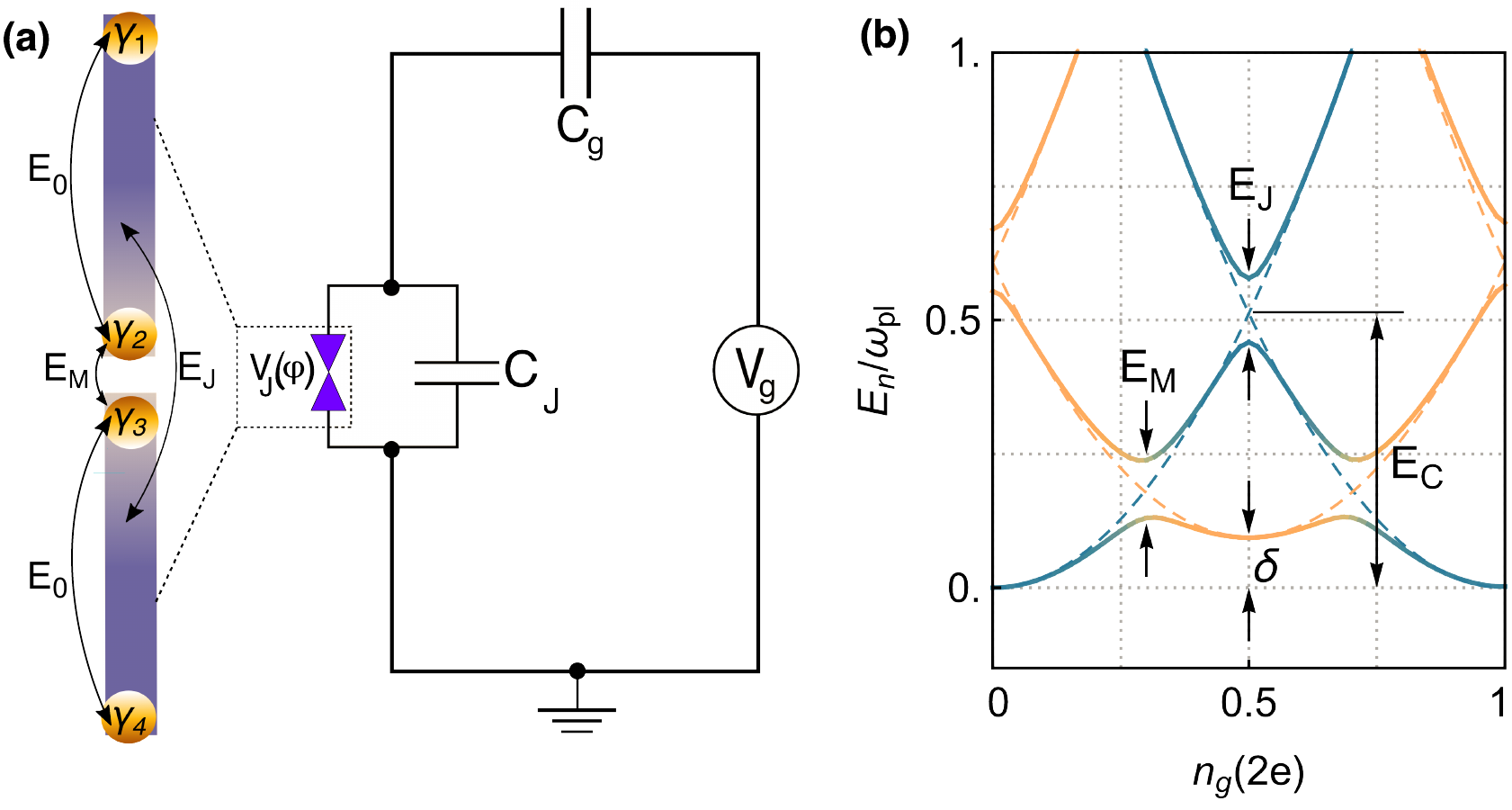}
\caption{\label{Fig1} \textbf{Sketch and spectrum of a NW-based superconducting island.} (a) Sketch of the NW-based superconducting island \editE{simplified} circuit. All the microscopic details of the NW junction (bowtie shape) are encoded in the Josephson potential $V_J(\varphi)$, while the combination of a shunting capacitor $C_J$ and the gate capacitance $C_g$ define the charging energy $E_C=e^2/2(C_J+C_g)$. The dashed region is a blow-up showing the schematics of the NW junction with all the relevant energies involved in the problem \editRR{(the Josephson coupling $E_J$, the Majorana coupling $E_M$ and the Majorana splitting $E_0$)}. All these energy scales are calculated microscopically from the BdG Hamiltonian of the junction, modeled as two Lutchyn-Oreg segments coupled through a weak link. Orange circles with $\gamma_i$ represent MBSs. (b) Spectrum of the island in the charging regime \editEE{($E_J\ll E_C$)} and for $B\sim B_c$ \editEE{(where $B_c$ is the field after which the two segments of the NW become topological)}, showing all the competing energy scales in the problem. Blue/orange dashed curves denote even/odd parity Coulomb parabolas in the absence of tunneling coupling across the junction. A finite coupling generates both standard Josephson coupling (avoided crossings $\sim E_J$ between same-color parabolas with minima differing by two electron charges $2e$ in gate space) and Majorana coupling (avoided crossings $\sim E_M$ between different-color parabolas with minima differing by one electron charge $e$ in gate space). \editRR{The energy difference between the odd and even parity sectors $\delta$ (which is in turn given by the Majorana splitting on each segment, $\delta=2E_0$) changes with $B$ field}. When it becomes smaller than $E_C$ (as in the case shown here) the ground state of the island around $n_g=0.5$ becomes odd.}
\end{figure}

We focus here on a specific proposal where the weak link is based on a semiconducting NW which is proximitized by a superconductor in \editE{its left and right} regions, thus forming a superconductor-normal-superconductor (SNS) junction \cite{PhysRevLett.115.127001,PhysRevLett.115.127002,PhysRevB.97.060508,PhysRevLett.116.150505,PhysRevLett.120.100502,PhysRevB.99.085434,Goffman_2017,PhysRevX.9.011010}, \editE{see Fig. \ref{Fig1}}. \editE{For the purposes of this work, the two regions are viewed as two Josephson-coupled superconducting islands}. Interestingly, an intrinsic Rashba spin-orbit (SO) coupling in the NW combined with an external Zeeman field $B$ generates, for a small chemical potential $\mu$ in the NW, helical bands with spin-momentum locking similar to that of topological insulators \cite{Fu:PRL08}. As demonstrated by Lutchyn {\it et al} \cite{Lutchyn:PRL10} and Oreg {\it et al} \cite{Oreg:PRL10}, when proximitizing such helical bands with a standard $s$-wave superconductor, this system is a physical realization of the Kitaev model for one-dimensional $p$-wave superconductivity \cite{Kitaev:PU01}. Similar to the Kitaev model, these Lutchyn-Oreg wires \editP{possess phases with non-trivial electronic topology}. In particular, they can be driven into a topological superconductor phase when the external Zeeman field $B$ exceeds a critical value $B_c\equiv\sqrt{\Delta^2+\mu^2}$, where $\Delta$ is the superconducting pairing term induced in the semiconducting NW owing to proximity effect \cite{Lutchyn:PRL10,Oreg:PRL10}. In NWs with finite length $L_S$, this topological superconductor phase is characterized by Majorana bound states (MBSs) emerging in pairs, one at either end of the wire. One pair of Majorana states forms a non-local fermion. The occupation of two such fermions, like in e.g. a SNS junction with two topological NW segments, defines the elementary qubit in proposals of topological quantum computers \cite{Kitaev:PU01,Nayak:RMP08,Sarma:NQI15}.

The goal of this paper is to present a comprehensive study of \editE{the} \editP{Josephson-coupled} superconducting islands described by a generalization of Eq. \eqref{transmonH} \editP{that incorporates the dynamics of Majoranas in the junction if present. The resulting Hamiltonian, \editE{which we will present} in Eq. \eqref{Majorana-Transmon hamiltonian2}, is derived as a low-energy projection of the full microscopic Hamiltonian for the two coupled islands,}
\begin{eqnarray}
\label{transmonH-NW}
H&=&4E_C(\hat N-n_g)^2 + V_J(\hat \varphi)\\
V_J(\hat \varphi) &=& \frac{1}{2}\check {\bm{c}}^\dagger H_\textrm{BdG}(\hat \varphi) \check {\bm{c}}\nonumber,
\end{eqnarray}
\editP{where $\hat N$ is now the \emph{relative} Cooper-pair number operator and its conjugate $\hat \varphi$ is the island superconducting phase difference. $E_C=e^2/2C_\Sigma$ is the relative charging energy, written in terms of a total capacitance that we denote generically as $C_\Sigma=C_J+C_g$ (with $C_J$ and $C_g$ the shunting and gate capacitances, respectively), \editE{see schematic circuit in Fig. \ref{Fig1}}.  This charging energy results from a combination of on-site charging energies of each island and the mutual charging energy between the islands \cite{RevModPhys.75.1,PhysRevB.98.205403}. $V_J$ is the full, microscopic, non-interacting Bogoliubov--de Gennes (BdG) Hamiltonian of the junction modeled as a single-mode Lutchyn-Oreg NW, to be projected onto the relevant low-energy fermionic $\check{\bm{c}}$ subspace}. While some previous studies have been presented in the literature, they are mostly based on either low-energy effective toy models \cite{PhysRevB.98.205403,ginossar2014microwave,PhysRevB.92.075143} or partial microscopic descriptions~\cite{10.21468/SciPostPhys.7.4.050}. Our paper is, to the best of our knowledge, the first full microscopic study of such junctions covering the full range of parameters: from the CPB to the transmon regimes in the island as the ratio $E_J/E_C$ increases; and from the trivial to topological regimes in the NW as the $B$ Zeeman field increases above $B_c$.

Our discussion is based on the simplest model that covers all these relevant regimes: two segments of a single-mode semiconductor NW that are proximitized by a conventional $s$-wave superconductor separated by a short normal region, thus forming a SNS junction with a weak link of normal transparency $T_N$. The BdG spectrum of such weak link creates the Josephson potential $V_J(\varphi)$ that enters the superconducting island Hamiltonian in Eq. \eqref{transmonH-NW}. \editP{This model adds another important energy scale $E_M$ corresponding to the junction coupling between MBSs localized at either side of the weak link, that may or may not dominate over $E_J$, \editE{see Fig. \ref{Fig1}}. For the single-channel short junction case considered in this work, the Josephson coupling $E_J$ may easily be smaller than $E_M$, since $E_M\sim\sqrt{T_N}\Delta_T$, while $E_J\sim T_N\Delta$ \cite{Fu:PRB09,San-Jose:PRL14,Tiira:NC17,Cayao:PRB17}, with $T_N$ denoting} the normal transmission of the junction and $\Delta_T$ the so-called ``topological \editE{mini}gap" separating MBSs from the rest of quasiparticle excitations in the system, \editP{see Fig. \ref{Fig2}. Thus, the Majorana-mediated Josephson energy $E_M$ introduces another important ratio $E_M/E_C$ into the problem.}  \editRR{Finally, the spatial overlap between Majoranas belonging to the same proximitized portions of the NW also introduces a new energy scale $E_0$, \editP{representing their hybridization splitting}, which depends on microscopic parameters of the NW such as length, magnetic field, chemical potential, etc. The  energy difference between odd and even fermionic parities in the junction, $\delta=2E_0$, follows the same dependence.}
The interplay of all these energy scales gives rise to a rich variety of novel regimes and physical phenomena, well beyond that of standard superconducting islands, as we shall describe.

The paper is organized as follows. Section \ref{Junction} is devoted to various relevant aspects of the NW-based JJs. After presenting the BdG model for the junction in Sec. \ref{Model}, we discuss in Sec. \ref{fourMBS} the 
basic phenomenology of the subgap spectrum and the relevant energy scales of the NW junction, with emphasis on $E_M$ and $\delta$ which, as argued above, give rise to novel regimes not discussed before. Section \ref{island} is devoted to the complete superconducting island problem. In Sec. \ref{projection} we discuss our projection method that allows to simplify the full island Hamiltonian in Eq. \eqref{transmonH-NW} and keep only the relevant, subgap, degrees of freedom. Section \ref{tight} presents the island Hamiltonian in tight-binding form, with Sec. \ref{parameters} devoted to the dependence of the superconducting island parameters on the microscopic parameters of the NW. Section \ref{benchmark} sets the stage before presenting the main results of the paper. We include a benchmark of the method against well-known limits (e.g. Majorana island limit) in Sec. \ref{known} and a discussion about the Josephson inductance and anharmonicity of the junction in Sec. \ref{inductance}. The ratio $E_M/E_J$ is analyzed in Sec. \ref{ratioMajorana-Josephson}. We finally present the main results of our work regarding the MW spectroscopy of a NW-based superconducting island in Sec. \ref{MW}. Different regimes are characterized in detail. This includes the $E_M/E_C<1$ regime (Sec. \ref{MW1}) and the opposite regime of non-negligible $E_M/E_C$ ratios, Sec. \ref{MW2}, which is of relevance to the experiments with junctions in the few-channel NW regime. In Sec. \ref{CharliePAT} we also discuss a regime with $E_M/E_C\lesssim 1$ and $E_J\to 0$, which is relevant to the recent experimental observation of parity mixing in superconducting islands owing to zero modes \cite{CharliePAT}. \editRR{Section \ref{splitjunction} considers a split junction geometry, where the coupling to an ancillary JJ allows to study the phase-dispersing subgap levels of the NW junction, including the $4\pi$ effect in the topological regime.} Finally, we conclude the paper with some final remarks in Sec. \ref{remarks}.

We finish the introduction by mentioning that, in parallel to this work, we present a companion paper, Ref. \onlinecite{Avila-accompanying}, with emphasis on the transmon limit with $E_M/E_C> 1$ and the role that parity crossings have on the transmon MW spectrum owing to Majorana oscillations (oscillations of $\delta$ as a function of magnetic field).


\section{\label{Junction} NW-based Josephson junction}
\subsection{\label{Model} Model}
\editP{With full generality, the Josephson potential is given by the BdG Hamiltonian. In a short SNS NW we can write it as}
\begin{equation}
H_\textrm{BdG}(\varphi)=\begin{pmatrix}
H_{\rm{NW}}&\Delta(x,\varphi)\\
\Delta(x,\varphi)^\dagger&-H_{\rm{NW}}^*
\end{pmatrix},
\end{equation}
where \editP{$H_{\rm{NW}} = H_L + H_R + V_{T_N}$ is the normal NW Hamiltonian. $H_{\rm{NW}}$ consists of the Hamiltonians for the two (left/right) segments $H_{L/R}$, coupled across a short weak link of transparency $T_N\in[0,1]$ by a $V_{T_N}$}. Each segment contains all the microscopic NW details (Rashba spin-orbit coupling $\alpha_{\rm{SO}}$, Zeeman field $B$ and chemical potential $\mu$) and is described by a single-band Lutchyn-Oreg model \cite{Lutchyn:PRL10,Oreg:PRL10}
\begin{equation}
\label{Lutchyn-Oreg}
\hat H_{L/R}=\frac{\hat p_x^2}{2m}-\mu-\frac{\alpha_{\rm{SO}}}{\hbar}\hat\sigma_y \hat{p}_x+B\hat\sigma_x ,
\end{equation}
with $\hat p_x=-i\hbar\partial_{\hat x}$ the momentum operator and $\hat\sigma_i$ Pauli matrices in spin space. $\hat\Delta(x,\varphi)=i\hat\sigma_y\Delta e^{\pm i\hat\varphi/2}$ (\editP{where the $\pm$ corresponds to $x\in L/R$, respectively}) is the induced pairing term. The discretized version of the above model reads:

\begin{figure}
\centering \includegraphics[width=\columnwidth]{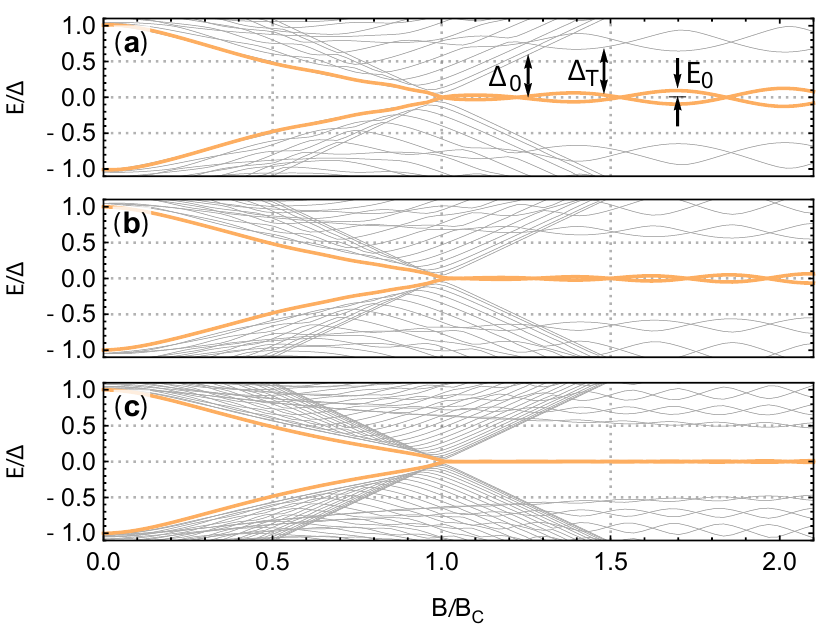}
\caption{\label{Fig2} \textbf{BdG spectrum as a function of Zeeman field $B$ and Majorana oscillations.} %
BdG spectrum of NWs of increasing lengths $L_S=2.2\mu m$ (a), $L_S=3\mu m$ (b) and $L_S=5\mu m$ (c), as a function of the ratio $B/B_c$. For $B>B_c$ the lowest mode (orange line), corresponding to weakly overlapping MBSs, \editRR{shows clear oscillations of amplitude $E_0$ around zero energy} in (a) and (b). These oscillations become progressively reduced as $L_S$ increases, (c). For $B\gg B_c$ the Majorana mode around zero energy is separated from the quasi-continuum formed by the rest of BdG excitations (grey lines) by a so-called topological gap $\Delta_T$, which varies with the $B$ field and depends on microscopic parameters of the wire ($\mu$ and $\alpha_{\rm{SO}}$). Near $B_c$ the relevant gap is the one that closes and reopens at the topological transition (the zero momentum gap $\Delta_0$).} 
\end{figure}
\begin{equation}\label{wiretb}
H_\textrm{BdG} = \editP{\frac{1}{2}}\sum_i \check c_i^\dagger h_i \check c_i + \editP{\frac{1}{2}}\sum_{\langle ij\rangle } \check c_i^\dagger v_{ij} \check c_j,
\end{equation}
where \editP{$\check c_i = (c_{i\uparrow},c_{i\downarrow},c^\dagger_{i\uparrow},c^\dagger_{i\downarrow})$ are Nambu spinors in spin ($\sigma_i$) and particle-hole ($\zeta_i$) sectors,} $\langle ij\rangle $ means sum over nearest neighbours, and $h_i$, $v_{ij}$ are onsite and hopping parts of the Hamiltonian:
\begin{equation}
\begin{split}
h_i&= (2t-\mu)\zeta_z\sigma_0+B\zeta_z\sigma_x+\Delta(\varphi)\zeta_y\sigma_y,\\
v_{ij}&= -t_{ij}\zeta_z\sigma_0-i\frac{\alpha_{\rm{SO}}}{2a}\zeta_z\sigma_y,\\
t_{ij}&=\begin{cases}
t & \textrm{within the same region}\\
\tau t & \textrm{at interface}
\end{cases}.
\end{split}
\end{equation}

\editP{The tight-binding hopping parameter above is $t=\hbar^2/2ma_0^2$, where $m=0.015m_e$ in InSb NWs,
$m_e$ is the electron's mass and $a_0$ a lattice discretization parameter. The finite weak link transparency $T_N$ is modeled through a renormalization of $t$ by a ``transparency factor'' $\tau\in[0,1]$ that is monotonous (although non-linear) with $T_N$ \cite{Cayao:PRB17}}.
Some other material parameters are fixed according to typical experimental values: $\mu=0.5$meV, $\Delta=0.25$meV, $\alpha_{\rm{SO}}=20\,\mathrm{meV\AA}$.  For simplicity, in what follows both NW segments are assumed to be equal and of the same length $L_S$.
\subsection{\label{fourMBS} Subgap spectrum and relevant energy scales}
To set the stage and for the sake of completeness, we now discuss some well-known results about the subgap spectrum of both single NWs and NW SNS junctions (for a recent review, see  Ref. \onlinecite{prada2019}). The aim of this subsection is to provide some estimates (notably of $E_M$), based on the single-band NW model in Eq. \eqref{Lutchyn-Oreg}, which will be of relevance to our superconducting island results in the next subsections. As we already mentioned, a proximitized NW undergoes a topological phase transition at $B_c\equiv\sqrt{\Delta^2+\mu^2}$ \editP{simultaneous} with the appearance of MBSs at their edges \cite{Lutchyn:PRL10,Oreg:PRL10}, \editE{see sketch in Fig. \ref{Fig1}}. This is illustrated in Fig. \ref{Fig2}, which plots the BdG spectrum different NWs for increasing lengths $L_S=2.2\mu $m (a), $L_S=3\mu $m (b) and $L_S=5\mu $m (c), as a function of the ratio $B/B_c$. As the magnetic field increases, the \editP{energy of the lowest mode (orange line) decreases} until it reaches zero at $B\sim B_c$ (\editP{with finite-$L_S$ corrections}). This smooth cross-over is the finite-size version of the predicted $L_S\to\infty$ topological transition at \editP{exactly} $B=B_c$, where the Zeeman-dominated gap at zero momentum $\Delta_{p_x=0}\equiv\Delta_0$ closes and reopens again \cite{Lutchyn:PRL10,Oreg:PRL10}. For $B>B_c$ \editP{and finite $L_S$, the lowest energy mode \editP{is the superposition of two weakly overlapping MBSs, which endows it} with a finite energy $E_0$ due to Majorana hybridization. This Majorana splitting} is of order $E_0\sim \frac{\hbar^2p_F}{m\xi_M} e^{-2L_S/\xi_M}\cos(p_FL_S)$ \cite{Das-Sarma:PRB12}, where $p_F$ is the Fermi momentum (that grows with $\mu$ and/or $B$), $\xi_M=\hbar v_F/\Delta$ is the Majorana superconducting coherence length, and $v_F$ denotes the Fermi velocity. For \editP{small-to-moderate} magnetic fields \editE{and small chemical potentials}, the Fermi velocity is well approximated by $v_F\sim\alpha_{\rm{SO}}/\hbar$ and thus  $\xi_M\sim\alpha_{\rm{SO}}/\Delta$.  For larger magnetic fields, the Majorana length acquires a prefactor that depends on the ratio between the Zeeman energy and the SO energy $E_{\rm{SO}}=m\alpha_{\rm{SO}}^2/2\hbar^2$, \editP{resulting in} a parametric dependence $\xi_M\sim 2(B/\Delta) l_{\rm{SO}}$ \cite{Mishmash:PRB16}, with the SO length given by \editE{$l_{\rm{SO}}=\hbar^2/(m\alpha_{\rm{SO}})$}. As it becomes evident from this discussion, the energy splitting $E_0$ of Majoranas has a rich dependence on the microscopic details of the NW and, importantly, \editP{oscillates around zero with an amplitude} that grows with $B$ \cite{PhysRevB.86.121103,Prada:PRB12, Das-Sarma:PRB12,Mishmash:PRB16,Rainis:PRB13}, see e.g Fig. \ref{Fig2}(a). \editP{For $B$ fields sufficiently above $B_c$, the near-zero Majorana mode becomes separated from the rest of excitations by a topological gap given by the superconducting pairing term at the Fermi momentum $\Delta_{p_x=p_F}\equiv\Delta_T$). The topological gap $\Delta_T$, the zero-momentum $\Delta_0$ and the Majorana splitting $E_0$ are all are marked by} arrows in  Fig. \ref{Fig2}. \editJ{For asymptotically infinite wires ($L_S\to\infty$) and }\editP{at $\mu=0$ we can write $\Delta_T$ analytically} as~\editJ{\cite{Cayao:PRB17}}:
\begin{equation}
\Delta_T=\frac{2\Delta E_{\rm{SO}}}{\sqrt{E_{\rm{SO}}(2E_{\rm{SO}}+\sqrt{B^2+4E^2_{\rm{SO}}})}}.
\end{equation}
It has a maximum value $\Delta_T=\Delta$ in the SO-dominated limit $E_{\rm{SO}}\gg B$ and can be much smaller in the opposite Zeeman-dominated limit $B\gg E_{\rm{SO}}$, where it decreases with $B$ as $\Delta_T\sim 2\Delta\sqrt{E_{\rm{SO}}/B}$ (of order $\sim 2\sqrt{\Delta E_{\rm{SO}}}$ near the \editP{$\mu=0$} critical Zeeman field $B_c\approx \Delta$). \editP{The value of the topological gap $\Delta_T$ is particularly important for this paper since it governs the Majorana Josephson coupling term $E_M$. Its dependence on microscopic parameters is therefore very relevant in superconducting island qubits} based on topological NWs with MBSs, \editP{and will be discussed in detail throughout} this paper.

 \begin{figure}
\centering \includegraphics[width=\columnwidth]{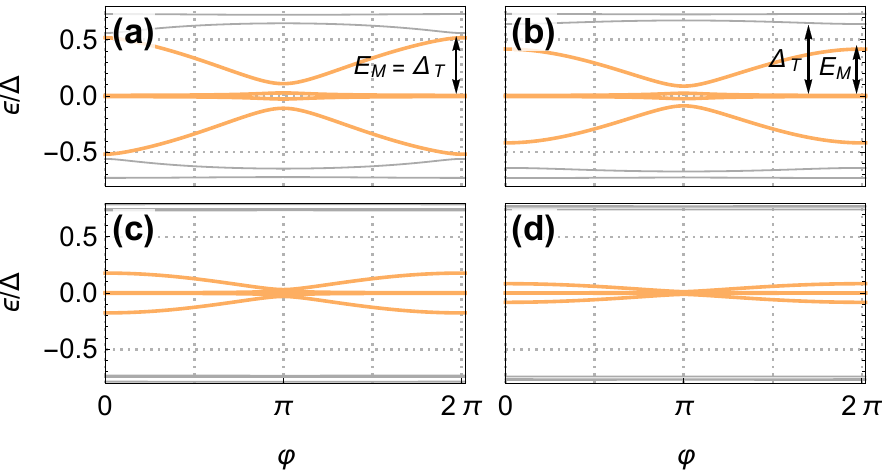}
\caption{\label{Fig3} \textbf{Low-energy Andreev spectrum as a function of $\varphi$.} %
Different phase-dependent low-energy spectra of a junction where each of the two segments is a NW like the one shown in Fig. \ref{Fig1}(a). Each panel shows decreasing $\tau=1, 0.8, 0.4, 0.2$, from (a) to (d). The Zeeman field is fixed at $B/B_c\approx 1.84$, corresponding to one of the minima of the Majorana oscillations. The two lowest modes (orange lines) are originated from the overlap of four MBSs, as discussed in the main text. For $\tau=1$, they touch the quasi-continuum formed by the rest of levels (grey lines) at the so-called topological gap $\Delta^T$, which provides an upper bound for $E_M$ (see the main text). For $\tau\neq 1$, the Majorana modes detach from this quasicontinuum of levels.}
\end{figure}
 \begin{figure}
\centering \includegraphics[width=\columnwidth]{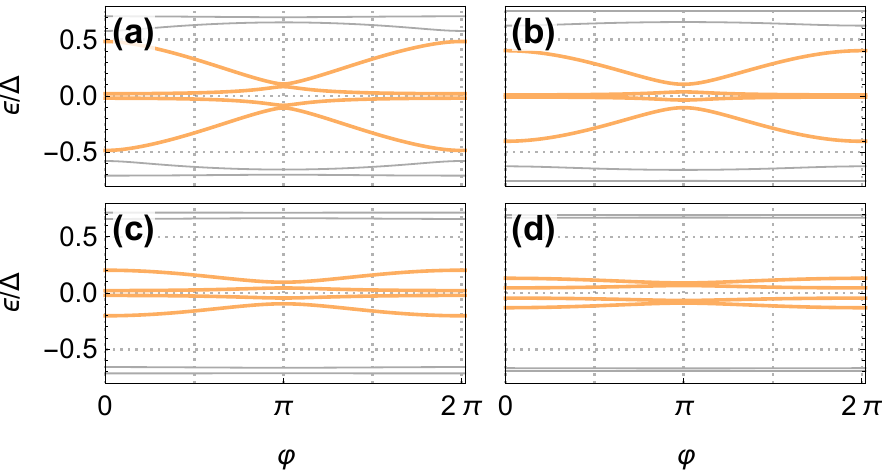}
\caption{\label{Fig4} \textbf{Low-energy Andreev spectrum as a function of $\varphi$.} %
Same as Fig.~\ref{Fig3} but with the Zeeman field fixed at $B/B_c=2$, corresponding to one of the maxima of the Majorana oscillations.}
\end{figure}

In Figs. \ref{Fig3}  and \ref{Fig4} we show various examples of typical $\varphi$-dispersing subgap spectra of a short NW SNS \editP{junction in the topological $B>B_c$ regime}. We focus in particular on how the subgap spectrum (orange lines) of the SNS junction changes for decreasing \editP{transparency factor} $\tau$. Quite generically, \editP{the short-junction} subgap spectrum can be \editP{expressed as an effective model of four Majorana operators, $\gamma_{1,2}\in L$ and  $\gamma_{3,4}\in R$, of the form}\cite{PhysRevLett.108.257001,PhysRevB.86.140504}:
 \begin{eqnarray}
\label{subgap}
&\hat H_\textrm{BdG}^{sub}(\varphi)=i\lambda_{12}\gamma_1\gamma_2+i\lambda_{13}(\varphi)\gamma_1\gamma_3%
+i\lambda_{14}(\varphi)\gamma_1\gamma_4\nonumber\\
&+i\lambda_{23}(\varphi)\gamma_2\gamma_3+%
i\lambda_{24}(\varphi)\gamma_2\gamma_4+i\lambda_{34}\gamma_3\gamma_4.%
\end{eqnarray}
The \editE{four} \editP{finite-energy subgap eigenstates (Andreev bound states) are different, $\varphi$-dependent superpositions of the four Majorana states, which couple pairwise through the $\lambda_{ij}$ terms.} \editE{Note that the four subgap energy levels correspond to \editP{empty/full occupations} of two independent fermions, one per proximitized wire,} \editP{as is standard in the BdG description. The Majorana basis $\gamma_i$ for said subgap states can be used for any value of $B$ or $L_S$, although in the case of long, decoupled topological NW segments ($B>B_c$, $L_S\rightarrow\infty$ and $\tau=0$) all $\lambda_{ij}$ become zero, making the four Majoranas zero energy topological eigenstates, each located at one end of the two NW segments. At finite transparency ($\tau>0$) the two ``inner'' Majoranas at either side of the junction ($\gamma_2$ and $\gamma_3$) hybridize to a finite energy 
\begin{equation}
\lambda_{23}(\varphi) = E_M \cos(\varphi/2),
\end{equation}
save at $\varphi=\pi$ where $\lambda_{23}=0$ \cite{PhysRevLett.108.257001,PhysRevB.86.140504}, while the ``outer'' Majoranas ($\gamma_1$ and $\gamma_4$) remain decoupled from the junction due to the large $L_S$ (only $\lambda_{23}\neq 0$). In this limit, the $\varphi=\pi$ crossing of the inner Majoranas gives rise to the so-called $4\pi$-periodic Josephson effect. This anomalous Josephson is destroyed by finite $L_S$ corrections, however, due to a lifting of the $\varphi=\pi$ crossing by the remaining $\lambda$ terms, as shown in Figs. \ref{Fig3} and \ref{Fig4}.}

The value of the inner Majorana coupling $E_M$ can be estimated from the above plots to be in the approximate range of a few tenths of $\Delta$ (depending on the tunneling coupling). The upper bound for $E_M$ is \editP{reached in} the transparent limit ($\tau=1$),
where \editP{$\lambda_{23}(\varphi = 2\pi m) = E_M$} touches the quasi-continuum formed by the rest of BdG levels (gray lines) \cite{San-Jose:NJP13,Cayao:Belstein18}, see Fig. \ref{Fig3}(a). For $B\gg B_c$, this happens at the topological gap, namely $E_M=\Delta_T$ (for the particular microscopic parameters of this plot $\Delta_T\sim 0.5\Delta$). For $\tau\neq 0$, the inner Majorana coupling is always $E_M<\Delta_T$, Fig. \ref{Fig3}(b), and can be approximated as $E_M\sim \sqrt{T_N}\Delta_T$, with $T_N$ the normal transmission of the junction. This reflects the fact that, at high $B$ fields, and still neglecting the role of the outer Majorana modes, the physics governing the NW low-energy subgap spectrum is that of a (single) proximitized helical channel, as described by the Fu-Kane model for a quantum spin Hall edge \cite{Fu:PRB09}. By considering the critical current supported by a single channel $I_c^0=eT_N\Delta/\hbar$ (at $B=0$ with spin degeneracy), and comparing it with the critical current resulting from the $4\pi$-periodic Majorana Josephson effect $I_c^M=e\sqrt{T_N}\Delta_T/2\hbar$\cite{Fu:PRB09,San-Jose:PRL14,Tiira:NC17,Cayao:PRB17}, the ratio between both couplings can be estimated \editP{as 
\begin{equation}\label{EMEJ}
E_M/E_J\sim \eta \Delta_T/\Delta\sim2\eta\sqrt{E_\mathrm{SO}/B} +\mathcal{O}(\mu),
\end{equation}
with} a prefactor $\eta\equiv\sqrt{T_N}/T_N>1$.
We will come back with more precise estimations of the ratio $E_M/E_J$ in Sec. \ref{ratioMajorana-Josephson}.

\section{\label{island} NW-based superconducting islands}
\subsection{\label{projection}Effective low-energy model and projection}
Our \editP{first} goal is to \editP{derive a quantitative but simple low-energy description of a short SNS NW junction that extends Eq. \eqref{transmonH-NW} by taking into account both} standard Josephson events due to Cooper pair tunneling, as well as anomalous Majorana-mediated events where a single electron is transferred across the junction. In order to do this, it is convenient to distinguish two contributions, \editP{$V_J = V_J^{bulk} + \hat H_\textrm{BdG}^{sub}$}. The first one takes into account \editRR{the bulk contribution of the BdG levels \emph{above} the gap to the ground state energy. We write this contribution as \footnote{While strictly correct, the definition of $V_J^{bulk}(\varphi)$ given in Eq. \eqref{VJbulk} is non-optimal (since finding the correct lowest states of the NW spectrum as a function of $\varphi$ for all magnetic fields can be quite a cumbersome task). \editP{Instead, we just compute it by subtracting the low energy sector ($\hat{\mathcal{H}}$ in Eq. \eqref{eq:lowenergy} from the total spectrum sum $\sum \epsilon_p$)}.} 
\begin{equation}\label{VJbulk}
V_J^{bulk}(\varphi)=-\sum_{\epsilon_p>\Delta}\epsilon_p(\varphi).
\end{equation}}
The second contribution corresponds to the subgap sector. As in the preceding subsection, we assume there are only two \editE{independent} spin-resolved \editP{fermionic subgap} states (short junction). \editP{Unlike for the states above the gap, we do not make further assumptions about them 
and instead treat their dynamics as fully coherent, governed by the $\hat H_\textrm{BdG}^{sub}(\varphi)$ Hamiltonian introduced in Eq. \eqref{subgap}. By extending $V_J$ with a contribution $\hat H_\textrm{BdG}^{sub}$ in this way, we are supplementing our relevant quantum degrees of freedom $N,\varphi$ with the $\gamma_i$ Majorana operators. The challenge remains of relating $\hat H_\textrm{BdG}^{sub}(\varphi)$ to the microscopic Hamiltonian $H_\mathrm{BdG}$ by projecting the latter onto the low-energy subspace of Majorana operators. The procedure, described below \cite{PhysRevLett.108.257001,PhysRevB.86.140504}, starts by defining the basis of left and right low-energy fermions $c_{L/R}$ and $c^\dagger_{L/R}$ of the decoupled NWs. These states are a basis to the four lowest BdG eigenstates of the microscopic $H_\textrm{BdG}$ with $\tau=0$, and are related to the $\gamma_i$ operators by a simple rotation}
\begin{equation}
	\begin{split}\label{eq:majorana_to_fermion}
\sqrt{2}\gamma_1&=c_L+c^\dagger_L,\quad%
\sqrt{2}\gamma_2=i(c_L-c^\dagger_L),\\
\sqrt{2}\gamma_3&=c_R+c^\dagger_R,\quad%
\sqrt{2}\gamma_4=i(c_R-c^\dagger_R).\quad%
\end{split}
\end{equation}
\editP{In terms of these operators, the fermion numbers on each segment are simply $\hat n_L=c_L^\dagger c_L=(1+i\gamma_1\gamma_2)/2$ and $\hat n_R=c^\dagger_Rc_R=(1+i\gamma_3\gamma_4)/2$. Next, we integrate out all states outside this low-energy decoupled subspace. This is done by computing the matrix elements of the resolvent of $H_\textrm{BdG}$, $G(\omega)=(\omega+i\varepsilon-H_\textrm{BdG})^{-1}$ at $\omega = 0$ on the $\psi^{0}=(c_L, c^\dagger_L, c_R, c^\dagger_R)$ state basis. This defines a $4\times 4$ matrix, whose inverse is the matrix $\mathcal{H}_{ij}$ of the $H_\textrm{BdG}$ projection we are after, 
\begin{eqnarray}
	(\mathcal{H}^{-1})_{ij}&=&\langle \psi^0_i|G(\omega=0)|\psi^0_j\rangle ,\\
	\hat{\mathcal{H}}&=&\frac{1}{2}\sum_{ij}\psi^{0\dagger}_i\mathcal{H}_{ij}\psi^{0}_j.
\label{eq:lowenergy}
\end{eqnarray}
Finally, we identify the above $\hat{\mathcal{H}}$ as the $\hat H_\textrm{BdG}^{sub}$ in Eq. (\ref{subgap}), from which we extract the dependence of $\lambda_{ij}(\varphi)$ on all microscopic parameter in $H_\textrm{BdG}$ using Eq. \eqref{eq:majorana_to_fermion}. This identification is an approximation, although we have checked that it is  a very accurate one in practice. \editRR{Also, it is important to mention that, strictly speaking, the separation between bulk and subgap contributions is only well defined if the subgap modes are well detached from the quasicontinuum, $E_M < \Delta_T$, namely $\tau\neq 1$. Also, it can break down right at the transition point $B\approx B_c$ when the gap closing may prevent from having a good separation between low and high energy BdG excitations. In all our calculations,  we make sure that the transparencies are such that subgap and quasicontinuum contributions can be clearly distinguished. Also, finite size effects (finite $L_S$) allow to have a controlled separation between subgap and  quasicontinuum degrees of freedom near $B\approx B_c$.}
} 

\editP{We can now write the matrix elements of $V_J = V_J^{bulk} + \hat H_\textrm{BdG}^{sub}$ in the parity basis $|n_L\,n_R\rangle$. 
The effective Josephson coupling reads:
\small
\begin{eqnarray}
\label{Majorana-Transmon hamiltonian1}
&& V_J(\varphi)=\\
&&\begin{pmatrix}V_J^{bulk}(\varphi)+\langle 00|\hat H^{sub}_\textrm{BdG}(\varphi)|00\rangle  & \langle 00|\hat H^{sub}_\textrm{BdG}(\varphi)|11\rangle \\
	\langle 11|\hat H^{sub}_\textrm{BdG}(\varphi)|00\rangle  & V_J^{bulk}(\varphi)+\langle 11|\hat H^{sub}_\textrm{BdG}(\varphi)|11\rangle \end{pmatrix}.\nonumber
\end{eqnarray}
\normalsize
}
\editP{The final low-energy Hamiltonian \editE{is thus a generalization} of Eq. \eqref{transmonH} \editE{to} a $2\times 2$ operator with the above $V_J$}
\begin{equation}
\label{Majorana-Transmon hamiltonian2}
	\hat{H}=[4E_C\left(-i\partial_{\varphi}-n_g\right)^2]\mathbb{1}+V_J(\varphi).
	\end{equation}
	
\editP{The above result vastly reduces the complexity of the original Eq. \eqref{transmonH-NW}, making it much easier to work with.} Note that the diagonal part of Eq. \eqref{Majorana-Transmon hamiltonian1} \editP{describes two copies with different $n_{L/R}$ parity} of a standard superconducting island, while the off-diagonal subgap contribution \editP{$\langle 00|\hat H^{sub}_\textrm{BdG}(\varphi)|11\rangle \sim E_M$} mixes them. We emphasize that, despite the superficial similarity with the effective low-energy model in Refs. \onlinecite{ginossar2014microwave,PhysRevB.92.075143}, \editP{$V_J$ in Eq. \eqref{Majorana-Transmon hamiltonian1} is obtained} by projecting the fully microscopic $H_\textrm{BdG}$.

\editRR{The eigenstates of Eq. \eqref{Majorana-Transmon hamiltonian2} are defined as a two-component spinor $\Psi_k=(f_k(\varphi),\,g_k(\varphi))^T$, owing to the pseudospin structure in the parity basis. The even/odd fermionic parity of each of the spinor's components translates into periodic/antiperiodic boundary conditions in phase space: $f(\varphi+2\pi)=f(\varphi)$ and $g(\varphi+2\pi)=-g(\varphi)$.
To make the Hamiltonian fully periodic, it is rotated according to $ H(\varphi)\rightarrow UH(\varphi)U^\dagger$, with $U=\textrm{diag}(1,e^{i\varphi/2})$. For a detailed discussion about the boundary conditions, see Refs. \onlinecite{ginossar2014microwave,10.21468/SciPostPhys.7.4.050}. }The fermionic $n_{L/R}$ parity content can be calculated by projecting each eigenstate onto the parity axis defined by $\hat\tau_z\equiv|00\rangle\langle 00|-|11\rangle\langle 11|$. Henceforth, we plot \editP{energy levels with a well-defined even/odd parity using blue/orange lines, while mixed parities are encoded using gradient colors between blue and orange, with a light-green midpoint color denoting a 50\% parity mixture.}

\begin{figure} 
\centering \includegraphics[width=\columnwidth]{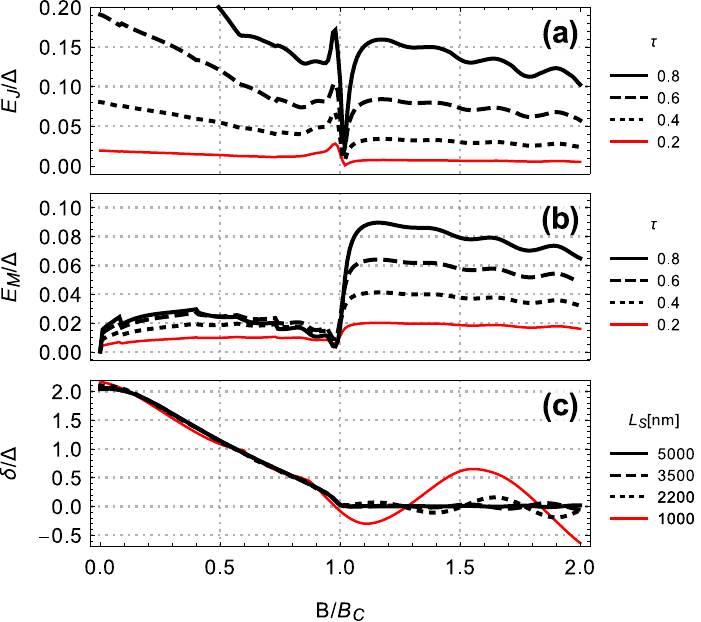}
\caption{\label{Fig5} \textbf{Dependence of the superconducting energy scales on $B$}. Apart from $E_C$, all the energy scales entering the superconducting-island Hamiltonian in Eq. \eqref{Majorana-Transmon hamiltonian2} are sensitive to various NW properties (including its topological transition/gap closing and reopening, the emergence of MBSs and their characteristic oscillatory pattern with $B$ due to finite overlap). The overall qualitative dependence of $E_J$ and $E_M$ on $B$  \editE{varies significantly with junction transparency \editP{factor} $\tau$ but} does not change significantly with $L_S$ and hence it is not shown (\editE{the opposite} holds for $\delta$ which is $\tau$-independent). Parameters (if not specified): $L_S=2.2\mu$m, $\tau=0.8$.}
\end{figure} 
 
 \subsection{\label{tight}NW-based superconducting-island model in tight-binding form}
\editP{Next} we want to solve the superconducting island Hamiltonian of Eq. \eqref{Majorana-Transmon hamiltonian2}.
This is accomplished by discretizing the phase space as
$\varphi_j=2\pi j/\ell^\varphi$, $j=1,2,...,\ell^\varphi$.
In so doing, the Hamiltonian acquires a tight-binding form, where the discretized phase may be seen as a set of sites arranged into a circular chain.
\editP{This discretization defines a finite} fermionic Hilbert space and operators $b_i^{(\dagger)}$ whose action on the ground state is defined as $b^\dagger_i|0\rangle =\Psi(\varphi_i)$, where $\Psi(\varphi)$ is the Hamiltonian eigenstate at phase $\varphi$.
The derivative \editP{$N=-i\partial_\varphi$} translates in this language into the usual hopping term $-i\partial_\varphi=-i(b^\dagger_{i+1}-b^\dagger_{i-1})b_i/(2a_\varphi)$, where $a_\varphi=2\sin(\pi/\ell^\varphi)$ is the phase lattice constant.
Using this tight-binding language, the Hamiltonian \eqref{Majorana-Transmon hamiltonian2} reads 
\begin{equation}
\begin{split}
 H(\varphi)&=\sum_i b_i^\dagger h^\varphi_i b_i + \sum_{\langle ij\rangle } b_i^\dagger v^\varphi_{ij} b_j,\\
	h_i^\varphi&= 4E_C(2a_\varphi^{-2}+n_g^2)+V_J(\varphi_i),\\
v_{ij}^\varphi&= 4E_C\left[\textrm{sgn}(j-i)\,\textrm{i}\,n_ga_\varphi^{-1}-a_\varphi^{-2}\right].
\end{split}
\end{equation}
Each site element $h_i^\varphi$, $v_{ij}^\varphi$ is a $2\times 2$ matrix, owing to the pseudospin structure from even-odd projection, \editP{Eq. \eqref{Majorana-Transmon hamiltonian1}}. This tight-binding model is numerically solved by means of the MathQ software~\cite{MathQ}.

\subsection{\label{parameters}Dependence of the superconducting-island parameters on microscopic parameters of the NWs}
The NW microscopic details enter this problem through the effective Josephson potential $V_J(\varphi)$. In particular, the three relevant NW energy scales that govern the superconducting island Hamiltonian in Eq. \eqref{Majorana-Transmon hamiltonian2}  (the Josephson coupling $E_J$, the energy difference between odd and even fermionic parities $\delta$, and the single-electron contribution to the Josephson coupling $E_M$) can be defined in terms of the projected Hamiltonian as:
\begin{align}
E_J &= \int_0^{2\pi}\frac{\textrm{d}\varphi}{\pi}\left[ V^{bulk}_J(\varphi)+\langle 00|\hat H^{sub}_\textrm{BdG}(\varphi)|00\rangle \right]\cos(\varphi),\nonumber\\
\delta &=  \langle 11|\hat H^{sub}_\textrm{BdG}(\varphi=0)|11\rangle - \langle 00|\hat H^{sub}_\textrm{BdG}(\varphi=0)|00\rangle ,\nonumber\\\
E_M &= \int_0^{2\pi}\frac{\textrm{d}\varphi}{\pi}\, \langle 00|\hat H^{sub}_\textrm{BdG}(\varphi)|11\rangle \,\cos(\varphi).
\end{align}
All these parameters depend on relevant quantities such as e.g. NW length and magnetic field. As defined above, $E_J$ refers to the \editP{energy contribution associated to Cooper-pair transfers across the junction, and hence to the critical current of the system.
$E_M$ on the other hand accounts for single-quasiparticle transfer through the subgap states of the spectrum}, either Andreev states (trivial) or Majorana states (topological phase). $\delta$ is the \editP{minimal energy cost for exciting one quasiparticle on each NW segment} above the ground state.  Importantly, the effective models in Refs. \onlinecite{ginossar2014microwave,PhysRevB.92.075143} \editP{assume a simplified} Josephson term of the form 
\begin{eqnarray}
\label{Majorana-Transmon hamiltonian-eff}
&&	V_J(\varphi)=
\begin{pmatrix}-E_J\cos(\varphi) & E_M\cos(\varphi/2)\\
	E_M\cos(\varphi/2) & -E_J\cos(\varphi)\end{pmatrix},
\end{eqnarray}
which is not able to capture \editP{the full $\varphi$-anharmonicity, or the various parameter regimes and their associated} phenomenology inherent to the microscopic description \editP{employed} here. \editP{This includes the trivial (Andreev) regime, the topological (Majorana) regime and the crossover/transition between the two with $B$ field. Our approach also yields} the detailed dependence of the junction $\delta$, $E_M$ and $E_J$ on various NW parameters (such as e.g. $\alpha_{\rm{SO}}$) and transparency of the junction, see Sec. \ref{ratioMajorana-Josephson}. These three quantities are plotted in Fig.~\ref{Fig5} as a function of $B$ before and after the topological transition. They inherit some important features of the NW behavior for finite $B$ fields. These include the closing and reopening of the gap and the characteristic oscillatory pattern due to finite-length splitting of Majorana excitations. \editRR{Right at the transition point $B\approx B_c$, a good numerical accuracy can only be obtained if there is a well defined separation between bulk and subgap contributions. For the typical parameters that we use in this work, we have checked that the projection starts to break down for $\tau\gtrsim 0.9$.}
\begin{figure}
\centering \includegraphics[width=\columnwidth]{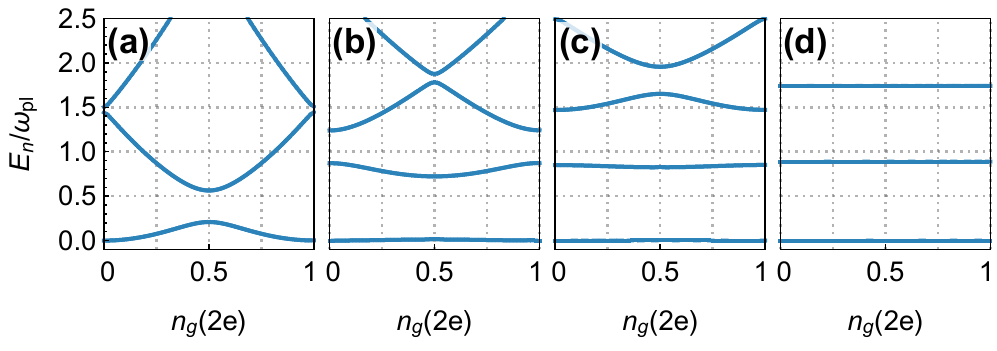}
\caption{\label{Fig6} \textbf{Spectrum of a SIS junction with no magnetic field.} %
\editE{For $B=0$ and $\tau\rightarrow 0$ we recover the standard SIS tunnel junction}. These four cases correspond to increasing $E_J/E_C$ values \editE{in the transmon regime},  $E_J/E_C=1,5,10,50$. %
$\omega_{pl}=\sqrt{8E_JE_C}/\hbar$ defines the plasma frequency. %
Parameters for each NW segment like in Fig. \ref{Fig2}(a).}
\end{figure}
\begin{figure}
\includegraphics[width=\columnwidth]{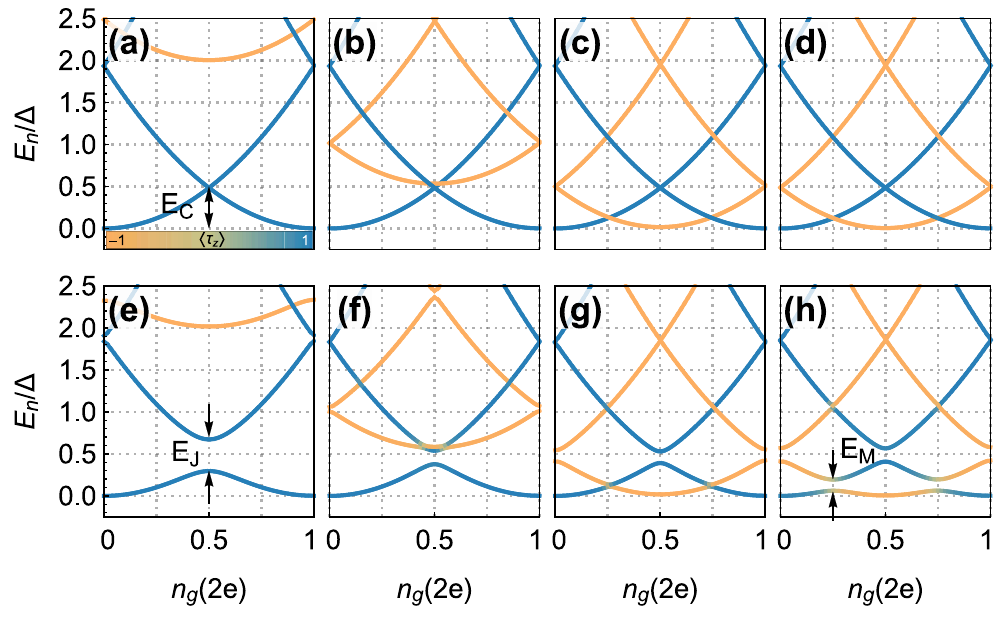}
\caption{\label{Fig7} \textbf{Spectrum of a NW-based superconducting island \editE{with magnetic field} (basic phenomenology).} %
\editE{The parity content of energy levels is calculated by projecting each eigenstate onto the parity axis defined by $\hat\tau_z\equiv|00\rangle\langle 00|-|11\rangle\langle 11|$ (see main text).} The even sector is represented by blue parabolas with minima at $n_g=m\in\mathbb Z$, while the odd sector is represented by orange parabolas with minima at $n_g=m+n_g^0$, $n_g^0=1/2$. 
(a-d) \editE{Coulomb} island: charging regime with  $E_C=0.5\Delta$ and $E_J/E_C = 10^{-4}$. Transparency factor $\tau=0.01$.
For zero magnetic field, odd parabolas are shifted in the energy axis by exactly $\delta=2\Delta$, panel (a). The spectrum is 2$e$-periodic.
This energy shift $\delta$ decreases for increasing $B$ fields, \editE{see panel (b) for $B=0.7B_{c}$, and  vanishes exactly at the topological transition at $B=B_c$, panel (c), where the spectrum becomes $e$-periodic. Panel (d) corresponds to $B=1.5B_{c}$.}
(e-h) Finite Josephson and Majorana coupling: same as top panels but with $\tau=0.8$. The finite Josephson coupling (here $E_J=0.8E_C$) results in avoided crossings between parabolas of the same parity. (h) For $B>B_c$ there is also a finite Majorana coupling that induces avoided crossings around $n_g=0.25$ and 0.75 between parabolas of opposite parity. Note that the Majorana-induced avoided crossing ($\sim E_M$) is non-negligible with respect to the maximum at $n_g=0.5$ ($\sim E_C$). Parameters for each NW segment like in Fig. \ref{Fig2}(c).}
\end{figure}
\section{\label{benchmark}Benchmark results}
\subsection{\label{known}Known limits}
As first checks of our procedure, we benchmark our method against well-known limits. This includes the standard superconducting island behavior in the $B\to 0$, $\tau\to 0$ limits, Fig. \ref{Fig6}. As expected, the charge dispersion $\partial E_n/\partial n_g$ \editP{of all energy levels $E_n$} gets progressively reduced by increasing the ratio $E_J/E_C$, and the island \editP{crosses over} from the CPB to the transmon regimes~\cite{koch2007charge}. \editP{The latter is characterized by a spectrum of a slightly anharmonic oscillator} with frequency given by the plasma frequency 
\begin{equation}
\omega_{pl}=\sqrt{8E_JE_C}/\hbar.
\end{equation}
Another important limit \editE{is} the \editE{Coulomb} island regime \editE{(sometimes called Majorana island \cite{Albrecht:N16})}, top panels in Fig. \ref{Fig7}, which we can reach by drastically reducing both the Josephson and Majorana couplings in the $\tau\to 0$ limit (i.e., for two isolated NWs with charging energy $E_C$). For $B=0$ (orange lines), odd parabolas are energy shifted from even ones (blue lines) by exactly $\delta=2\Delta$, panel (a). As the Zeeman field increases, $\delta$ becomes progressively reduced until it becomes of the order of $E_C$ (panel (b)) or smaller, which results in \editP{a transition from an even to} an odd ground state around \editP{half-integer $n_g=m+1/2$, with $m\in\mathbb Z$}. For $B=B_c$, \editE{panel (c)}, both parity sectors have minima at zero energy and the periodicity becomes $e$ \cite{Albrecht:N16,Shen2018}, as opposed to the $2e$-periodicity of the standard superconducting island at $B=0$. Increasing $\tau$ (lower panels), results in finite Josephson coupling which leads to avoided crossings between parabolas of the same parity \editP{due to $E_J$}. For $B>B_c$, panel (h), \editP{there appear} also avoided crossings between parabolas of the opposite parity owing to \editP{a finite Majorana coupling $E_M$.}

%
\begin{figure} 
\centering \includegraphics[width=\columnwidth]{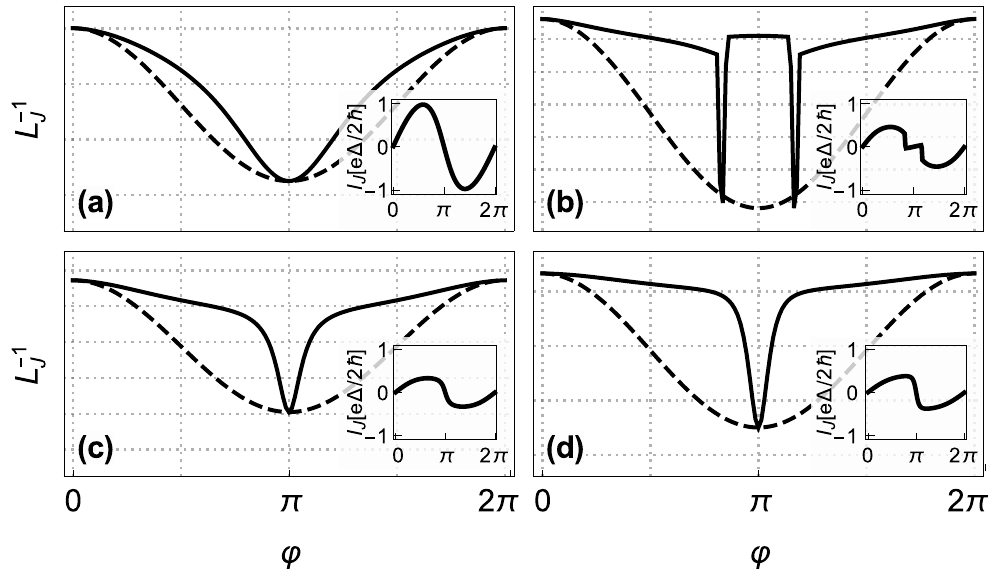}
\caption{\label{Fig8} \textbf{Deviation of SNS junctions from standard Josephson behavior.} Josephson inductance $L_J$ of SNS junctions (solid lines) provides an useful tool to measure deviation from idealized conditions ($L_J^{-1}\sim\cos\varphi$, dashed lines). Panels describe phase dependence of $L^{-1}_J$ for increasing magnetic fields. \editRRR{The insets on each panel show the corresponding $I_J(\varphi)$. The spikes in panel b) correspond to a Zeeman-induced $0-\pi$ transition in the junction}. Parameters: $L_S=2.2\mu$m, $\tau=0.8$, $B/B_c=0,0.8,1,1.2$ from (a) to (d).}
\end{figure}
\begin{figure} 
\centering \includegraphics[width=\columnwidth]{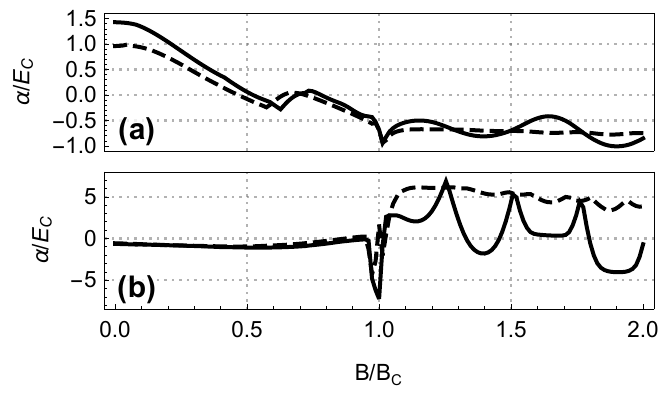}
\caption{\label{Fig9} \textbf{Anharmonicity $\alpha$ of a NW-based superconducting qubit at $n_g=0.5$.} Solid (dashed) curves show transmon dependence on $B$ for short (long) wires. Both CPB (a) and transmon limits (b) are displayed. Anharmonicity provides a precise smoking gun to detect topological transitions and Majorana oscillations, specially for the transmon limit. Parameters: $L_S=2.2\mu$m (a), $5\mu$m (b), $\tau=0.8$, $E_J/E_C=0.5,25$.}
\end{figure}

\subsection{\label{inductance}Josephson inductance and anharmonicity}
In a standard superconducting island, \editE{as mentioned in the introduction}, the SIS JJ is well described by an energy-phase relation of the form $V_J^{\rm{SIS}}(\varphi)=-E_J\cos(\varphi)$. Its corresponding inverse Josephson inductance reads $L_J^{-1}(\varphi)=(2e^2/\hbar)^2d^2V_J^{\rm{SIS}}(\varphi)/d\varphi^2=(2e^2/\hbar)^2E_J\cos(\varphi)$. The NW-based JJ that we discuss here strongly differs from this cosine form (which is only valid in the tunneling limit and in the absence of external magnetic fields, $\tau\rightarrow 0$, $B=0$). These deviations from a cosine form have relevant consequences when e.g. using superconducting islands as qubits. An important figure of merit is the anharmonicity parameter $\alpha\equiv E_{12}-E_{01}$, where $E_{mn}$ is the energy difference between $m$ and $n$ energy states, which controls the \editP{leakage rate} into noncomputational states (\editP{the high-energy states} out of the two-level qubit Hilbert space) \cite{PhysRevA.69.062320,PhysRevB.97.060508}.

Figure~\ref{Fig8} illustrates this by looking at the Josephson inductance $L_J$ for NWs with transparent links at several $B$ values, before and after the topological transition. Even for low magnetic fields, it becomes manifest that a standard, cosine-like SIS inductance \editP{(dashed line)} is by no means sufficient to study the qubit evolution with magnetic field and \editP{across} topological transitions in NW-based islands in the few-channel regime.

As for \editP{the anharmonicity parameter} $\alpha$, Fig.~\ref{Fig9} shows that such parameter strongly depends on the external magnetic field for both the CPB [panel (a)] and transmon [panel (b)] limits. The NW's topological regime is mostly evident in the transmon limit. Here, the anharmonicity remains approximately constant \editP{throughout} the trivial regime, \editP{pinned roughly at a value $\alpha\approx -E_C$}, as expected \cite{PhysRevA.69.062320}. \editP{In contrast, the} topological transition \editP{is characterized by an abrupt dip in $\alpha$, followed by finite-length Majorana oscillations $\delta(B)$  in the NW spectrum, which become also visible as oscillations in the anharmonicity}. Such imprints suggest that $\alpha$ may \editE{be} a relevant parameter to trace topology in real NW platforms in the transmon regime.

  \begin{figure}
\centering \includegraphics[width=\columnwidth]{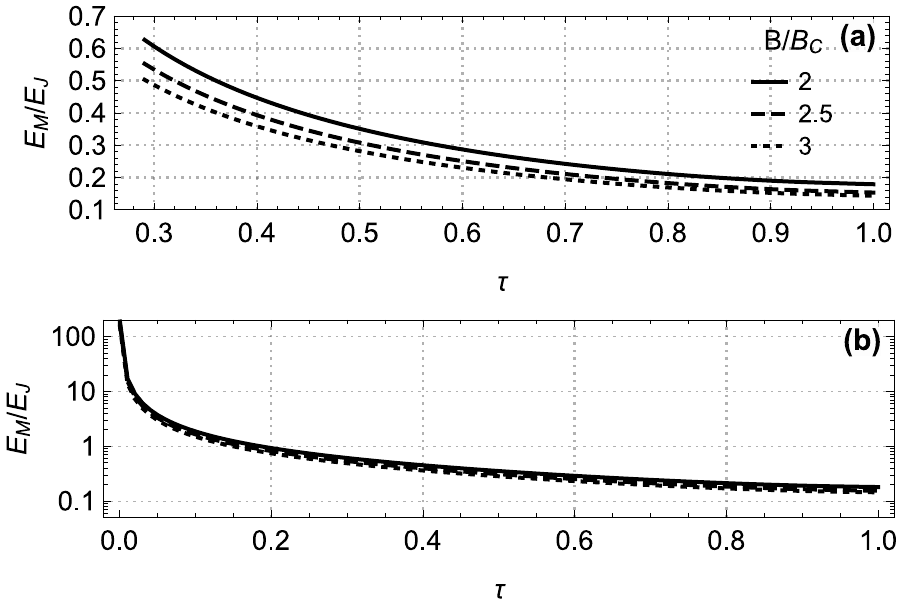}
\caption{\label{Fig10} \textbf{Majorana versus standard Josephson coupling.} %
Both panels show the evolution of the ratio $E_M/E_J$ as the transparency factor $\tau\in[0,1]$ is increased for different $B$ fields in the topological regime. For clarity, both linear (a) and logarithmic (b) plot scales are provided. For small transparencies below $\tau\approx 0.2$, the Majorana coupling $E_M$ becomes larger than $E_J$. Parameters for each NW segment like in Fig. \ref{Fig2}(a).}
\end{figure}

\subsection{\label{ratioMajorana-Josephson}Ratio $E_M/E_J$ for NW-based single-channel JJs in the topological regime.}
\editP{An important} figure of merit that governs the different physical regimes of Eq. \eqref{Majorana-Transmon hamiltonian2} is the ratio $E_M/E_J$, \editP{which controls the relative amplitude of different-parity and same-parity anticrossings in the CPB\editE{/transmon} spectrum (Fig. \ref{Fig7}), as well as the ratio $E_M/E_C$ at fixed $E_J/E_C$}. 
\editP{This subsection elaborates on this aspect of the problem, and shows that while $E_M/E_J$ depends on various model parameters, it is in general not small}. 

\begin{figure}
\centering \includegraphics[width=\columnwidth]{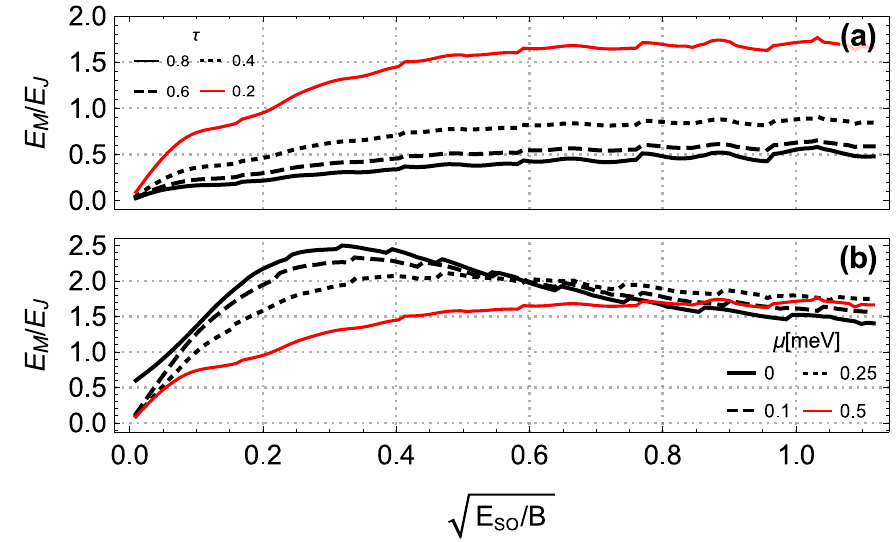}
\caption{\label{Fig11} \textbf{Majorana versus standard Josephson coupling.} %
$E_M/E_J$ against $\sqrt{E_{\rm{SO}}/B}$. Panel (a), calculated with the NW parameters of Fig. \ref{Fig2}(a) ($\mu=2\Delta=0.5\textrm{meV}$), illustrates how by decreasing $\tau$ one gets progressively larger values  $E_M/E_J$ for all $E_{\rm{SO}}/B$. Panel (b) shows the behavior at fixed $\tau=0.2$ for decreasing chemical potentials. The estimation $E_M/E_J\propto \sqrt {E_{\rm{SO}}/B}$ \editP{of Eq. \eqref{EMEJ}} is recovered for small chemical potentials in the $E_{\rm{SO}}/B\to 0$ limit.}
\end{figure}

In Fig. \ref{Fig10} we plot the dependence of the ratio $E_M/E_J$ with $\tau$ for different $B$ fields in the topological regime. For small $\tau$, this ratio can be much larger than unity, while it is of order $E_M/E_J\sim 0.1$ for $\tau\to 1$. \editP{This behavior is consistent with the different expected dependence on transparency $T_N$ of $E_J$ and $E_M$, as discussed before. In Fig. \ref{Fig11} we further plot} the ratio $E_M/E_J$ against $\sqrt{E_{\rm{SO}}/B}$ \edit{at fixed, finite $\mu$. As expected, it} deviates from the $\mu=0$ estimations \editP{of Eq. \eqref{EMEJ}}. Panel (a) illustrates \editP{the deviation} by plotting results for $\mu=2\Delta$.  In contrast, this ratio starts to approach the $E_M/E_J\propto \sqrt {E_{\rm{SO}}/B}$ estimation \editP{as $\mu$ and $E_{\rm{SO}}/B$ decrease, panel (b). The main} conclusion that can be drawn from this discussion is that, in general, $E_M/E_J$ is not a small number.  For typical islands in the $E_J/E_C>1$ regime, this also implies that $E_M$ is not small as compared with $E_C$. 

\editP{These calculations and estimations are based on the single channel junction. We} can expect that multichannel systems will show an overall increase of the Josephson coupling $E_J$. Considering a simple formula for the multichannel Josephson potential $V_J(\varphi)=-\Delta\sum_{i=1}^M\sqrt{1-T_i \sin^2(\varphi/2)}$ (which just assumes a short junction \editE{in the Andreev limit} containing $M$ channels with normal transmission probabilities $T_{i=1,..,M}$), the Josephson coupling is $E_J=\Delta/4\sum_{i=1}^M T_i$. If the junction contains $m$ highly transmitting channels, we can expect an overall reduction of the above estimation for $E_M/E_J$  of order  $\sim1/m$. Even in these multichannel cases, we argue that the parameter regimes explored in previous papers using low-energy effective toy models \cite{ginossar2014microwave,PhysRevB.92.075143} (with very small ratios $E_M/E_J\sim10^{-3}$) are somewhat unphysical since this would imply hundreds of highly transmitting channels (i. e. hundreds of trivial subbands contributing to $E_J$ on top of a topological single band contributing to $E_M$). Another limiting case in which $E_M/E_J$ is small, is the $E_{\rm{SO}}/B\ll 1$ limit in a few-channel junction. \editP{This case, however, corresponds to a very small topological gap  $\Delta_T\sim 2\Delta\sqrt{E_{\rm{SO}}/B}\ll \Delta$, which would naturally} hinder the observation of any Majorana-related physics in the superconducting island properties. 

Even in the opposite strong CPB regime with $E_J/E_C\ll 1$, the first experimental evidence of hybridization between different parity sectors owing to the $E_M$ coupling \cite{CharliePAT} gives \editP{estimated} ratios of the order $E_M/E_C\approx 0.25$, which, again, is much larger than the previously explored regimes in Refs. \onlinecite{ginossar2014microwave,PhysRevB.92.075143}. This regime seems to suggest that the superconducting islands used in the experiments of Ref. \onlinecite{CharliePAT} are based on few-channel junctions in the small \editP{transparency} regime (see Fig. \ref{Fig10}). 

We will discuss in full the implications of the different ratios $E_M/E_C$ in the next subsections \editJ{(in order to compare different situations, the ratios $E_M/E_C$ will be taken at a fixed field in the topological phase $B/B_c=1.4$. Therefore, for the same NW segments forming the JJ, different $E_M/E_C$ will correspond to different values of $E_C$).}  The novel regime $E_M/E_C>1$, with focus on transmons and parity crossings, is discussed \editP{extensively} in a companion paper, see Ref. \onlinecite{Avila-accompanying}. The $E_M/E_C\lesssim 1$ regime with $E_J\to 0$, which as we just mentioned is relevant for the experiments in  Ref. \onlinecite{CharliePAT}, will be discussed in Sec. \ref{CharliePAT}.

\section{\label{MW}MW spectroscopy of NW-based superconducting islands} 
Having discussed various relevant aspects of NW-based superconducting islands we are now ready to analyze in detail the MW response of such junctions. Using the solutions of  Eq. \eqref{Majorana-Transmon hamiltonian2}, \editP{the MW absorption} spectrum of the \editE{islands} can be written in linear response as
\begin{equation}
\label{microwave}
	S(\omega)=\sum_k \left|\langle k|\hat N|0\rangle \right|^2 \delta(\omega - (\omega_k-\omega_0)),
\end{equation}
\editJ{which in turn is convolved with a Cauchy-Lorentz distribution with a finite-line broadening for graphical purposes.} 
This response measures the energy transitions  $\omega_{0k}=\omega_k-\omega_0$ between the $k=0$ ground state $E_0=\hbar\omega_0$ and the excited states $E_k=\hbar\omega_k$ of the junction with a probability weighted by the matrix elements of the \editE{relative} number operator 

\begin{equation}
\label{microwave2}
	\langle k|\hat N|0\rangle = \int_0^{2\pi}\,\dd\varphi\, \Psi^\dagger_k \begin{pmatrix}-i\partial_\varphi&\editP{0}\\\editP{0}&-i\partial_\varphi+1/2\end{pmatrix}\Psi_{0}.
\end{equation}
A detailed discussion about the spectral weights of relevant MW transitions in terms of the above matrix elements, and their dependence on various island parameters, is included in Secs. \ref{matrix-elements1} and \ref{matrix-elements2}. 

\editP{We note that} the above notation $\omega_{0k}=\omega_k-\omega_0$ in terms of energy differences between the ground state and the excited states, \editP{with index $k$} ordered in terms of increasing energies, may lead to some confusion in the context of this paper since \editP{parity conservation (i.e., negligible $E_M$) can render some of these transitions invisible. For example, at $B=0$ the first transition $\omega_{01}$ is even-even (allowed), see e.g. Fig. \ref{Fig7}(a), and hence a standard qubit transition (assuming $2\Delta\gg\hbar\omega_{pl}$, so that odd states are at higher energies). However, for $B=B_c$, the equivalent parity-preserving transition is now  $\omega_{03}$, see e.g. Fig. \ref{Fig7}(c), with a strongly suppressed $\omega_{01}$ and $\omega_{02}$.}

When $B>B_c$ \editE{and $E_M$ is finite}, we can have other situations, \editP{such as $\omega_{01}$ (a transition within the ground state manifold) becoming visible thanks to the Majorana-induced parity mixing, particularly close to the $n_g=0.25$ and $n_g=0.75$ anticrossings, see e.g. Fig. \ref{Fig7}(h)}. 

When needed, and to avoid ambiguities, we will use, together with the above notation, a notation drawn from the superconducting qubit literature, based on the solutions of Eq. \eqref{transmonH} in terms of Mathieu functions~\cite{koch2007charge,PhysRevB.92.075143,10.21468/SciPostPhys.7.4.050}. This notation assumes decoupled even-odd sectors, essentially the diagonal part of Eq. \eqref{Majorana-Transmon hamiltonian2}, whose eigenstates are labelled as $|m,e/o\rangle$, with $m$ denoting the bosonic mode index of the island and e/o denoting even/odd parity. For example, using this notation, a transition  
$\omega_{01}$ at $B=0$ corresponds to a standard interband qubit transition $|0,e\rangle\rightarrow |1,e\rangle$, with $\omega_{01}=4E_C/\hbar$ at $n_g=0$, or $\omega_{01}=E_J/\hbar$ at $n_g=0.5$, see Fig. \ref{Fig7}(e). For $B>B_c$, the transition $\omega_{01}$ corresponds now to an intraband transition between two states (of \emph{approximately} well defined parity) within the ground state manifold, namely $|0,e\rangle\rightarrow |0,o\rangle$ at $n_g=0$, or viceversa $|0,o\rangle\rightarrow |0,e\rangle$ at $n_g=0.5$,  Fig. \ref{Fig7}(h).  These intraband transitions depend on the charge dispersion of the island and are of order $\hbar\omega_{01}\approx E_C(E_C/E_J)^{3/4}\exp(-\sqrt{8E_J/E_C})$ in the $E_J/E_C\gg 1$ limit~\cite{koch2007charge}. Obviously, this notation in terms of well defined parity is strictly valid only in the $E_M\to 0$ limit (namely, in the basis of $\hat\tau_z\equiv|00\rangle\langle 00|-|11\rangle\langle 11|$). In the opposite limit, parity is not well defined and we will rather use $|m,\pm\rangle$, \editP{denoting the two} mixed-parity eigenstates that diagonalize Eq. \eqref{Majorana-Transmon hamiltonian2}. Using the previous example for  $B>B_c$, the intraband transition within the ground state manifold around  $n_g=0.25$ is better described by the notation $|0,-\rangle\rightarrow |0,+\rangle$ and is of order $\omega_{01}\approx E_M/\hbar$.

\subsection{\label{MW1} NW-based superconducting islands in the $E_M/E_C\ll 1$ regime}
\subsubsection{MW spectroscopy in the \editP{$E_M/E_C\ll 1$} regime}

\begin{figure} 
\centering \includegraphics[width=\columnwidth]{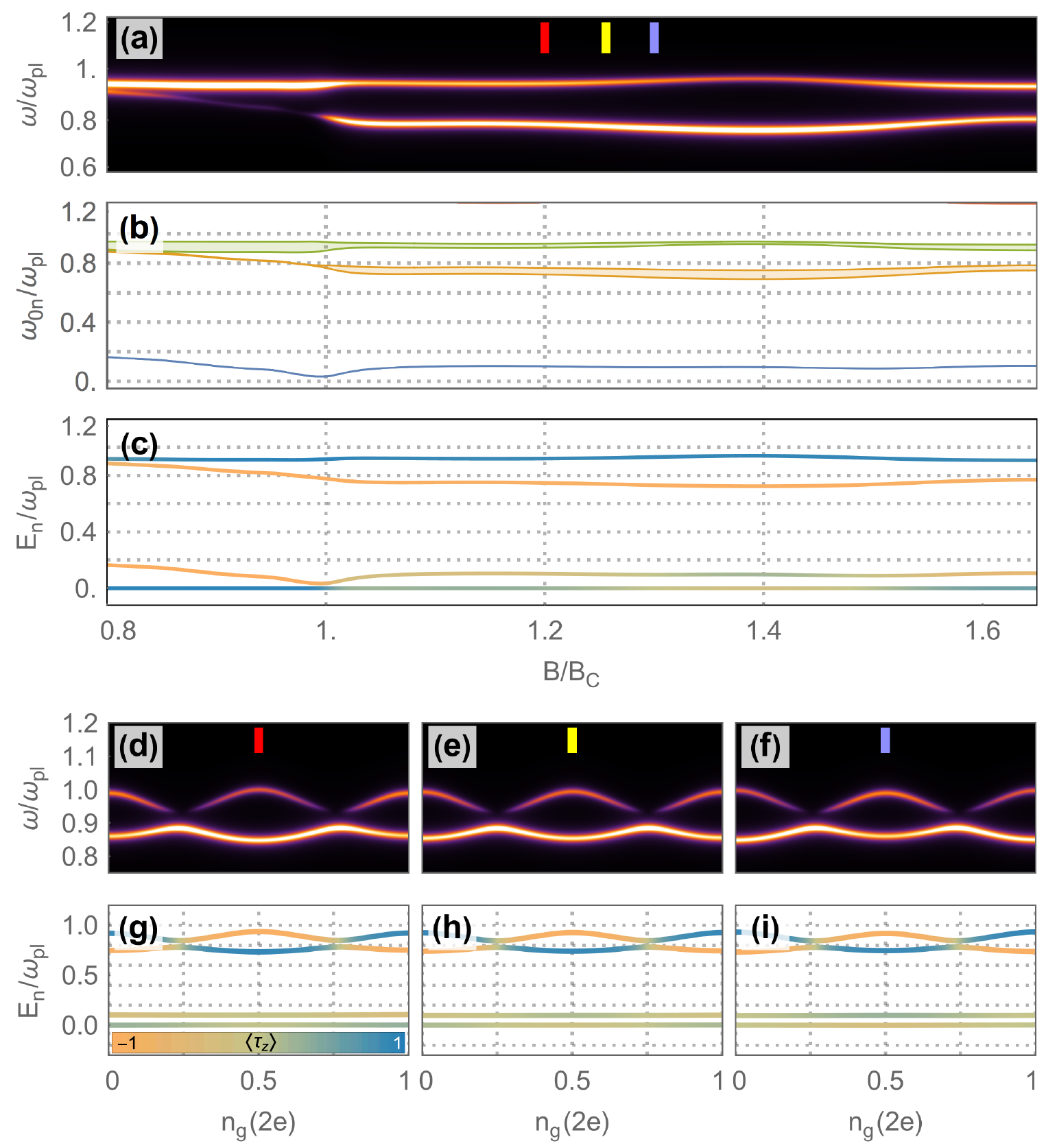}
\caption{\label{Fig12} \textbf{MW spectroscopy of a NW-based superconducting island in the \editP{$E_M/E_C\ll 1$} regime (with $E_J/E_C\approx 5$)}. Here, starting from a single channel calculation we artificially increase $E_J$ (simulating many channels) while keeping $E_M$ and $E_C$ constant. The ratios used in these plots are $E_M/E_C\approx 0.17$ and $E_M/E_J\approx 0.036$. \editEE{The superconducting island is formed by two coupled NW segments, each of them like in Fig. \ref{Fig2}(a),} and $\tau=0.8$. (a) Contour plot of MW absorption spectrum $S_N(\omega)$ versus $\omega$ and $B/B_c$ at $n_g=0$. Bright lines signal allowed transitions in the MW response. (b) Transition frequencies and spectral weights (shadowed widths). (c) Spectrum versus $B/B_c$ at $n_g=0$. (d-f) Gate dependence of $S_N(\omega)$ at three magnetic fields (color bars) marked in (a). (g-i) Spectra corresponding to panels (d-f).}
\end{figure}

To make connection with \editP{published} literature, we first analyze the \editP{$E_M/E_C\ll 1$} regime. In order to artificially enhance the ratio $E_J/E_M$, which models a multichannel situation, as discussed before, we add by hand a Josephson term $-E_J\cos\varphi$ \editP{to $V_J(\varphi)$}, such that the total $E_J$ used in the calculations is much larger than the one we obtain from the single-band NW calculation [the BdG spectrum employed here \editEE{for each segment of the NW junction} corresponds to that of Fig. \ref{Fig2}(a)]. 
The MW spectra of a paradigmatic case with $E_J/E_C\approx 5$ in this \editP{$E_M/E_C\ll 1$} regime are shown in Fig. \ref{Fig12}. We first plot the overall magnetic field dependence \editP{of transitions $\omega_{0n}$ and energy levels $E_n$ at fixed $n_g=0$, panels (a-c)}. The $B>B_c$ MW spectrum in this limit is just that of a transmon with a split line: owing to the Majorana coupling $E_M$, the original ground state splits into a doublet  $|0,\pm\rangle$, while \editP{there appear} two possible interband qubit transitions from the $|0,-\rangle$ ground state to two excited states of approximately good parity $|1,o/e\rangle$ [Fig. \ref{Fig12}(c)]. 
These split lines give rise to two possible transitions $\omega_{02}$ and $\omega_{03}$. We also show the corresponding matrix elements, \editEE{represented} as the thickness of the transition frequencies, in panel (b).  
Apart from the odd state that goes down in energy and reaches zero at $B\sim B_c$, an important aspect of the overall magnetic field dependence of the three panels in Figs. \ref{Fig12}(a-c) is the complete absence of $B>B_c$ parity crossings [originating from the Majorana oscillations in the NW spectrum of Fig. 1(a)]. This can be understood as a consequence of the large $E_C$, as compared to $E_M$, which largely prevents \editP{the} changes in the ground state fermionic parity \editP{that are associated to Majorana oscillations}. As a result, all the lines for $B>B_c$ are almost independent of $B$ field [including both curvature and parity, see colors of the lines \editE{in \ref{Fig12}(c)}]. Importantly, the intraband transition within the ground state doublet $\omega_{01}$ ($|0,-\rangle\rightarrow |0,+\rangle$) has no spectral weight in this regime. \editE{Thus, it is not visible in the absorption spectrum of \ref{Fig12}(a).} Different magnetic field cuts [colored bars in \ref{Fig12}(a)] are shown in panels \ref{Fig12}(d-f), with the corresponding \editE{energy} states in \ref{Fig12}(g-i). The splitting of the lines shows dispersion as a function of $n_g$ (as expected for this particular $E_J/E_C$ ratio) while having very little dependence on $B$ field (the three $B$-field cuts are essentially the same). The \editP{visible effect} of Majoranas in the NW is the appearance of ``spectral holes'' \editP{in the $\omega_{03}$ transition} near $n_g=0.25$ and  $n_g=0.75$ \editP{(namely, a zero of the transition matrix element at that point)}. This happens as the small $E_M$ coupling weakly removes the degeneracy of the even and odd parity sectors in the $E_M\to 0$ limit (occurring at $n_g=0.25$ and  $n_g=0.75$). All the above results are in full agreement with Refs. \onlinecite{ginossar2014microwave,PhysRevB.92.075143,10.21468/SciPostPhys.7.4.050}. 

\begin{figure} 
\centering \includegraphics[width=\columnwidth]{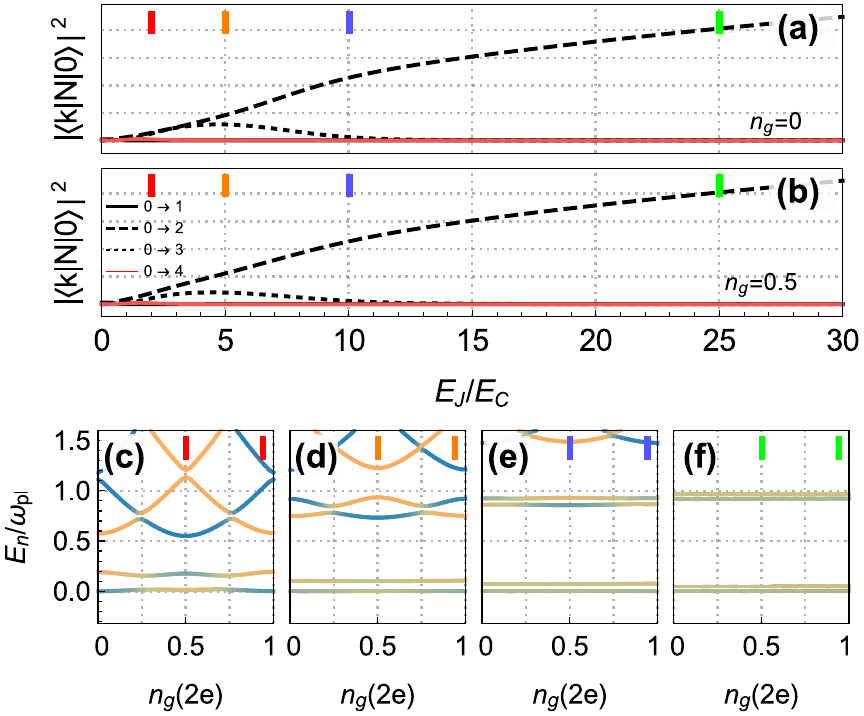}
\caption{\label{Fig13} \editRR{\textbf{Spectral weights and spectra in the \editP{$E_M/E_C\ll 1$} regime for increasing $E_J/E_C$ ratios}. The $E_J/E_C$ ratio is tuned by artificially increasing the $E_J$ amplitude, keeping $E_C$ constant. This regime imposes $E_C>E_M$, similarly to the cases considered in previous references \cite{ginossar2014microwave,PhysRevB.92.075143,10.21468/SciPostPhys.7.4.050}. (a,b) Spectral weights of the first transitions as a function of $E_J/E_C$ and fixed $B/B_c = 1.2$ 
 at different $n_g=0$ (a) and $n_g=0.5$ (b). (c-f) Spectra for increasing ratios $E_J/E_C$ from the CPB to the transmon regime ($E_J/E_C=2, 5,10, 25$ from (c) to (f), corresponding to the colored bars at the top). 
For transmon regimes $E_J/E_C\gg 1$ we recover the almost doubly degenerate transmon spectrum that is expected for $E_M\ll E_C$. \editEE{Note that} only the first transmon transitions $\omega_{02}$, $\omega_{03}$ are allowed. Parameters for each NW segment like in Fig. \ref{Fig2}(a) and $\tau=0.8$.}}
\end{figure}
\begin{figure} 
\centering \includegraphics[width=\columnwidth]{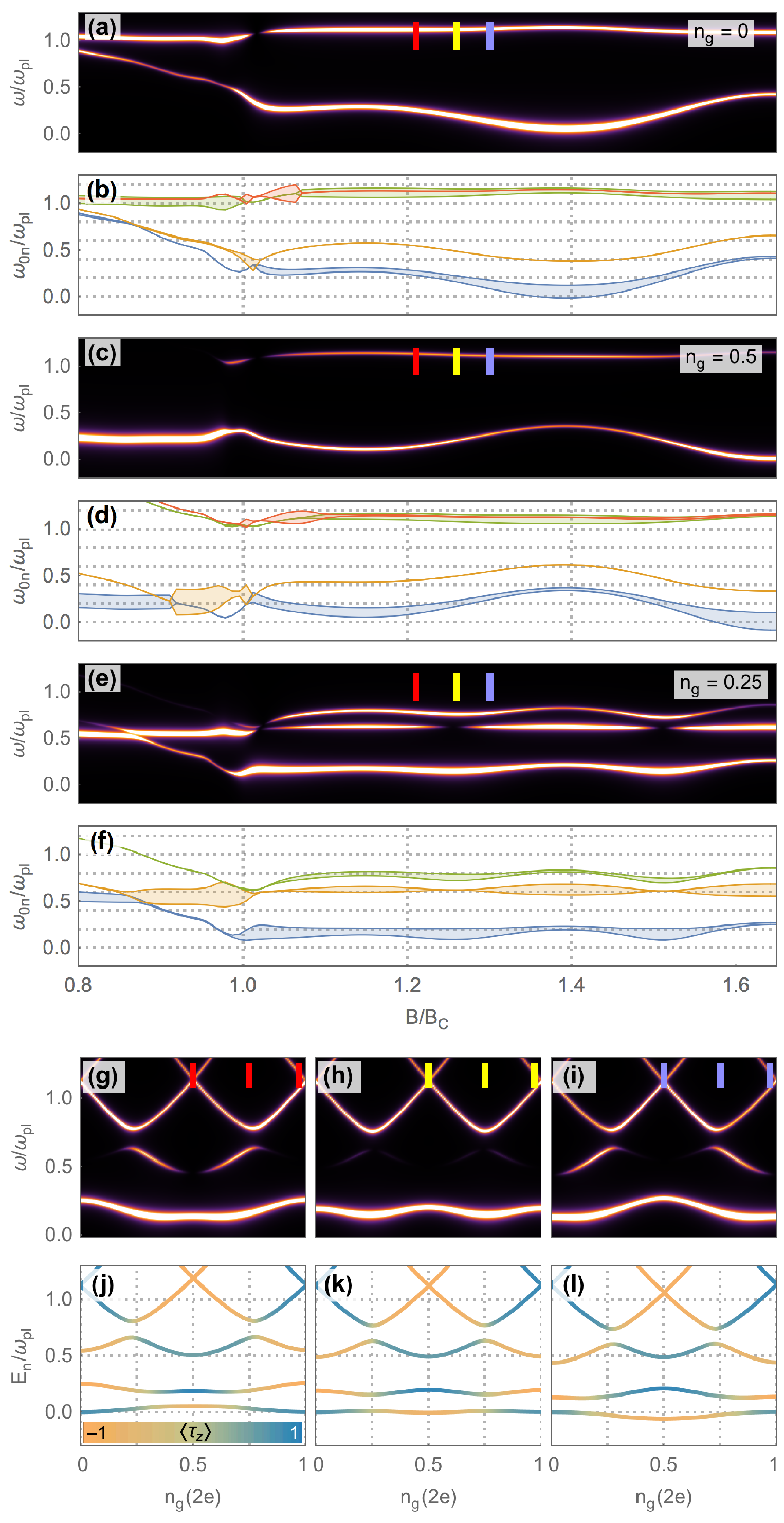}
\caption{\label{Fig14} \textbf{MW spectroscopy of a NW-based superconducting island in the $E_M\approx E_C$ regime (with $E_J/E_C\approx 2$)}. The ratios used in these plots are $E_M/E_C\approx 0.56$ and $E_M/E_J\approx 0.28$. (a) Contour plot of MW absorption spectrum $S_N(\omega)$ versus $\omega$ and $B/B_c$ at $n_g=0$. (b) Transition frequencies and spectral weights (shadowed widths). (c,d) Same as (a,b) but at $n_g=0.5$. (e,f) Same as (a,b) but at $n_g=0.25$. (g-i) Gate dependence of $S_N(\omega)$ at the three magnetic fields (color bars) marked in (a), (c) and (d). (j-l) Spectra corresponding to panels (g-i). Parameters for each NW segment as in Fig. \ref{Fig2}(a) and $\tau=0.8$.}
\end{figure}

\subsubsection{\label{matrix-elements1}Dependence of the spectral weights on the ratio $E_J/E_C$ for different $n_g$}

Before \editP{proceeding to the discussion of} the $E_M/E_C\gtrsim 1$ regime, we will analyze the above \editP{$E_M/E_C\ll 1$} results in terms of the spectral weights of the involved transitions. Figure \ref{Fig13} shows these matrix elements \editP{as we cross over from the CPB to the transmon regime by increasing $E_J/E_C$. It also shows} the corresponding spectra versus $n_g$ at specific values of $E_J/E_C$. 
 The upper panels \editJ{\ref{Fig13}(a,b)} show the spectral weights of the first transitions as a function of $E_J/E_C$ and fixed $B/B_c = 1.2$ [before the first parity crossing in the NW spectrum of Fig. \ref{Fig2}(a)], and for different $n_g=0$ (a) and $n_g=0.5$ (b). The overall behavior changes very little for different gates, with a dominant $\omega_{02}$ transition and a weaker $\omega_{03}$ transition. These transitions can be understood by looking at the spectra for different $E_J/E_C$ cuts, which are shown in the lower panels \ref{Fig13}(c-f). For increasing ratios  $E_J/E_C$ ($E_J/E_C=2, 5,10, 25$ from \ref{Fig13}(c) to \ref{Fig13}(f), corresponding to the colored bars at the top), the spectra evolve from the CPB to the transmon regimes, reaching an almost doubly degenerate transmon spectrum, as expected for $E_M\ll E_C$.
\editJ{This picture remains almost unchanged for magnetic fields $B$ after the topological transition, $B>B_c$.} This comes at no surprise since the spectra \editJ{within this $B$ region} are essentially the same as for $B/B_c = 1.2$.
  The overall \editP{$E_M/E_C\ll 1$ behavior therefore shows} little dependence with $B$, except for the split transmon lines for $B>B_c$, as discussed in Fig. \ref{Fig12}. As we discuss in the next subsections, a larger $E_M/E_C$ ratio completely changes this picture.

\begin{figure*} 
\centering \includegraphics[width=\textwidth]{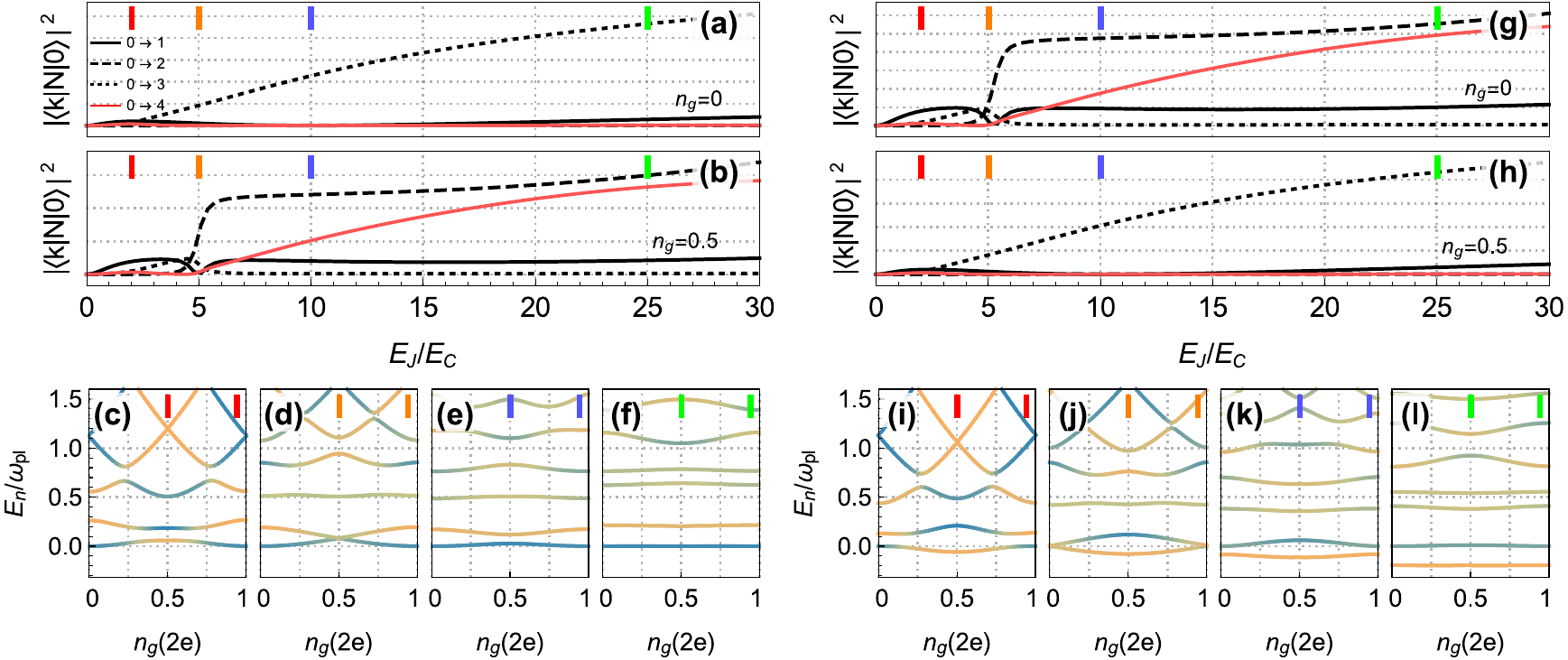}
\caption{\label{Fig15} \editRR{\textbf{Spectral weights and spectra for non-negligible \editP{$E_M/E_C\gtrsim 1$} ratios and increasing $E_J/E_C$ ratios} . (a,b) Spectral weights of the first transitions as a function of $E_J/E_C$ and fixed $B/B_c = 1.2$ (before a parity crossing) at different $n_g=0$ (a) and $n_g=0.5$ (b). (c-f) Different spectra at this magnetic field, $B/B_c = 1.2$, for increasing ratios $E_J/E_C$ from the CPB to the transmon regime ($E_J/E_C=2, 5,10, 25$ from (c) to (f), corresponding to the colored bars at the top). 
(g,h) and (i-l): Same as before but for $B/B_c = 1.3$ (after a parity crossing). In the CPB regime the $\omega_{01}$ transition has some weight, but deep in the transmon regime the only allowed transition is a transmon line $\omega_{03}$. Note how the matrix elements are fully exchanged between $n_g=0$ and $n_g=0.5$ after a parity crossing [i. e., compare panel (a) with (h) and (b) with (g)]. The exact cancellation at $E_J/E_C=5$ of the $\omega_{01}$ transition in (b) results from parity degeneracy at $n_g=0.5$ [panel (d)]. After a parity crossing, the full spectrum is shifted by one $e$ unit, while flipping parity, and the parity degeneracy occurs now at $n_g=0$ and $n_g=1$ [panel (j)]. Consequently, the exact cancellation at $E_J/E_C=5$ of $\omega_{01}$ in (g) occurs now at $n_g=0$. Parameters for each NW segment like in Fig. \ref{Fig2}(a) and $\tau=0.8$.}}
\end{figure*}

\subsection{\label{MW2}NW-based superconducting islands with non-negligible $E_M/E_C$ ratios}
\subsubsection{MW spectroscopy in the \editP{$E_M/E_C\gtrsim 1$} regime}
As soon as the ratio $E_M/E_C$ becomes \editP{non-negligible}, the overall MW spectral response \editP{becomes very different and substantially more complex than that of the preceding subsection, exhibiting} a stronger dependences with gate and Zeeman fields.  In Fig. \ref{Fig14} we \editP{plot} the MW spectra of a superconducting island nominally in an intermediate \editP{CPB-transmon} regime with $E_J/E_C=2$, but with a larger $E_M/E_C$ ratio ($E_M/E_C\approx 0.56$, \editP{to be compared to the $E_M/E_C\approx 0.17$ case shown} in Fig. \ref{Fig12}).  
 
In Figs. \ref{Fig14}(a-f) we plot the $B$ dependence of the MW response for different gates voltages. Fig. \ref{Fig14}(a) shows this magnetic field dependence at $n_g=0$ [with the corresponding transition \editP{peaks widened by} their corresponding spectral weights, as represented in Fig. \ref{Fig14}(b)]. The overall response is seemingly similar to the one discussed in  Fig. \ref{Fig12}, including the split lines for $B>B_c$. Note, however, that the splitting in Fig. \ref{Fig12} comes from interband transitions, as we mentioned, while here the lowest line corresponds to a $\omega_{01}$ transition $|0,e\rangle\rightarrow |0,o\rangle$ (namely, an intraband transition flipping parity). This MW resonance directly reflects Majorana coupling within the lowest-energy doublet. This explains why this lowest line lies near zero frequency and shows a sizable modulation with $B$-field [compare with Fig. \ref{Fig12}(a)]. 
The upper line here is a standard qubit transition  $\omega_{03}$ (($|0,e\rangle\rightarrow |1,e\rangle$) which conserves parity.  

At $n_g=0.5$, Figs. \ref{Fig14}(c-d), the MW spectrum is similar to the previous case but with all transitions \editP{with inverted parities respect to $n_g=0$} (i.e., the $\omega_{01}$ transition corresponds now to the process $|0,o\rangle\rightarrow |0,e\rangle$, while the $\omega_{03}$ to the process $|0,o\rangle\rightarrow |1,o\rangle$). This is expected since the ground state at $n_g=0.5$ is now odd [there is a 1$e$ shift with respect to the previous $n_g=0$ case, see the spectra in Figs. \ref{Fig14}(j-l)]. Interestingly, the magnetic field dependence of the $\omega_{01}$ transition is the opposite to the one at $n_g=0$, with exchanged maxima and minima.

At $n_g=0.25$, Fig. \ref{Fig14}(e-f), the response is richer with more transition lines \editP{becoming visible}. This originates from the strong parity mixing induced by the $E_M$ Majorana coupling at this gate value. Apart from the previous lines, the spectrum now shows another line originating from an allowed interband transition between states of mixed parity $\omega_{02}$ ($|0,-\rangle\rightarrow |1,-\rangle$). Note that, as opposed to Fig. \ref{Fig12}(a), the $\omega_{02}$ transition shows spectral holes precisely at $B$ fields where the $\omega_{01}$ transition has minima. This phenomenon is related to parity crossings in the NW spectrum owing to Majorana oscillations and can be understood by looking at the $n_g$ dependence at different magnetic fields across one of such minima  (red, yellow and blue bars). The bottom panels of Fig. \ref{Fig14} show this gate dependence [both for the MW spectra, (g-i), and for the energy spectrum, (j-l)]. If we compare the spectra for $B$ fields before and after a parity crossing, \editP{panels (j) and (l)}, the two lowest energy states $E_0$ and $E_1$ are shifted in gate voltage \emph{by exactly one electron} (a shift $n_g\rightarrow n_g+0.5$) while flipping parity. We explain this phenomenon in full in the next subsection.

\subsubsection{\label{matrix-elements2}Dependence of the spectral weights on the ratio $E_J/E_C$ for different $n_g$}
Our previous findings for non-negligible $E_M/E_C$ ratios can be understood by analyzing in detail the corresponding spectral weights for increasing $E_J/E_C$. These results are summarized in Fig. \ref{Fig15}. They clearly demonstrate that the phenomenology in this \editP{$E_M/E_C\gtrsim 1$} regime is different from \editE{the} one shown before in Fig. \ref{Fig13}. In Fig. \ref{Fig15}(a-b) we present the spectral weights as a function of $E_J/E_C$ and fixed $B/B_c = 1.2$ [namely, before the first parity crossing in the NW spectrum of Fig. \ref{Fig2}(a)]. The different panels show different gates, $n_g=0$ (a) and $n_g=0.5$ (b). The lower-left panels (c-f) show the full $n_g$ dependence of the spectrum at specific values of $E_J/E_C$ marked by colored bars in the upper panels. The spectral weights of different transitions have now a strong dependence on $n_g$ (as opposed to the results in Figs. \ref{Fig12} and \ref{Fig13}). For $n_g=0$, the dominant transition is the standard qubit transition in the even parity sector [$\omega_{03}$, corresponding to $|0,e\rangle\rightarrow |1,e\rangle$, see e.g  Fig. \ref{Fig15}(c)]. 
 \editJ{At $n_g=0.5$, Fig. \ref{Fig15}(b), and for large $E_J/E_C\gtrsim 5$, the transitions  $\omega_{02}$ and $\omega_{04}$, which signal Majorana-mediated parity mixing, are dominant.}
 Notably, the $\omega_{03}$ transition is now completely absent [compare to the $n_g=0$ case of Fig. \ref{Fig15}(a)]. Importantly, the transfer of spectral weight between the $\omega_{03}$  and the $\omega_{02}$ transitions occurs precisely at $E_J/E_C=5$, where $\omega_{01}$ has an exact minimum. Since this transition corresponds to an intraband transition within the lowest energy manifold (i. e,  the transition $|0,e\rangle\rightarrow |0,o\rangle$ between the lowest energy states with opposite fermionic parity), this minimum should be related to a parity crossing. Indeed, for $E_J/E_C=5$ there is an exact parity crossing at $n_g=0.5$, Fig. \ref{Fig15}(d), which occurs as $\delta$ becomes of order $E_C$. Other representative $E_J/E_C$ ratios are shown in Figs. \ref{Fig15}(c-f). Before and after the $n_g=0.5$ parity crossing at $E_J/E_C=5$, the ground state changes parity from odd, Fig. \ref{Fig15}(c), to even, Fig. \ref{Fig15}(e), which explains the transfer of spectral weights discussed above. All this phenomenology depends on magnetic field. The top-right panels (g-h) show the same matrix elements as in (a-b) but \emph{after} the parity crossing of the Majorana oscillation in the NW spectrum (here at $B/B_c = 1.3$). Remarkably, at this magnetic field all the matrix elements \editP{at $n_g$ and $n_g + 0.5$ gate voltages become interchanged, relative to those at fields} \emph{before} the crossing of the Majorana oscillation. Namely, all the matrix elements that we find for $n_g=0$ correspond now to $n_g=0.5$, and viceversa [compare panel (a) with (h) and (b) with (g)]. If we now compare the spectra at this magnetic field, panels (i-l), with the ones corresponding to the magnetic field before the crossing, panels (c-f), we find that there is an exact shift of one electron in the low energy sector \editP{(recall that a $0.5$ shift in $n_g$ corresponds to a single electron). This is consistent with} our explanation of the results of Fig. \ref{Fig14} and demonstrates that, indeed, the MW response of NW-based superconducting islands is sensitive to the underlying Majorana physics (including finite-length Majorana oscillations and the resulting fermion parity crossings of the ground state energy). This novel \editP{$E_M/E_C\gtrsim 1$} result, with emphasis on the deep transmon regime, is the focus of a companion paper in Ref. \onlinecite{Avila-accompanying}.  
\subsection{\label{CharliePAT}MW spectroscopy in the regime $E_J\ll E_C, E_M$}
\begin{figure} 
\centering \includegraphics[width=\columnwidth]{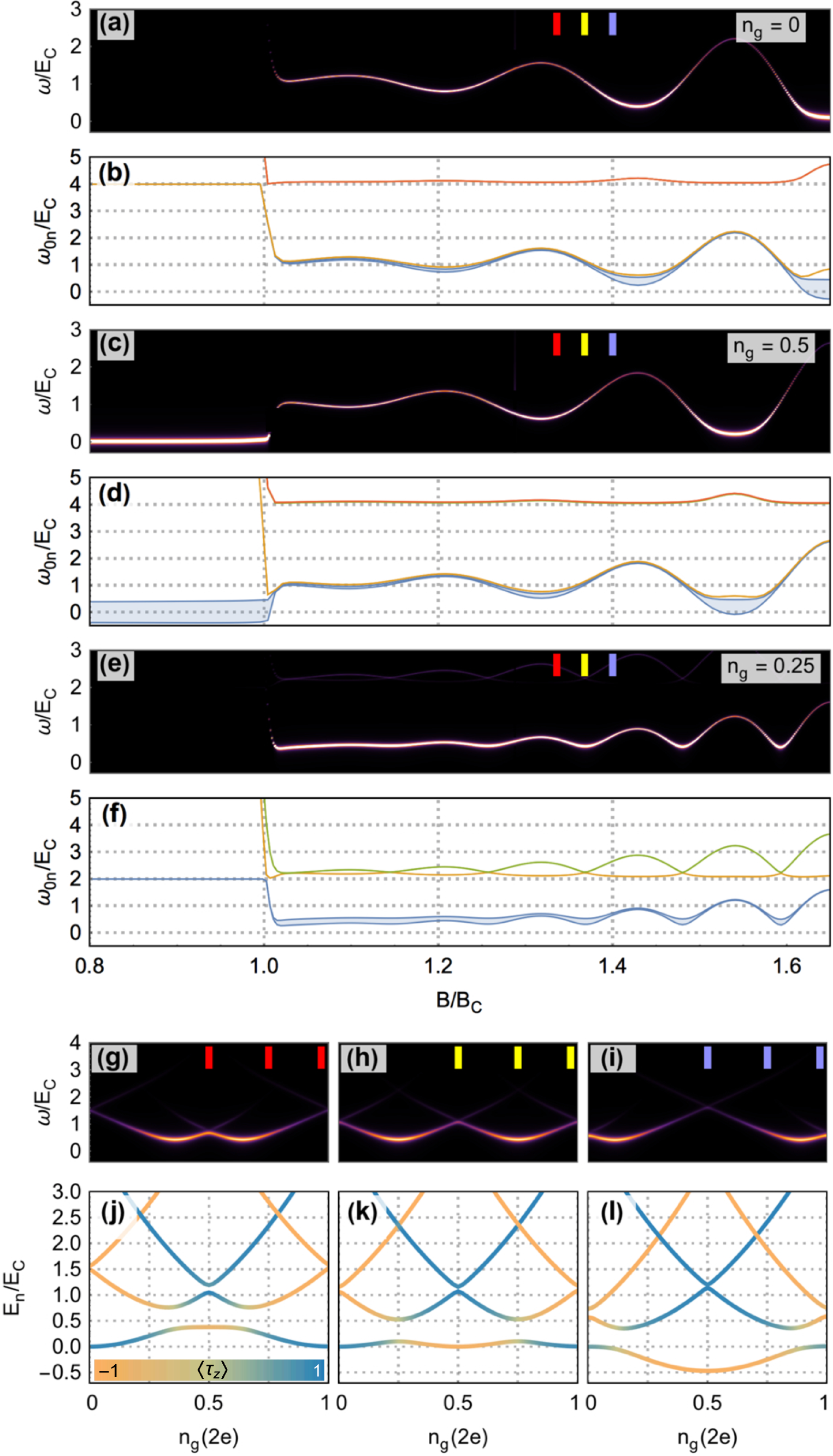}
\caption{\label{Fig16} \textbf{MW spectroscopy of a NW-based superconducting island in the $E_J\to 0$ regime, for long wires.} In contrast to previous cases, here we consider a tunnel junction for the wire, $\tau\ll 1$, so that the Josephson term $E_J$ almost vanishes. This in turn makes the Majorana coupling $E_M$ much larger that $E_J$ (see Fig.~\ref{Fig10} for $E_M/E_J$ vs $\tau$ dependence). We use, in particular, $\tau\simeq 0.01$, which translates into $E_M/E_J\sim 20$. This results in $E_C$ being the dominant energy scale of the island, as can be seen from ratios $E_M/E_C\approx 0.2$ and $E_J/E_C\approx 0.01$ (this regime is relevant for the experiments reported in Ref. \onlinecite{CharliePAT}). Contour plots of $S_N(\omega, B)$ and transition frequencies are alternatively shown for different gates: $n_g=0$ (a,b), $n_g=0.5$ (c,d), and $n_g=0.25$ (e,f). Transition frequency lines are shadowed according to their spectral weight. Panels (g-l) render gate dependence of $S_N(\omega)$ (g-i) and spectra (j-l) before, at and after a parity crossing  \editJ{as functions of $n_g$.}
 Parameters for each NW segment like in Fig. \ref{Fig2}(c) (with $L_S=5\mu$m).}
\end{figure}
\begin{figure} 
\centering \includegraphics[width=\columnwidth]{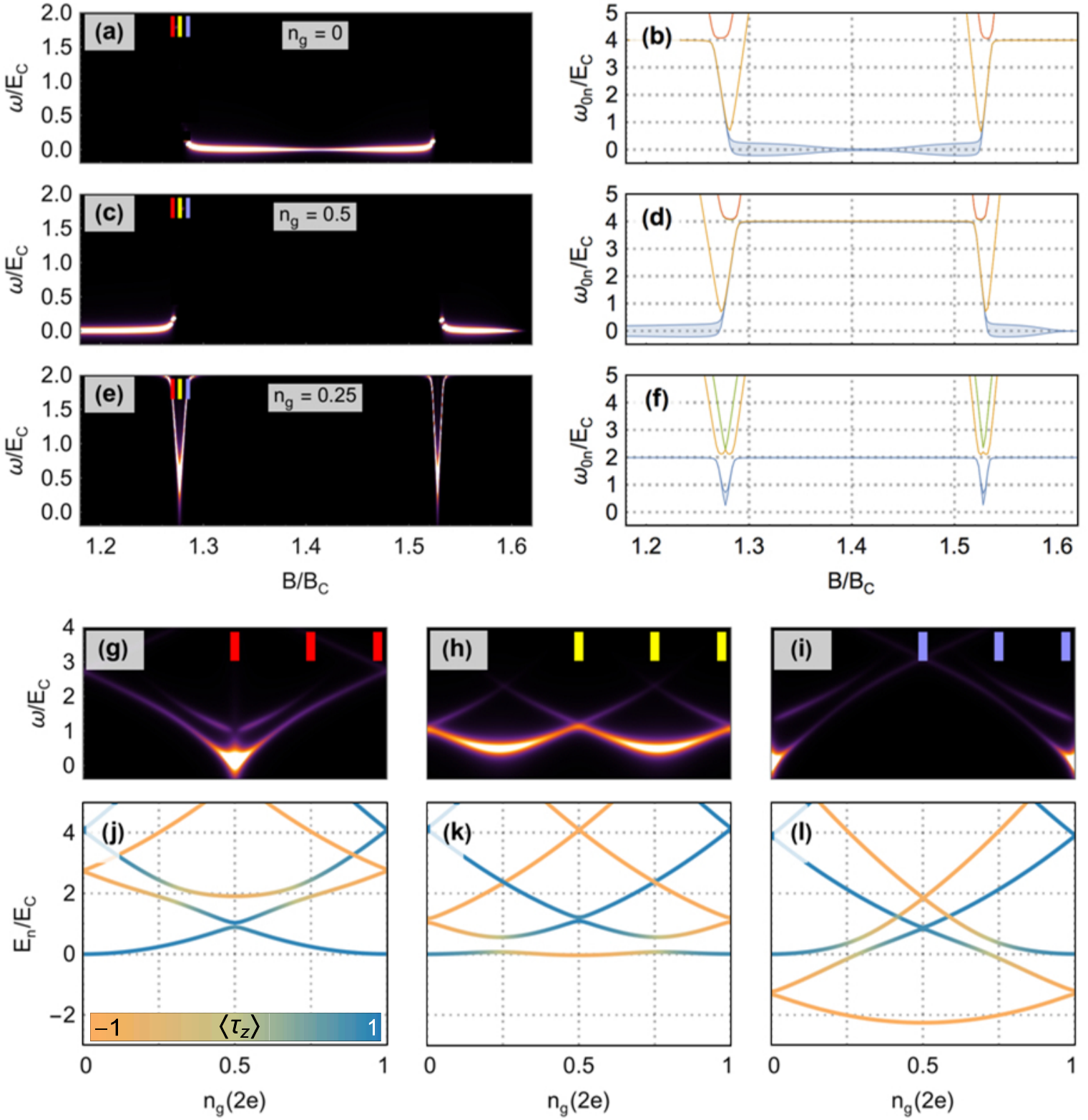}
\caption{\label{Fig17} \textbf{MW spectroscopy of a NW-based superconducting island in the $E_J\to 0$ regime, for short wires}. Similarly to Fig.~\ref{Fig16}, a strong charging regime is considered owing to a tunnel junction factor $\tau\ll 1$. \editJ{Energy scales are similar to those of Fig.~\ref{Fig16},} $E_M/E_C\approx 0.25$, $E_J/E_C\approx 0.01$. Now we focus on a shorter wire with $L_S=2.2\mu$m. Majorana splitting increases significantly over that of Fig.~\ref{Fig16} such that $\delta$ can be greater than $E_C$ in large ranges of $B$ field. As a consequence, transitions are a little more involved regarding parity mixing, except in quite narrow $B$ windows around parity switches of the ground state. Almost always, both sets of even-odd parabolas are strongly shifted from each other, making it impossible to observe parity events at lower transitions. (a-f) MW response and transition frequency of the lowest transition as a function of $B$, at three gates $n_g=0, 0.5, 0.25$. Gate dependence of MW response $S_N(\omega)$ (g-i) and of the corresponding spectra (j-l). Parameters for each NW segment as in Fig. \ref{Fig2}(a).}
\end{figure}
In this subsection we explore another novel regime relevant for the experimental data reported in Ref. \onlinecite{CharliePAT}. In these experiments, avoided crossings between even and odd parity sectors at high magnetic fields are estimated to be in the $E_M\approx 10$GHz range. Considering that the charging energies of the superconducting islands are of order $E_C\approx 40$GHz, this gives a ratio $E_M/E_C\approx 0.25$. Interestingly, these islands are in a very strong charging regime with negligible Josephson coupling $E_J$, which defines a completely new operation regime \editE{$E_J\ll E_C, E_M$}. We study this novel regime in Figs. \ref{Fig16} and \ref{Fig17}. In the first case, we concentrate on the MW response of a long NW [see Fig. \ref{Fig2}(c)] such that \editRR{the energy difference $\delta$ owing to Majorana overlap is always $|\delta|\leq E_C$ for all magnetic fields}.  In this case, the main transition line for the three relevant gates is always the intraband transition $\omega_{01}$ within the ground state manifold, Figs. \ref{Fig16}(a-f). The full gate dependence of $\omega_{01}$ for the three magnetic fields marked in the upper panels is shown in Figs. \ref{Fig16}(g-l). Again, a clear $n_g\rightarrow n_g+0.5$ shift occurs as magnetic field increases. Panels \ref{Fig16}(j-l) show the corresponding spectra. Considering that $|\delta|\leq E_C$ for all gates, this parameter regime is optimal for Majorana detection, since $\omega_{01}$ faithfully maps Majorana oscillations for all $B$.  This is no longer the case for shorter wires, where we can find realistic situations \editRR{with $|\delta|>E_C$}. In such cases, Majorana hybridization does not always occur within the ground state manifold, which gives rise to rather involved spectra. We illustrate one of these cases in Fig. \ref{Fig17}. Similar to the previous figures, we also plot the magnetic field dependence for the three relevant gates of the problem. In this regime, the magnetic field dependence is patchy, with large regions in magnetic field where a sharp resonance in the MW response at a given $n_g$ implies no response at the others. This is clearly seen in Figs. \ref{Fig17}(a,b), corresponding to $n_g=0$, where no low-frequency response is observed until we reach the magnetic field marked with the yellow bar (where 
$\delta\approx 0$). At lower magnetic fields, \editRR{$\delta$ is positive and typically larger than $E_C$}, which prevents from having Majorana-induced parity mixing in the ground state manifold (hence the absence of low $\omega$ response). After the magnetic field marked with the blue bar, the response is flat with $\omega_{01}\approx 0$. At $n_g=0.5$, Figs. \ref{Fig17}(c,d), we obtain \editP{an approximate} mirror image of the previous case: the only response occurs \editP{for fields below the yellow bar, saturating to} with $\omega_{01}\approx 0$ before the red bar. At $n_g=0.25$, Figs. \ref{Fig17}(e,f), the only finite response occurs \editP{within the narrow field window} between red and blue bars. This peculiar MW response can be fully understood by analyzing the $n_g$ dependence of the energy spectra at these three magnetic fields, \editP{see} Figs. \ref{Fig17}(j-l) [the corresponding MW responses are plotted in (g-i)]. The magnetic field at the red bar corresponds to a situation with $\delta>E_C$. In this case, the ground state has well-defined even parity and the only possible transition is a standard interband qubit transition (of $\omega\approx 4E_C$ at $n_g=0$ and of $\omega\approx 0$, owing to $E_J\to 0$, at $n_g=0.5$). At  $n_g=0.25$ the only parity mixing occurs at higher bands, but not within the ground state manifold. This residual mixing is weakly visible as a small splitting of the main qubit transition, see Fig. \ref{Fig17}(g). The central panels,  Figs. \ref{Fig17}(h,k), correspond to a $\delta\approx 0$ situation (yellow bar). This is the only magnetic field region where Majorana-mediated mixing within the ground state manifold is possible for all $n_g$. \editRR{Larger magnetic fields where $\delta$ is negative such that $-\delta>E_C$ (blue bar) induce a change of ground state parity}, which is now odd for all $n_g$, Fig. \ref{Fig17}(l). The only allowed transitions occur now near $n_g=0$ and $n_g=1$ and correspond to a $\omega_{01}\approx 0$ within the odd parity sector. Again, weak Majorana mixing occurs for higher bands and is seen as faint splittings near $n_g=0$ and $n_g=1$, Fig. \ref{Fig17}(i). This gate dependence explains the peculiar MW response as a function of increasing magnetic fields. 
We finish by noting that this seemingly extreme regime with a $2e$-periodic odd-parity ground state has been reported in the experiments discussed in Ref. \onlinecite{Shen2018}. 

\editRR{\subsection{\label{splitjunction} Split junction geometry} Finally, we consider a split-junction where a standard ancillary JJ in parallel with the NW JJ forms a loop which is threaded by an external flux $\Phi$, see Fig.~\ref{fig:18}. The Josephson potential is then split into two terms $-E_L\cos(\hat\varphi)+V_J(\phi-\hat\varphi)$, where $\phi\equiv 2\pi\Phi/\Phi_0$ with $\Phi_0=h/2e$ the flux quantum. When $E_L\gg E_C$, fluctuations of $\hat\varphi$ are small and centered around zero, while the external phase mostly drops over the NW JJ. In this limit, the dependence on $n_g$ is irrelevant and one is left with an effective LC harmonic oscillator
\begin{equation} 
H_b\sim E_C \hat N^2+\frac{E_L}{2}\hat\varphi^2=\hbar \omega_{pl}(b^\dagger b+\frac{1}{2}), 
\end{equation}
with plasma frequency $\omega_{pl}=\sqrt{8E_LE_C}/\hbar$, which interacts with the phase-dispersing levels of the NW JJ through an inductive term \cite{10.21468/SciPostPhys.7.4.050,PhysRevB.64.140511} 
\begin{equation}
H_I=-\frac{\Phi_0}{2\pi}\left(\frac{E_C}{8E_L}\right)^{1/4}(b^\dagger+b)\hat I.
\end{equation}
In practice, we calculate the current operator in the eigenbasis that diagonalizes the NW JJ effective Hamiltonian, 
\begin{equation}
\hat I=\frac{\partial V_J(\varphi)}{\partial\varphi}=\frac{\partial }{\partial\varphi}\left(\sum_k E_k |k\rangle\langle k|\right). 
\end{equation}
\begin{figure}
\includegraphics[width=0.6\columnwidth]{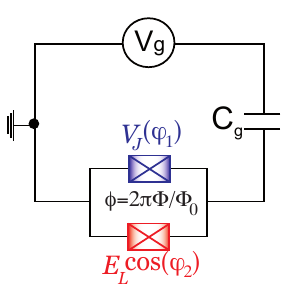}
\caption{\label{fig:18} \editRR{\textbf{Sketch of a NW-based superconducting island in a split-junction geometry.} The total phase across the split junction $\phi=\varphi_1+\varphi_2$ is fixed by an external applied flux $\phi \equiv 2\pi\frac{\Phi}{\Phi_0}$, where $\Phi_0=h/2e$ is the flux quantum. In strongly asymmetric junctions with $E_L\gg E_C$, quantum fluctuations of phase are small and the external phase mostly drops on the NW JJ with Josephson potential $V_J$.}}
\end{figure}
\begin{figure}
\includegraphics[width=\columnwidth]{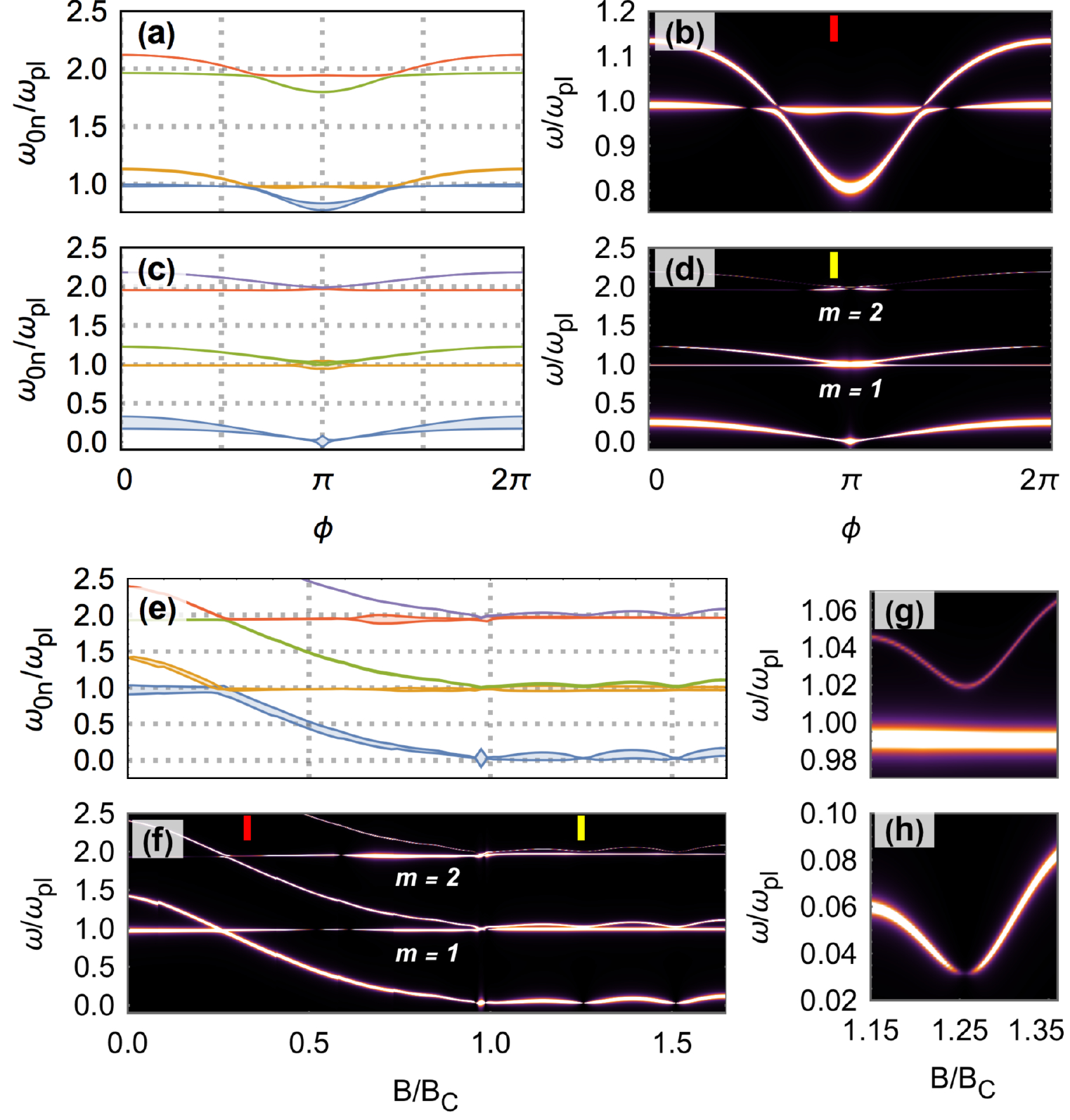}
\caption{\label{fig:19} \editRR{\textbf{MW spectroscopy in a split-junction geometry} with $E_L/E_J=60$. Parameters for each NW segment as in Fig. \ref{Fig2}(a). (a) Phase dispersion of the transition frequencies and their spectral weights (widths of the lines) in the trivial regime $B=0.33B_c$. (b) Blowup of the corresponding MW spectrum showing the avoided crossing between the plasma mode and the Andreev level in the NW junction. (c,d) Same in the topological regime, $B=1.25B_c$, showing plasma replicas of the underlying $4\pi$-periodic Josephson effect in the NW junction. (e,f) Magnetic field dependence near $\phi=\pi$. (g,h) Blowup near the first minimum at $B\approx 1.25B_c$.}}
\end{figure}
In this configuration, the visibility of the allowed MW transitions is just given by the matrix elements $\langle n|\hat I |0\rangle$. The above model generalizes the Jaynes--Cummings model that results from considering only one Andreev level in the junction \cite{PhysRevB.85.180506,PhysRevB.90.134506,vanWoerkom:NP17,PhysRevLett.121.047001}. Indeed, the topologically trivial $B<B_c$ regime captures this Jaynes--Cummings physics where an Andreev level strongly dispersing with phase $E_A(\phi)$ anticrosses with the plasma mode [Figs. \ref{fig:19}(a,b)], in good agreement with previous experiments \cite{Bretheau:N13,vanWoerkom:NP17,PhysRevLett.121.047001}. When $B>B_c$ the levels show a characteristic phase dispersion with a zero-energy crossing at $\phi=\pi$, see Figs. \ref{fig:19}(c,d). \editJJ{For $L_S\gg\xi_M$, these zero-energy crossings (kinks) in the MW response are a strong signature of the so-called $4\pi$-periodic Josephson effect (a similar response with kinks at $\phi=\pi$ has been discussed in topological JJs based on proximitized helical edge modes in a two-dimensional topological insulator~\cite{PhysRevB.92.134508}).} Apart from the fundamental transition, the MW response shows higher order processes where $m$ plasma modes are excited, which results in transitions occurring at $m\omega_{pl}$ and $E_A(\phi)+m\omega_{pl}$ \cite{Bretheau:N13}, giving rise to multiple replicas of the  $4\pi$-periodic Josephson effect. Interestingly, the residual splittings at $\phi=\pi$ owing to Majorana overlaps \cite{PhysRevLett.108.257001,PhysRevB.86.140504} are imprinted on each of these replicas that, near $\phi=\pi$, mimic the oscillatory Majorana behavior as a function of $B$ [Figs. \ref{fig:19} (e,f)]. Therefore, multiple replicas of the plasma frequency oscillating against magnetic field are a strong indication of Majoranas in the NW island. As before, minima of the Majorana oscillations result in spectral holes.}

\section{\label{remarks}Final remarks and conclusions}
We have presented a detailed analysis of the MW response of superconducting islands where the weak link in the Josephson element is formed by a proximitized semiconducting NW. Specifically, we describe the JJ as two segments of a single-mode semiconductor NW that are proximitized by a conventional s-wave superconductor (the so-called Lutchyn-Oreg model) separated by a short normal region. The BdG spectrum of such weak link creates the Josephson potential $V_J(\varphi)$ that enters the superconducting island Hamiltonian substituting the standard $V_J(\varphi)=-E_J\cos(\varphi)$ in conventional superconducting islands. Our description allows to uncover all the relevant regimes (from the trivial to the topological one) as the external Zeeman field increases. It takes into account both standard Josephson events due to Cooper pair tunneling, as well as anomalous Majorana-mediated events where a single electron is transferred across the junction. This anomalous single-electron Josephson tunneling is governed by the subgap excitations of the BdG Hamiltonian, whose dynamics are fully taken into account by means of a projection technique. Quite naturally, the superconducting island properties depend on important microscopic NW parameters, \editRR{notably the energy splitting between different fermionic parities on the junction, $\delta$ (twice the so-called Majorana splitting $E_0$ due to finite length on each NW segment)}, and the single-electron contribution to the Josephson coupling, $E_M$. These new scales in the problem, together with $E_J$ and $E_C$, define novel regimes such as e.g. $E_M/E_C\gtrsim 1$ and/or  $\delta/E_C\gtrsim 1$, hitherto unexplored in the literature and relevant for current experiments using NW JJs. 

Our results demonstrate that the MW response is a very useful tool to study Majorana physics in such junctions. \editE{Being a global measurement, it allows to avoid some of the issues that challenge the interpretation of zero-bias anomalies in tunneling spectroscopy \cite{prada2019}.} As we discuss, the typical experimental knobs in standard transport experiments for Majorana detection (i. e., the external Zeeman field), can be supplemented with other knobs that characterize the island ($n_g$, $E_C$, $E_J$) in order to unveil Majorana physics in the junction. This, in particular, allows to fully characterize Majorana oscillations and their concomitant fermion parity crossings. 

The discussion presented in this paper is based on the simplest model that \editP{can describe} all the relevant regimes. The \editP{analysis performed} here may be readily extended to other relevant NW regimes not discussed here, like multiband NWs \cite{San-Jose:PRL14}, the role of the electrostatic environment \cite{Vuik:NJP16,Dominguez:NQM17,Antipov:18,Escribano:BJN18,PhysRevB.99.245408} and orbital effects \cite{PhysRevB.93.235434,PhysRevLett.119.037701,PhysRevB.97.045419,PhysRevB.97.235445}. Other geometries currently under intensive experimental study, including junctions with quantum dots \cite{bargerbos2019,kringhj2019}, superconducting islands in the fluxonium regime \cite{pitavidal2019} and gatemons based on full-shell NWs \cite{Vaitiekenas:S20,Penaranda:19}, can be also studied using our method. While the focus of the paper is on semiconductor-NW junctions, our procedure is general and can be applied to other weak links and gate-tunable JJs where \editP{the subgap BdG spectrum is a crucial contribution to the Josephson potential}. Novel systems where our method could be extremely useful include two-dimensional semiconductor gases proximitized by superconductors \cite{Casparis2018} and Van der Waals heterostructures \cite{Kroll2018,Schmidt2018,Wang2019,Tahan19}.

\acknowledgements

Research supported by the Spanish Ministry of Science, Innovation and Universities through Grants PGC2018-097018-B-I00, FIS2016-80434-P (AEI/FEDER, EU), BES-2016-078122 (FPI program), RYC-2011-09345 (Ram\'on y Cajal program) and the Mar\'ia de Maeztu Program for Units of Excellence in R\&D (CEX2018-000805-M).
We also acknowledge support from the EU Horizon 2020 research and innovation program under the FETOPEN Grant Agreement No. 828948 and from the CSIC Research Platform on Quantum Technologies PTI-001.

\bibliography{biblio}

\begin{thebibliography}{85}%
\makeatletter
\providecommand \@ifxundefined [1]{%
 \@ifx{#1\undefined}
}%
\providecommand \@ifnum [1]{%
 \ifnum #1\expandafter \@firstoftwo
 \else \expandafter \@secondoftwo
 \fi
}%
\providecommand \@ifx [1]{%
 \ifx #1\expandafter \@firstoftwo
 \else \expandafter \@secondoftwo
 \fi
}%
\providecommand \natexlab [1]{#1}%
\providecommand \enquote  [1]{``#1''}%
\providecommand \bibnamefont  [1]{#1}%
\providecommand \bibfnamefont [1]{#1}%
\providecommand \citenamefont [1]{#1}%
\providecommand \href@noop [0]{\@secondoftwo}%
\providecommand \href [0]{\begingroup \@sanitize@url \@href}%
\providecommand \@href[1]{\@@startlink{#1}\@@href}%
\providecommand \@@href[1]{\endgroup#1\@@endlink}%
\providecommand \@sanitize@url [0]{\catcode `\\12\catcode `\$12\catcode
  `\&12\catcode `\#12\catcode `\^12\catcode `\_12\catcode `\%12\relax}%
\providecommand \@@startlink[1]{}%
\providecommand \@@endlink[0]{}%
\providecommand \url  [0]{\begingroup\@sanitize@url \@url }%
\providecommand \@url [1]{\endgroup\@href {#1}{\urlprefix }}%
\providecommand \urlprefix  [0]{URL }%
\providecommand \Eprint [0]{\href }%
\providecommand \doibase [0]{http://dx.doi.org/}%
\providecommand \selectlanguage [0]{\@gobble}%
\providecommand \bibinfo  [0]{\@secondoftwo}%
\providecommand \bibfield  [0]{\@secondoftwo}%
\providecommand \translation [1]{[#1]}%
\providecommand \BibitemOpen [0]{}%
\providecommand \bibitemStop [0]{}%
\providecommand \bibitemNoStop [0]{.\EOS\space}%
\providecommand \EOS [0]{\spacefactor3000\relax}%
\providecommand \BibitemShut  [1]{\csname bibitem#1\endcsname}%
\let\auto@bib@innerbib\@empty
\bibitem [{\citenamefont {Blais}\ \emph {et~al.}(2004)\citenamefont {Blais},
  \citenamefont {Huang}, \citenamefont {Wallraff}, \citenamefont {Girvin},\
  and\ \citenamefont {Schoelkopf}}]{PhysRevA.69.062320}%
  \BibitemOpen
  \bibfield  {author} {\bibinfo {author} {\bibfnamefont {A.}~\bibnamefont
  {Blais}}, \bibinfo {author} {\bibfnamefont {R.-S.}\ \bibnamefont {Huang}},
  \bibinfo {author} {\bibfnamefont {A.}~\bibnamefont {Wallraff}}, \bibinfo
  {author} {\bibfnamefont {S.~M.}\ \bibnamefont {Girvin}}, \ and\ \bibinfo
  {author} {\bibfnamefont {R.~J.}\ \bibnamefont {Schoelkopf}},\ }\href
  {\doibase 10.1103/PhysRevA.69.062320} {\bibfield  {journal} {\bibinfo
  {journal} {Phys. Rev. A}\ }\textbf {\bibinfo {volume} {69}},\ \bibinfo
  {pages} {062320} (\bibinfo {year} {2004})}\BibitemShut {NoStop}%
\bibitem [{\citenamefont {Devoret}\ and\ \citenamefont
  {Schoelkopf}(2013)}]{Devoret1169}%
  \BibitemOpen
  \bibfield  {author} {\bibinfo {author} {\bibfnamefont {M.~H.}\ \bibnamefont
  {Devoret}}\ and\ \bibinfo {author} {\bibfnamefont {R.~J.}\ \bibnamefont
  {Schoelkopf}},\ }\href {\doibase 10.1126/science.1231930} {\bibfield
  {journal} {\bibinfo  {journal} {Science}\ }\textbf {\bibinfo {volume}
  {339}},\ \bibinfo {pages} {1169} (\bibinfo {year} {2013})}\BibitemShut
  {NoStop}%
\bibitem [{\citenamefont {Wendin}(2017)}]{Wendin_2017}%
  \BibitemOpen
  \bibfield  {author} {\bibinfo {author} {\bibfnamefont {G.}~\bibnamefont
  {Wendin}},\ }\href {\doibase 10.1088/1361-6633/aa7e1a} {\bibfield  {journal}
  {\bibinfo  {journal} {Reports on Progress in Physics}\ }\textbf {\bibinfo
  {volume} {80}},\ \bibinfo {pages} {106001} (\bibinfo {year}
  {2017})}\BibitemShut {NoStop}%
\bibitem [{\citenamefont {Bouchiat}\ \emph {et~al.}(1998)\citenamefont
  {Bouchiat}, \citenamefont {Vion}, \citenamefont {Joyez}, \citenamefont
  {Esteve},\ and\ \citenamefont {Devoret}}]{Bouchiat_1998}%
  \BibitemOpen
  \bibfield  {author} {\bibinfo {author} {\bibfnamefont {V.}~\bibnamefont
  {Bouchiat}}, \bibinfo {author} {\bibfnamefont {D.}~\bibnamefont {Vion}},
  \bibinfo {author} {\bibfnamefont {P.}~\bibnamefont {Joyez}}, \bibinfo
  {author} {\bibfnamefont {D.}~\bibnamefont {Esteve}}, \ and\ \bibinfo {author}
  {\bibfnamefont {M.~H.}\ \bibnamefont {Devoret}},\ }\href {\doibase
  10.1238/physica.topical.076a00165} {\bibfield  {journal} {\bibinfo  {journal}
  {Physica Scripta}\ }\textbf {\bibinfo {volume} {T76}},\ \bibinfo {pages}
  {165} (\bibinfo {year} {1998})}\BibitemShut {NoStop}%
\bibitem [{\citenamefont {Koch}\ \emph {et~al.}(2007)\citenamefont {Koch},
  \citenamefont {Terri}, \citenamefont {Gambetta}, \citenamefont {Houck},
  \citenamefont {Schuster}, \citenamefont {Majer}, \citenamefont {Blais},
  \citenamefont {Devoret}, \citenamefont {Girvin},\ and\ \citenamefont
  {Schoelkopf}}]{koch2007charge}%
  \BibitemOpen
  \bibfield  {author} {\bibinfo {author} {\bibfnamefont {J.}~\bibnamefont
  {Koch}}, \bibinfo {author} {\bibfnamefont {M.~Y.}\ \bibnamefont {Terri}},
  \bibinfo {author} {\bibfnamefont {J.}~\bibnamefont {Gambetta}}, \bibinfo
  {author} {\bibfnamefont {A.~A.}\ \bibnamefont {Houck}}, \bibinfo {author}
  {\bibfnamefont {D.}~\bibnamefont {Schuster}}, \bibinfo {author}
  {\bibfnamefont {J.}~\bibnamefont {Majer}}, \bibinfo {author} {\bibfnamefont
  {A.}~\bibnamefont {Blais}}, \bibinfo {author} {\bibfnamefont {M.~H.}\
  \bibnamefont {Devoret}}, \bibinfo {author} {\bibfnamefont {S.~M.}\
  \bibnamefont {Girvin}}, \ and\ \bibinfo {author} {\bibfnamefont {R.~J.}\
  \bibnamefont {Schoelkopf}},\ }\href@noop {} {\bibfield  {journal} {\bibinfo
  {journal} {Physical Review A}\ }\textbf {\bibinfo {volume} {76}},\ \bibinfo
  {pages} {042319} (\bibinfo {year} {2007})}\BibitemShut {NoStop}%
\bibitem [{\citenamefont {Larsen}\ \emph {et~al.}(2015)\citenamefont {Larsen},
  \citenamefont {Petersson}, \citenamefont {Kuemmeth}, \citenamefont
  {Jespersen}, \citenamefont {Krogstrup}, \citenamefont {Nyg\aa{}rd},\ and\
  \citenamefont {Marcus}}]{PhysRevLett.115.127001}%
  \BibitemOpen
  \bibfield  {author} {\bibinfo {author} {\bibfnamefont {T.~W.}\ \bibnamefont
  {Larsen}}, \bibinfo {author} {\bibfnamefont {K.~D.}\ \bibnamefont
  {Petersson}}, \bibinfo {author} {\bibfnamefont {F.}~\bibnamefont {Kuemmeth}},
  \bibinfo {author} {\bibfnamefont {T.~S.}\ \bibnamefont {Jespersen}}, \bibinfo
  {author} {\bibfnamefont {P.}~\bibnamefont {Krogstrup}}, \bibinfo {author}
  {\bibfnamefont {J.}~\bibnamefont {Nyg\aa{}rd}}, \ and\ \bibinfo {author}
  {\bibfnamefont {C.~M.}\ \bibnamefont {Marcus}},\ }\href {\doibase
  10.1103/PhysRevLett.115.127001} {\bibfield  {journal} {\bibinfo  {journal}
  {Phys. Rev. Lett.}\ }\textbf {\bibinfo {volume} {115}},\ \bibinfo {pages}
  {127001} (\bibinfo {year} {2015})}\BibitemShut {NoStop}%
\bibitem [{\citenamefont {de~Lange}\ \emph {et~al.}(2015)\citenamefont
  {de~Lange}, \citenamefont {van Heck}, \citenamefont {Bruno}, \citenamefont
  {van Woerkom}, \citenamefont {Geresdi}, \citenamefont {Plissard},
  \citenamefont {Bakkers}, \citenamefont {Akhmerov},\ and\ \citenamefont
  {DiCarlo}}]{PhysRevLett.115.127002}%
  \BibitemOpen
  \bibfield  {author} {\bibinfo {author} {\bibfnamefont {G.}~\bibnamefont
  {de~Lange}}, \bibinfo {author} {\bibfnamefont {B.}~\bibnamefont {van Heck}},
  \bibinfo {author} {\bibfnamefont {A.}~\bibnamefont {Bruno}}, \bibinfo
  {author} {\bibfnamefont {D.~J.}\ \bibnamefont {van Woerkom}}, \bibinfo
  {author} {\bibfnamefont {A.}~\bibnamefont {Geresdi}}, \bibinfo {author}
  {\bibfnamefont {S.~R.}\ \bibnamefont {Plissard}}, \bibinfo {author}
  {\bibfnamefont {E.~P. A.~M.}\ \bibnamefont {Bakkers}}, \bibinfo {author}
  {\bibfnamefont {A.~R.}\ \bibnamefont {Akhmerov}}, \ and\ \bibinfo {author}
  {\bibfnamefont {L.}~\bibnamefont {DiCarlo}},\ }\href {\doibase
  10.1103/PhysRevLett.115.127002} {\bibfield  {journal} {\bibinfo  {journal}
  {Phys. Rev. Lett.}\ }\textbf {\bibinfo {volume} {115}},\ \bibinfo {pages}
  {127002} (\bibinfo {year} {2015})}\BibitemShut {NoStop}%
\bibitem [{\citenamefont {Kringh\o{}j}\ \emph {et~al.}(2018)\citenamefont
  {Kringh\o{}j}, \citenamefont {Casparis}, \citenamefont {Hell}, \citenamefont
  {Larsen}, \citenamefont {Kuemmeth}, \citenamefont {Leijnse}, \citenamefont
  {Flensberg}, \citenamefont {Krogstrup}, \citenamefont {Nyg\aa{}rd},
  \citenamefont {Petersson},\ and\ \citenamefont
  {Marcus}}]{PhysRevB.97.060508}%
  \BibitemOpen
  \bibfield  {author} {\bibinfo {author} {\bibfnamefont {A.}~\bibnamefont
  {Kringh\o{}j}}, \bibinfo {author} {\bibfnamefont {L.}~\bibnamefont
  {Casparis}}, \bibinfo {author} {\bibfnamefont {M.}~\bibnamefont {Hell}},
  \bibinfo {author} {\bibfnamefont {T.~W.}\ \bibnamefont {Larsen}}, \bibinfo
  {author} {\bibfnamefont {F.}~\bibnamefont {Kuemmeth}}, \bibinfo {author}
  {\bibfnamefont {M.}~\bibnamefont {Leijnse}}, \bibinfo {author} {\bibfnamefont
  {K.}~\bibnamefont {Flensberg}}, \bibinfo {author} {\bibfnamefont
  {P.}~\bibnamefont {Krogstrup}}, \bibinfo {author} {\bibfnamefont
  {J.}~\bibnamefont {Nyg\aa{}rd}}, \bibinfo {author} {\bibfnamefont {K.~D.}\
  \bibnamefont {Petersson}}, \ and\ \bibinfo {author} {\bibfnamefont {C.~M.}\
  \bibnamefont {Marcus}},\ }\href {\doibase 10.1103/PhysRevB.97.060508}
  {\bibfield  {journal} {\bibinfo  {journal} {Phys. Rev. B}\ }\textbf {\bibinfo
  {volume} {97}},\ \bibinfo {pages} {060508} (\bibinfo {year}
  {2018})}\BibitemShut {NoStop}%
\bibitem [{\citenamefont {Casparis}\ \emph {et~al.}(2016)\citenamefont
  {Casparis}, \citenamefont {Larsen}, \citenamefont {Olsen}, \citenamefont
  {Kuemmeth}, \citenamefont {Krogstrup}, \citenamefont {Nyg\aa{}rd},
  \citenamefont {Petersson},\ and\ \citenamefont
  {Marcus}}]{PhysRevLett.116.150505}%
  \BibitemOpen
  \bibfield  {author} {\bibinfo {author} {\bibfnamefont {L.}~\bibnamefont
  {Casparis}}, \bibinfo {author} {\bibfnamefont {T.~W.}\ \bibnamefont
  {Larsen}}, \bibinfo {author} {\bibfnamefont {M.~S.}\ \bibnamefont {Olsen}},
  \bibinfo {author} {\bibfnamefont {F.}~\bibnamefont {Kuemmeth}}, \bibinfo
  {author} {\bibfnamefont {P.}~\bibnamefont {Krogstrup}}, \bibinfo {author}
  {\bibfnamefont {J.}~\bibnamefont {Nyg\aa{}rd}}, \bibinfo {author}
  {\bibfnamefont {K.~D.}\ \bibnamefont {Petersson}}, \ and\ \bibinfo {author}
  {\bibfnamefont {C.~M.}\ \bibnamefont {Marcus}},\ }\href {\doibase
  10.1103/PhysRevLett.116.150505} {\bibfield  {journal} {\bibinfo  {journal}
  {Phys. Rev. Lett.}\ }\textbf {\bibinfo {volume} {116}},\ \bibinfo {pages}
  {150505} (\bibinfo {year} {2016})}\BibitemShut {NoStop}%
\bibitem [{\citenamefont {Luthi}\ \emph {et~al.}(2018)\citenamefont {Luthi},
  \citenamefont {Stavenga}, \citenamefont {Enzing}, \citenamefont {Bruno},
  \citenamefont {Dickel}, \citenamefont {Langford}, \citenamefont {Rol},
  \citenamefont {Jespersen}, \citenamefont {Nyg\aa{}rd}, \citenamefont
  {Krogstrup},\ and\ \citenamefont {DiCarlo}}]{PhysRevLett.120.100502}%
  \BibitemOpen
  \bibfield  {author} {\bibinfo {author} {\bibfnamefont {F.}~\bibnamefont
  {Luthi}}, \bibinfo {author} {\bibfnamefont {T.}~\bibnamefont {Stavenga}},
  \bibinfo {author} {\bibfnamefont {O.~W.}\ \bibnamefont {Enzing}}, \bibinfo
  {author} {\bibfnamefont {A.}~\bibnamefont {Bruno}}, \bibinfo {author}
  {\bibfnamefont {C.}~\bibnamefont {Dickel}}, \bibinfo {author} {\bibfnamefont
  {N.~K.}\ \bibnamefont {Langford}}, \bibinfo {author} {\bibfnamefont {M.~A.}\
  \bibnamefont {Rol}}, \bibinfo {author} {\bibfnamefont {T.~S.}\ \bibnamefont
  {Jespersen}}, \bibinfo {author} {\bibfnamefont {J.}~\bibnamefont
  {Nyg\aa{}rd}}, \bibinfo {author} {\bibfnamefont {P.}~\bibnamefont
  {Krogstrup}}, \ and\ \bibinfo {author} {\bibfnamefont {L.}~\bibnamefont
  {DiCarlo}},\ }\href {\doibase 10.1103/PhysRevLett.120.100502} {\bibfield
  {journal} {\bibinfo  {journal} {Phys. Rev. Lett.}\ }\textbf {\bibinfo
  {volume} {120}},\ \bibinfo {pages} {100502} (\bibinfo {year}
  {2018})}\BibitemShut {NoStop}%
\bibitem [{\citenamefont {Casparis}\ \emph {et~al.}(2019)\citenamefont
  {Casparis}, \citenamefont {Pearson}, \citenamefont {Kringh\o{}j},
  \citenamefont {Larsen}, \citenamefont {Kuemmeth}, \citenamefont {Nyg\aa{}rd},
  \citenamefont {Krogstrup}, \citenamefont {Petersson},\ and\ \citenamefont
  {Marcus}}]{PhysRevB.99.085434}%
  \BibitemOpen
  \bibfield  {author} {\bibinfo {author} {\bibfnamefont {L.}~\bibnamefont
  {Casparis}}, \bibinfo {author} {\bibfnamefont {N.~J.}\ \bibnamefont
  {Pearson}}, \bibinfo {author} {\bibfnamefont {A.}~\bibnamefont
  {Kringh\o{}j}}, \bibinfo {author} {\bibfnamefont {T.~W.}\ \bibnamefont
  {Larsen}}, \bibinfo {author} {\bibfnamefont {F.}~\bibnamefont {Kuemmeth}},
  \bibinfo {author} {\bibfnamefont {J.}~\bibnamefont {Nyg\aa{}rd}}, \bibinfo
  {author} {\bibfnamefont {P.}~\bibnamefont {Krogstrup}}, \bibinfo {author}
  {\bibfnamefont {K.~D.}\ \bibnamefont {Petersson}}, \ and\ \bibinfo {author}
  {\bibfnamefont {C.~M.}\ \bibnamefont {Marcus}},\ }\href {\doibase
  10.1103/PhysRevB.99.085434} {\bibfield  {journal} {\bibinfo  {journal} {Phys.
  Rev. B}\ }\textbf {\bibinfo {volume} {99}},\ \bibinfo {pages} {085434}
  (\bibinfo {year} {2019})}\BibitemShut {NoStop}%
\bibitem [{\citenamefont {Casparis}\ \emph {et~al.}(2018)\citenamefont
  {Casparis}, \citenamefont {Connolly}, \citenamefont {Kjaergaard},
  \citenamefont {Pearson}, \citenamefont {Kringh{\o}j}, \citenamefont {Larsen},
  \citenamefont {Kuemmeth}, \citenamefont {Wang}, \citenamefont {Thomas},
  \citenamefont {Gronin}, \citenamefont {Gardner}, \citenamefont {Manfra},
  \citenamefont {Marcus},\ and\ \citenamefont {Petersson}}]{Casparis2018}%
  \BibitemOpen
  \bibfield  {author} {\bibinfo {author} {\bibfnamefont {L.}~\bibnamefont
  {Casparis}}, \bibinfo {author} {\bibfnamefont {M.~R.}\ \bibnamefont
  {Connolly}}, \bibinfo {author} {\bibfnamefont {M.}~\bibnamefont
  {Kjaergaard}}, \bibinfo {author} {\bibfnamefont {N.~J.}\ \bibnamefont
  {Pearson}}, \bibinfo {author} {\bibfnamefont {A.}~\bibnamefont
  {Kringh{\o}j}}, \bibinfo {author} {\bibfnamefont {T.~W.}\ \bibnamefont
  {Larsen}}, \bibinfo {author} {\bibfnamefont {F.}~\bibnamefont {Kuemmeth}},
  \bibinfo {author} {\bibfnamefont {T.}~\bibnamefont {Wang}}, \bibinfo {author}
  {\bibfnamefont {C.}~\bibnamefont {Thomas}}, \bibinfo {author} {\bibfnamefont
  {S.}~\bibnamefont {Gronin}}, \bibinfo {author} {\bibfnamefont {G.~C.}\
  \bibnamefont {Gardner}}, \bibinfo {author} {\bibfnamefont {M.~J.}\
  \bibnamefont {Manfra}}, \bibinfo {author} {\bibfnamefont {C.~M.}\
  \bibnamefont {Marcus}}, \ and\ \bibinfo {author} {\bibfnamefont {K.~D.}\
  \bibnamefont {Petersson}},\ }\href@noop {} {\bibfield  {journal} {\bibinfo
  {journal} {Nature Nanotechnology}\ }\textbf {\bibinfo {volume} {13}},\
  \bibinfo {pages} {915} (\bibinfo {year} {2018})}\BibitemShut {NoStop}%
\bibitem [{\citenamefont {Kroll}\ \emph {et~al.}(2018)\citenamefont {Kroll},
  \citenamefont {Uilhoorn}, \citenamefont {van~der Enden}, \citenamefont
  {de~Jong}, \citenamefont {Watanabe}, \citenamefont {Taniguchi}, \citenamefont
  {Goswami}, \citenamefont {Cassidy},\ and\ \citenamefont
  {Kouwenhoven}}]{Kroll2018}%
  \BibitemOpen
  \bibfield  {author} {\bibinfo {author} {\bibfnamefont {J.~G.}\ \bibnamefont
  {Kroll}}, \bibinfo {author} {\bibfnamefont {W.}~\bibnamefont {Uilhoorn}},
  \bibinfo {author} {\bibfnamefont {K.~L.}\ \bibnamefont {van~der Enden}},
  \bibinfo {author} {\bibfnamefont {D.}~\bibnamefont {de~Jong}}, \bibinfo
  {author} {\bibfnamefont {K.}~\bibnamefont {Watanabe}}, \bibinfo {author}
  {\bibfnamefont {T.}~\bibnamefont {Taniguchi}}, \bibinfo {author}
  {\bibfnamefont {S.}~\bibnamefont {Goswami}}, \bibinfo {author} {\bibfnamefont
  {M.~C.}\ \bibnamefont {Cassidy}}, \ and\ \bibinfo {author} {\bibfnamefont
  {L.~P.}\ \bibnamefont {Kouwenhoven}},\ }\href@noop {} {\bibfield  {journal}
  {\bibinfo  {journal} {Nature Communications}\ }\textbf {\bibinfo {volume}
  {9}},\ \bibinfo {pages} {4615} (\bibinfo {year} {2018})}\BibitemShut
  {NoStop}%
\bibitem [{\citenamefont {Schmidt}\ \emph {et~al.}(2018)\citenamefont
  {Schmidt}, \citenamefont {Jenkins}, \citenamefont {Watanabe}, \citenamefont
  {Taniguchi},\ and\ \citenamefont {Steele}}]{Schmidt2018}%
  \BibitemOpen
  \bibfield  {author} {\bibinfo {author} {\bibfnamefont {F.~E.}\ \bibnamefont
  {Schmidt}}, \bibinfo {author} {\bibfnamefont {M.~D.}\ \bibnamefont
  {Jenkins}}, \bibinfo {author} {\bibfnamefont {K.}~\bibnamefont {Watanabe}},
  \bibinfo {author} {\bibfnamefont {T.}~\bibnamefont {Taniguchi}}, \ and\
  \bibinfo {author} {\bibfnamefont {G.~A.}\ \bibnamefont {Steele}},\
  }\href@noop {} {\bibfield  {journal} {\bibinfo  {journal} {Nature
  Communications}\ }\textbf {\bibinfo {volume} {9}},\ \bibinfo {pages} {4069}
  (\bibinfo {year} {2018})}\BibitemShut {NoStop}%
\bibitem [{\citenamefont {Wang}\ \emph {et~al.}(2019)\citenamefont {Wang},
  \citenamefont {Rodan-Legrain}, \citenamefont {Bretheau}, \citenamefont
  {Campbell}, \citenamefont {Kannan}, \citenamefont {Kim}, \citenamefont
  {Kjaergaard}, \citenamefont {Krantz}, \citenamefont {Samach}, \citenamefont
  {Yan}, \citenamefont {Yoder}, \citenamefont {Watanabe}, \citenamefont
  {Taniguchi}, \citenamefont {Orlando}, \citenamefont {Gustavsson},
  \citenamefont {Jarillo-Herrero},\ and\ \citenamefont {Oliver}}]{Wang2019}%
  \BibitemOpen
  \bibfield  {author} {\bibinfo {author} {\bibfnamefont {J.~I.-J.}\
  \bibnamefont {Wang}}, \bibinfo {author} {\bibfnamefont {D.}~\bibnamefont
  {Rodan-Legrain}}, \bibinfo {author} {\bibfnamefont {L.}~\bibnamefont
  {Bretheau}}, \bibinfo {author} {\bibfnamefont {D.~L.}\ \bibnamefont
  {Campbell}}, \bibinfo {author} {\bibfnamefont {B.}~\bibnamefont {Kannan}},
  \bibinfo {author} {\bibfnamefont {D.}~\bibnamefont {Kim}}, \bibinfo {author}
  {\bibfnamefont {M.}~\bibnamefont {Kjaergaard}}, \bibinfo {author}
  {\bibfnamefont {P.}~\bibnamefont {Krantz}}, \bibinfo {author} {\bibfnamefont
  {G.~O.}\ \bibnamefont {Samach}}, \bibinfo {author} {\bibfnamefont
  {F.}~\bibnamefont {Yan}}, \bibinfo {author} {\bibfnamefont {J.~L.}\
  \bibnamefont {Yoder}}, \bibinfo {author} {\bibfnamefont {K.}~\bibnamefont
  {Watanabe}}, \bibinfo {author} {\bibfnamefont {T.}~\bibnamefont {Taniguchi}},
  \bibinfo {author} {\bibfnamefont {T.~P.}\ \bibnamefont {Orlando}}, \bibinfo
  {author} {\bibfnamefont {S.}~\bibnamefont {Gustavsson}}, \bibinfo {author}
  {\bibfnamefont {P.}~\bibnamefont {Jarillo-Herrero}}, \ and\ \bibinfo {author}
  {\bibfnamefont {W.~D.}\ \bibnamefont {Oliver}},\ }\href@noop {} {\bibfield
  {journal} {\bibinfo  {journal} {Nature Nanotechnology}\ }\textbf {\bibinfo
  {volume} {14}},\ \bibinfo {pages} {120} (\bibinfo {year} {2019})}\BibitemShut
  {NoStop}%
\bibitem [{\citenamefont {Tahan}(2019)}]{Tahan19}%
  \BibitemOpen
  \bibfield  {author} {\bibinfo {author} {\bibfnamefont {C.}~\bibnamefont
  {Tahan}},\ }\href@noop {} {\bibfield  {journal} {\bibinfo  {journal} {Nature
  Nanotechnology}\ }\textbf {\bibinfo {volume} {14}},\ \bibinfo {pages} {102}
  (\bibinfo {year} {2019})}\BibitemShut {NoStop}%
\bibitem [{\citenamefont {Leijnse}\ and\ \citenamefont
  {Flensberg}(2012)}]{Leijnse:SSAT12}%
  \BibitemOpen
  \bibfield  {author} {\bibinfo {author} {\bibfnamefont {M.}~\bibnamefont
  {Leijnse}}\ and\ \bibinfo {author} {\bibfnamefont {K.}~\bibnamefont
  {Flensberg}},\ }\href {http://stacks.iop.org/0268-1242/27/i=12/a=124003}
  {\bibfield  {journal} {\bibinfo  {journal} {Semicond. Sci. Technol.}\
  }\textbf {\bibinfo {volume} {27}},\ \bibinfo {pages} {124003} (\bibinfo
  {year} {2012})}\BibitemShut {NoStop}%
\bibitem [{\citenamefont {Alicea}(2012)}]{Alicea:RPP12}%
  \BibitemOpen
  \bibfield  {author} {\bibinfo {author} {\bibfnamefont {J.}~\bibnamefont
  {Alicea}},\ }\href@noop {} {\bibfield  {journal} {\bibinfo  {journal} {Rep.
  Prog. Phys.}\ }\textbf {\bibinfo {volume} {75}},\ \bibinfo {pages} {076501}
  (\bibinfo {year} {2012})}\BibitemShut {NoStop}%
\bibitem [{\citenamefont {Beenakker}(2013)}]{Beenakker:ARCMP13}%
  \BibitemOpen
  \bibfield  {author} {\bibinfo {author} {\bibfnamefont {C.}~\bibnamefont
  {Beenakker}},\ }\href {\doibase 10.1146/annurev-conmatphys-030212-184337}
  {\bibfield  {journal} {\bibinfo  {journal} {Annu. Rev. Cond. Mat. Phys.}\
  }\textbf {\bibinfo {volume} {4}},\ \bibinfo {pages} {113} (\bibinfo {year}
  {2013})}\BibitemShut {NoStop}%
\bibitem [{\citenamefont {Sato}\ and\ \citenamefont
  {Fujimoto}(2016)}]{Sato:JPSJ16}%
  \BibitemOpen
  \bibfield  {author} {\bibinfo {author} {\bibfnamefont {M.}~\bibnamefont
  {Sato}}\ and\ \bibinfo {author} {\bibfnamefont {S.}~\bibnamefont
  {Fujimoto}},\ }\href {\doibase 10.7566/JPSJ.85.072001} {\bibfield  {journal}
  {\bibinfo  {journal} {J. Phys. Soc. Jpn}\ }\textbf {\bibinfo {volume} {85}},\
  \bibinfo {pages} {072001} (\bibinfo {year} {2016})}\BibitemShut {NoStop}%
\bibitem [{\citenamefont {Aguado}(2017)}]{Aguado:RNC17}%
  \BibitemOpen
  \bibfield  {author} {\bibinfo {author} {\bibfnamefont {R.}~\bibnamefont
  {Aguado}},\ }\href {\doibase 10.1393/ncr/i2017-10141-9} {\bibfield  {journal}
  {\bibinfo  {journal} {Riv. Nuovo Cimento}\ }\textbf {\bibinfo {volume}
  {40}},\ \bibinfo {pages} {523} (\bibinfo {year} {2017})}\BibitemShut
  {NoStop}%
\bibitem [{\citenamefont {Sato}\ and\ \citenamefont
  {Ando}(2017)}]{Sato:ROPIP17}%
  \BibitemOpen
  \bibfield  {author} {\bibinfo {author} {\bibfnamefont {M.}~\bibnamefont
  {Sato}}\ and\ \bibinfo {author} {\bibfnamefont {Y.}~\bibnamefont {Ando}},\
  }\href {http://stacks.iop.org/0034-4885/80/i=7/a=076501} {\bibfield
  {journal} {\bibinfo  {journal} {Rep. Prog. Phys.}\ }\textbf {\bibinfo
  {volume} {80}},\ \bibinfo {pages} {076501} (\bibinfo {year}
  {2017})}\BibitemShut {NoStop}%
\bibitem [{\citenamefont {Lutchyn}\ \emph {et~al.}(2018)\citenamefont
  {Lutchyn}, \citenamefont {Bakkers}, \citenamefont {Kouwenhoven},
  \citenamefont {Krogstrup}, \citenamefont {Marcus},\ and\ \citenamefont
  {Oreg}}]{Lutchyn:NRM18}%
  \BibitemOpen
  \bibfield  {author} {\bibinfo {author} {\bibfnamefont {R.~M.}\ \bibnamefont
  {Lutchyn}}, \bibinfo {author} {\bibfnamefont {E.~P. A.~M.}\ \bibnamefont
  {Bakkers}}, \bibinfo {author} {\bibfnamefont {L.~P.}\ \bibnamefont
  {Kouwenhoven}}, \bibinfo {author} {\bibfnamefont {P.}~\bibnamefont
  {Krogstrup}}, \bibinfo {author} {\bibfnamefont {C.~M.}\ \bibnamefont
  {Marcus}}, \ and\ \bibinfo {author} {\bibfnamefont {Y.}~\bibnamefont
  {Oreg}},\ }\href {\doibase 10.1038/s41578-018-0003-1} {\bibfield  {journal}
  {\bibinfo  {journal} {Nature Reviews Materials}\ }\textbf {\bibinfo {volume}
  {3}},\ \bibinfo {pages} {52} (\bibinfo {year} {2018})}\BibitemShut {NoStop}%
\bibitem [{\citenamefont {Hassler}\ \emph {et~al.}(2011)\citenamefont
  {Hassler}, \citenamefont {Akhmerov},\ and\ \citenamefont
  {Beenakker}}]{Hassler_2011}%
  \BibitemOpen
  \bibfield  {author} {\bibinfo {author} {\bibfnamefont {F.}~\bibnamefont
  {Hassler}}, \bibinfo {author} {\bibfnamefont {A.~R.}\ \bibnamefont
  {Akhmerov}}, \ and\ \bibinfo {author} {\bibfnamefont {C.~W.~J.}\ \bibnamefont
  {Beenakker}},\ }\href@noop {} {\bibfield  {journal} {\bibinfo  {journal} {New
  Journal of Physics}\ }\textbf {\bibinfo {volume} {13}},\ \bibinfo {pages}
  {095004} (\bibinfo {year} {2011})}\BibitemShut {NoStop}%
\bibitem [{\citenamefont {M\"uller}\ \emph {et~al.}(2013)\citenamefont
  {M\"uller}, \citenamefont {Bourassa},\ and\ \citenamefont
  {Blais}}]{PhysRevB.88.235401}%
  \BibitemOpen
  \bibfield  {author} {\bibinfo {author} {\bibfnamefont {C.}~\bibnamefont
  {M\"uller}}, \bibinfo {author} {\bibfnamefont {J.}~\bibnamefont {Bourassa}},
  \ and\ \bibinfo {author} {\bibfnamefont {A.}~\bibnamefont {Blais}},\ }\href
  {\doibase 10.1103/PhysRevB.88.235401} {\bibfield  {journal} {\bibinfo
  {journal} {Phys. Rev. B}\ }\textbf {\bibinfo {volume} {88}},\ \bibinfo
  {pages} {235401} (\bibinfo {year} {2013})}\BibitemShut {NoStop}%
\bibitem [{\citenamefont {Pekker}\ \emph {et~al.}(2013)\citenamefont {Pekker},
  \citenamefont {Hou}, \citenamefont {Manucharyan},\ and\ \citenamefont
  {Demler}}]{PhysRevLett.111.107007}%
  \BibitemOpen
  \bibfield  {author} {\bibinfo {author} {\bibfnamefont {D.}~\bibnamefont
  {Pekker}}, \bibinfo {author} {\bibfnamefont {C.-Y.}\ \bibnamefont {Hou}},
  \bibinfo {author} {\bibfnamefont {V.~E.}\ \bibnamefont {Manucharyan}}, \ and\
  \bibinfo {author} {\bibfnamefont {E.}~\bibnamefont {Demler}},\ }\href
  {\doibase 10.1103/PhysRevLett.111.107007} {\bibfield  {journal} {\bibinfo
  {journal} {Phys. Rev. Lett.}\ }\textbf {\bibinfo {volume} {111}},\ \bibinfo
  {pages} {107007} (\bibinfo {year} {2013})}\BibitemShut {NoStop}%
\bibitem [{\citenamefont {Virtanen}\ and\ \citenamefont
  {Recher}(2013)}]{PhysRevB.88.144507}%
  \BibitemOpen
  \bibfield  {author} {\bibinfo {author} {\bibfnamefont {P.}~\bibnamefont
  {Virtanen}}\ and\ \bibinfo {author} {\bibfnamefont {P.}~\bibnamefont
  {Recher}},\ }\href {\doibase 10.1103/PhysRevB.88.144507} {\bibfield
  {journal} {\bibinfo  {journal} {Phys. Rev. B}\ }\textbf {\bibinfo {volume}
  {88}},\ \bibinfo {pages} {144507} (\bibinfo {year} {2013})}\BibitemShut
  {NoStop}%
\bibitem [{\citenamefont {Ginossar}\ and\ \citenamefont
  {Grosfeld}(2014)}]{ginossar2014microwave}%
  \BibitemOpen
  \bibfield  {author} {\bibinfo {author} {\bibfnamefont {E.}~\bibnamefont
  {Ginossar}}\ and\ \bibinfo {author} {\bibfnamefont {E.}~\bibnamefont
  {Grosfeld}},\ }\href@noop {} {\bibfield  {journal} {\bibinfo  {journal}
  {Nature communications}\ }\textbf {\bibinfo {volume} {5}},\ \bibinfo {pages}
  {4772} (\bibinfo {year} {2014})}\BibitemShut {NoStop}%
\bibitem [{\citenamefont {Yavilberg}\ \emph {et~al.}(2015)\citenamefont
  {Yavilberg}, \citenamefont {Ginossar},\ and\ \citenamefont
  {Grosfeld}}]{PhysRevB.92.075143}%
  \BibitemOpen
  \bibfield  {author} {\bibinfo {author} {\bibfnamefont {K.}~\bibnamefont
  {Yavilberg}}, \bibinfo {author} {\bibfnamefont {E.}~\bibnamefont {Ginossar}},
  \ and\ \bibinfo {author} {\bibfnamefont {E.}~\bibnamefont {Grosfeld}},\
  }\href {\doibase 10.1103/PhysRevB.92.075143} {\bibfield  {journal} {\bibinfo
  {journal} {Phys. Rev. B}\ }\textbf {\bibinfo {volume} {92}},\ \bibinfo
  {pages} {075143} (\bibinfo {year} {2015})}\BibitemShut {NoStop}%
\bibitem [{\citenamefont {Dmytruk}\ \emph {et~al.}(2015)\citenamefont
  {Dmytruk}, \citenamefont {Trif},\ and\ \citenamefont
  {Simon}}]{PhysRevB.92.245432}%
  \BibitemOpen
  \bibfield  {author} {\bibinfo {author} {\bibfnamefont {O.}~\bibnamefont
  {Dmytruk}}, \bibinfo {author} {\bibfnamefont {M.}~\bibnamefont {Trif}}, \
  and\ \bibinfo {author} {\bibfnamefont {P.}~\bibnamefont {Simon}},\ }\href
  {\doibase 10.1103/PhysRevB.92.245432} {\bibfield  {journal} {\bibinfo
  {journal} {Phys. Rev. B}\ }\textbf {\bibinfo {volume} {92}},\ \bibinfo
  {pages} {245432} (\bibinfo {year} {2015})}\BibitemShut {NoStop}%
\bibitem [{\citenamefont {V\"ayrynen}\ \emph {et~al.}(2015)\citenamefont
  {V\"ayrynen}, \citenamefont {Rastelli}, \citenamefont {Belzig},\ and\
  \citenamefont {Glazman}}]{PhysRevB.92.134508}%
  \BibitemOpen
  \bibfield  {author} {\bibinfo {author} {\bibfnamefont {J.~I.}\ \bibnamefont
  {V\"ayrynen}}, \bibinfo {author} {\bibfnamefont {G.}~\bibnamefont
  {Rastelli}}, \bibinfo {author} {\bibfnamefont {W.}~\bibnamefont {Belzig}}, \
  and\ \bibinfo {author} {\bibfnamefont {L.~I.}\ \bibnamefont {Glazman}},\
  }\href {\doibase 10.1103/PhysRevB.92.134508} {\bibfield  {journal} {\bibinfo
  {journal} {Phys. Rev. B}\ }\textbf {\bibinfo {volume} {92}},\ \bibinfo
  {pages} {134508} (\bibinfo {year} {2015})}\BibitemShut {NoStop}%
\bibitem [{\citenamefont {Peng}\ \emph {et~al.}(2016)\citenamefont {Peng},
  \citenamefont {Pientka}, \citenamefont {Berg}, \citenamefont {Oreg},\ and\
  \citenamefont {von Oppen}}]{PhysRevB.94.085409}%
  \BibitemOpen
  \bibfield  {author} {\bibinfo {author} {\bibfnamefont {Y.}~\bibnamefont
  {Peng}}, \bibinfo {author} {\bibfnamefont {F.}~\bibnamefont {Pientka}},
  \bibinfo {author} {\bibfnamefont {E.}~\bibnamefont {Berg}}, \bibinfo {author}
  {\bibfnamefont {Y.}~\bibnamefont {Oreg}}, \ and\ \bibinfo {author}
  {\bibfnamefont {F.}~\bibnamefont {von Oppen}},\ }\href {\doibase
  10.1103/PhysRevB.94.085409} {\bibfield  {journal} {\bibinfo  {journal} {Phys.
  Rev. B}\ }\textbf {\bibinfo {volume} {94}},\ \bibinfo {pages} {085409}
  (\bibinfo {year} {2016})}\BibitemShut {NoStop}%
\bibitem [{\citenamefont {Dartiailh}\ \emph {et~al.}(2017)\citenamefont
  {Dartiailh}, \citenamefont {Kontos}, \citenamefont
  {Dou\ifmmode~\mbox{\c{c}}\else \c{c}\fi{}ot},\ and\ \citenamefont
  {Cottet}}]{PhysRevLett.118.126803}%
  \BibitemOpen
  \bibfield  {author} {\bibinfo {author} {\bibfnamefont {M.~C.}\ \bibnamefont
  {Dartiailh}}, \bibinfo {author} {\bibfnamefont {T.}~\bibnamefont {Kontos}},
  \bibinfo {author} {\bibfnamefont {B.}~\bibnamefont
  {Dou\ifmmode~\mbox{\c{c}}\else \c{c}\fi{}ot}}, \ and\ \bibinfo {author}
  {\bibfnamefont {A.}~\bibnamefont {Cottet}},\ }\href {\doibase
  10.1103/PhysRevLett.118.126803} {\bibfield  {journal} {\bibinfo  {journal}
  {Phys. Rev. Lett.}\ }\textbf {\bibinfo {volume} {118}},\ \bibinfo {pages}
  {126803} (\bibinfo {year} {2017})}\BibitemShut {NoStop}%
\bibitem [{\citenamefont {Trif}\ \emph {et~al.}(2018)\citenamefont {Trif},
  \citenamefont {Dmytruk}, \citenamefont {Bouchiat}, \citenamefont {Aguado},\
  and\ \citenamefont {Simon}}]{PhysRevB.97.041415}%
  \BibitemOpen
  \bibfield  {author} {\bibinfo {author} {\bibfnamefont {M.}~\bibnamefont
  {Trif}}, \bibinfo {author} {\bibfnamefont {O.}~\bibnamefont {Dmytruk}},
  \bibinfo {author} {\bibfnamefont {H.}~\bibnamefont {Bouchiat}}, \bibinfo
  {author} {\bibfnamefont {R.}~\bibnamefont {Aguado}}, \ and\ \bibinfo {author}
  {\bibfnamefont {P.}~\bibnamefont {Simon}},\ }\href {\doibase
  10.1103/PhysRevB.97.041415} {\bibfield  {journal} {\bibinfo  {journal} {Phys.
  Rev. B}\ }\textbf {\bibinfo {volume} {97}},\ \bibinfo {pages} {041415}
  (\bibinfo {year} {2018})}\BibitemShut {NoStop}%
\bibitem [{\citenamefont {Keselman}\ \emph {et~al.}(2019)\citenamefont
  {Keselman}, \citenamefont {Murthy}, \citenamefont {van Heck},\ and\
  \citenamefont {Bauer}}]{10.21468/SciPostPhys.7.4.050}%
  \BibitemOpen
  \bibfield  {author} {\bibinfo {author} {\bibfnamefont {A.}~\bibnamefont
  {Keselman}}, \bibinfo {author} {\bibfnamefont {C.}~\bibnamefont {Murthy}},
  \bibinfo {author} {\bibfnamefont {B.}~\bibnamefont {van Heck}}, \ and\
  \bibinfo {author} {\bibfnamefont {B.}~\bibnamefont {Bauer}},\ }\href
  {\doibase 10.21468/SciPostPhys.7.4.050} {\bibfield  {journal} {\bibinfo
  {journal} {SciPost Phys.}\ }\textbf {\bibinfo {volume} {7}},\ \bibinfo
  {pages} {50} (\bibinfo {year} {2019})}\BibitemShut {NoStop}%
\bibitem [{\citenamefont {Goffman}\ \emph {et~al.}(2017)\citenamefont
  {Goffman}, \citenamefont {Urbina}, \citenamefont {Pothier}, \citenamefont
  {Nygard}, \citenamefont {Marcus},\ and\ \citenamefont
  {Krogstrup}}]{Goffman_2017}%
  \BibitemOpen
  \bibfield  {author} {\bibinfo {author} {\bibfnamefont {M.~F.}\ \bibnamefont
  {Goffman}}, \bibinfo {author} {\bibfnamefont {C.}~\bibnamefont {Urbina}},
  \bibinfo {author} {\bibfnamefont {H.}~\bibnamefont {Pothier}}, \bibinfo
  {author} {\bibfnamefont {J.}~\bibnamefont {Nygard}}, \bibinfo {author}
  {\bibfnamefont {C.~M.}\ \bibnamefont {Marcus}}, \ and\ \bibinfo {author}
  {\bibfnamefont {P.}~\bibnamefont {Krogstrup}},\ }\href {\doibase
  10.1088/1367-2630/aa7641} {\bibfield  {journal} {\bibinfo  {journal} {New
  Journal of Physics}\ }\textbf {\bibinfo {volume} {19}},\ \bibinfo {pages}
  {092002} (\bibinfo {year} {2017})}\BibitemShut {NoStop}%
\bibitem [{\citenamefont {Tosi}\ \emph {et~al.}(2019)\citenamefont {Tosi},
  \citenamefont {Metzger}, \citenamefont {Goffman}, \citenamefont {Urbina},
  \citenamefont {Pothier}, \citenamefont {Park}, \citenamefont {Yeyati},
  \citenamefont {Nyg\aa{}rd},\ and\ \citenamefont
  {Krogstrup}}]{PhysRevX.9.011010}%
  \BibitemOpen
  \bibfield  {author} {\bibinfo {author} {\bibfnamefont {L.}~\bibnamefont
  {Tosi}}, \bibinfo {author} {\bibfnamefont {C.}~\bibnamefont {Metzger}},
  \bibinfo {author} {\bibfnamefont {M.~F.}\ \bibnamefont {Goffman}}, \bibinfo
  {author} {\bibfnamefont {C.}~\bibnamefont {Urbina}}, \bibinfo {author}
  {\bibfnamefont {H.}~\bibnamefont {Pothier}}, \bibinfo {author} {\bibfnamefont
  {S.}~\bibnamefont {Park}}, \bibinfo {author} {\bibfnamefont {A.~L.}\
  \bibnamefont {Yeyati}}, \bibinfo {author} {\bibfnamefont {J.}~\bibnamefont
  {Nyg\aa{}rd}}, \ and\ \bibinfo {author} {\bibfnamefont {P.}~\bibnamefont
  {Krogstrup}},\ }\href {\doibase 10.1103/PhysRevX.9.011010} {\bibfield
  {journal} {\bibinfo  {journal} {Phys. Rev. X}\ }\textbf {\bibinfo {volume}
  {9}},\ \bibinfo {pages} {011010} (\bibinfo {year} {2019})}\BibitemShut
  {NoStop}%
\bibitem [{\citenamefont {Fu}\ and\ \citenamefont {Kane}(2008)}]{Fu:PRL08}%
  \BibitemOpen
  \bibfield  {author} {\bibinfo {author} {\bibfnamefont {L.}~\bibnamefont
  {Fu}}\ and\ \bibinfo {author} {\bibfnamefont {C.~L.}\ \bibnamefont {Kane}},\
  }\href {\doibase 10.1103/PhysRevLett.100.096407} {\bibfield  {journal}
  {\bibinfo  {journal} {Phys. Rev. Lett.}\ }\textbf {\bibinfo {volume} {100}},\
  \bibinfo {pages} {096407} (\bibinfo {year} {2008})}\BibitemShut {NoStop}%
\bibitem [{\citenamefont {Lutchyn}\ \emph {et~al.}(2010)\citenamefont
  {Lutchyn}, \citenamefont {Sau},\ and\ \citenamefont
  {Das~Sarma}}]{Lutchyn:PRL10}%
  \BibitemOpen
  \bibfield  {author} {\bibinfo {author} {\bibfnamefont {R.~M.}\ \bibnamefont
  {Lutchyn}}, \bibinfo {author} {\bibfnamefont {J.~D.}\ \bibnamefont {Sau}}, \
  and\ \bibinfo {author} {\bibfnamefont {S.}~\bibnamefont {Das~Sarma}},\ }\href
  {\doibase 10.1103/PhysRevLett.105.077001} {\bibfield  {journal} {\bibinfo
  {journal} {Phys. Rev. Lett.}\ }\textbf {\bibinfo {volume} {105}},\ \bibinfo
  {pages} {077001} (\bibinfo {year} {2010})}\BibitemShut {NoStop}%
\bibitem [{\citenamefont {Oreg}\ \emph {et~al.}(2010)\citenamefont {Oreg},
  \citenamefont {Refael},\ and\ \citenamefont {von Oppen}}]{Oreg:PRL10}%
  \BibitemOpen
  \bibfield  {author} {\bibinfo {author} {\bibfnamefont {Y.}~\bibnamefont
  {Oreg}}, \bibinfo {author} {\bibfnamefont {G.}~\bibnamefont {Refael}}, \ and\
  \bibinfo {author} {\bibfnamefont {F.}~\bibnamefont {von Oppen}},\ }\href
  {\doibase 10.1103/PhysRevLett.105.177002} {\bibfield  {journal} {\bibinfo
  {journal} {Phys. Rev. Lett.}\ }\textbf {\bibinfo {volume} {105}},\ \bibinfo
  {pages} {177002} (\bibinfo {year} {2010})}\BibitemShut {NoStop}%
\bibitem [{\citenamefont {Kitaev}(2001)}]{Kitaev:PU01}%
  \BibitemOpen
  \bibfield  {author} {\bibinfo {author} {\bibfnamefont {A.~Y.}\ \bibnamefont
  {Kitaev}},\ }\href {http://stacks.iop.org/1063-7869/44/i=10S/a=S29}
  {\bibfield  {journal} {\bibinfo  {journal} {Phys. Usp.}\ }\textbf {\bibinfo
  {volume} {44}},\ \bibinfo {pages} {131} (\bibinfo {year} {2001})}\BibitemShut
  {NoStop}%
\bibitem [{\citenamefont {Nayak}\ \emph {et~al.}(2008)\citenamefont {Nayak},
  \citenamefont {Simon}, \citenamefont {Stern}, \citenamefont {Freedman},\ and\
  \citenamefont {Das~Sarma}}]{Nayak:RMP08}%
  \BibitemOpen
  \bibfield  {author} {\bibinfo {author} {\bibfnamefont {C.}~\bibnamefont
  {Nayak}}, \bibinfo {author} {\bibfnamefont {S.}~\bibnamefont {Simon}},
  \bibinfo {author} {\bibfnamefont {A.}~\bibnamefont {Stern}}, \bibinfo
  {author} {\bibfnamefont {M.}~\bibnamefont {Freedman}}, \ and\ \bibinfo
  {author} {\bibfnamefont {S.}~\bibnamefont {Das~Sarma}},\ }\href@noop {}
  {\bibfield  {journal} {\bibinfo  {journal} {Rev. Mod. Phys.}\ }\textbf
  {\bibinfo {volume} {80}},\ \bibinfo {pages} {1083} (\bibinfo {year}
  {2008})}\BibitemShut {NoStop}%
\bibitem [{\citenamefont {Sarma}\ \emph {et~al.}(2015)\citenamefont {Sarma},
  \citenamefont {Freedman},\ and\ \citenamefont {Nayak}}]{Sarma:NQI15}%
  \BibitemOpen
  \bibfield  {author} {\bibinfo {author} {\bibfnamefont {S.~D.}\ \bibnamefont
  {Sarma}}, \bibinfo {author} {\bibfnamefont {M.}~\bibnamefont {Freedman}}, \
  and\ \bibinfo {author} {\bibfnamefont {C.}~\bibnamefont {Nayak}},\ }\href
  {http://dx.doi.org/10.1038/npjqi.2015.1} {\bibfield  {journal} {\bibinfo
  {journal} {Npj Quantum Information}\ }\textbf {\bibinfo {volume} {1}},\
  \bibinfo {pages} {15001 EP } (\bibinfo {year} {2015})}\BibitemShut {NoStop}%
\bibitem [{\citenamefont {van~der Wiel}\ \emph {et~al.}(2002)\citenamefont
  {van~der Wiel}, \citenamefont {De~Franceschi}, \citenamefont {Elzerman},
  \citenamefont {Fujisawa}, \citenamefont {Tarucha},\ and\ \citenamefont
  {Kouwenhoven}}]{RevModPhys.75.1}%
  \BibitemOpen
  \bibfield  {author} {\bibinfo {author} {\bibfnamefont {W.~G.}\ \bibnamefont
  {van~der Wiel}}, \bibinfo {author} {\bibfnamefont {S.}~\bibnamefont
  {De~Franceschi}}, \bibinfo {author} {\bibfnamefont {J.~M.}\ \bibnamefont
  {Elzerman}}, \bibinfo {author} {\bibfnamefont {T.}~\bibnamefont {Fujisawa}},
  \bibinfo {author} {\bibfnamefont {S.}~\bibnamefont {Tarucha}}, \ and\
  \bibinfo {author} {\bibfnamefont {L.~P.}\ \bibnamefont {Kouwenhoven}},\
  }\href {\doibase 10.1103/RevModPhys.75.1} {\bibfield  {journal} {\bibinfo
  {journal} {Rev. Mod. Phys.}\ }\textbf {\bibinfo {volume} {75}},\ \bibinfo
  {pages} {1} (\bibinfo {year} {2002})}\BibitemShut {NoStop}%
\bibitem [{\citenamefont {Li}\ \emph {et~al.}(2018)\citenamefont {Li},
  \citenamefont {Coish}, \citenamefont {Hell}, \citenamefont {Flensberg},\ and\
  \citenamefont {Leijnse}}]{PhysRevB.98.205403}%
  \BibitemOpen
  \bibfield  {author} {\bibinfo {author} {\bibfnamefont {T.}~\bibnamefont
  {Li}}, \bibinfo {author} {\bibfnamefont {W.~A.}\ \bibnamefont {Coish}},
  \bibinfo {author} {\bibfnamefont {M.}~\bibnamefont {Hell}}, \bibinfo {author}
  {\bibfnamefont {K.}~\bibnamefont {Flensberg}}, \ and\ \bibinfo {author}
  {\bibfnamefont {M.}~\bibnamefont {Leijnse}},\ }\href {\doibase
  10.1103/PhysRevB.98.205403} {\bibfield  {journal} {\bibinfo  {journal} {Phys.
  Rev. B}\ }\textbf {\bibinfo {volume} {98}},\ \bibinfo {pages} {205403}
  (\bibinfo {year} {2018})}\BibitemShut {NoStop}%
\bibitem [{\citenamefont {Fu}\ and\ \citenamefont {Kane}(2009)}]{Fu:PRB09}%
  \BibitemOpen
  \bibfield  {author} {\bibinfo {author} {\bibfnamefont {L.}~\bibnamefont
  {Fu}}\ and\ \bibinfo {author} {\bibfnamefont {C.~L.}\ \bibnamefont {Kane}},\
  }\href {\doibase 10.1103/PhysRevB.79.161408} {\bibfield  {journal} {\bibinfo
  {journal} {Phys. Rev. B}\ }\textbf {\bibinfo {volume} {79}},\ \bibinfo
  {pages} {161408} (\bibinfo {year} {2009})}\BibitemShut {NoStop}%
\bibitem [{\citenamefont {San-Jose}\ \emph {et~al.}(2014)\citenamefont
  {San-Jose}, \citenamefont {Prada},\ and\ \citenamefont
  {Aguado}}]{San-Jose:PRL14}%
  \BibitemOpen
  \bibfield  {author} {\bibinfo {author} {\bibfnamefont {P.}~\bibnamefont
  {San-Jose}}, \bibinfo {author} {\bibfnamefont {E.}~\bibnamefont {Prada}}, \
  and\ \bibinfo {author} {\bibfnamefont {R.}~\bibnamefont {Aguado}},\ }\href
  {\doibase 10.1103/PhysRevLett.112.137001} {\bibfield  {journal} {\bibinfo
  {journal} {Phys. Rev. Lett.}\ }\textbf {\bibinfo {volume} {112}},\ \bibinfo
  {pages} {137001} (\bibinfo {year} {2014})}\BibitemShut {NoStop}%
\bibitem [{\citenamefont {Tiira}\ \emph {et~al.}(2017)\citenamefont {Tiira},
  \citenamefont {Strambini}, \citenamefont {Amado}, \citenamefont {Roddaro},
  \citenamefont {San-Jose}, \citenamefont {Aguado}, \citenamefont {Bergeret},
  \citenamefont {Ercolani}, \citenamefont {Sorba},\ and\ \citenamefont
  {Giazotto}}]{Tiira:NC17}%
  \BibitemOpen
  \bibfield  {author} {\bibinfo {author} {\bibfnamefont {J.}~\bibnamefont
  {Tiira}}, \bibinfo {author} {\bibfnamefont {E.}~\bibnamefont {Strambini}},
  \bibinfo {author} {\bibfnamefont {M.}~\bibnamefont {Amado}}, \bibinfo
  {author} {\bibfnamefont {S.}~\bibnamefont {Roddaro}}, \bibinfo {author}
  {\bibfnamefont {P.}~\bibnamefont {San-Jose}}, \bibinfo {author}
  {\bibfnamefont {R.}~\bibnamefont {Aguado}}, \bibinfo {author} {\bibfnamefont
  {F.~S.}\ \bibnamefont {Bergeret}}, \bibinfo {author} {\bibfnamefont
  {D.}~\bibnamefont {Ercolani}}, \bibinfo {author} {\bibfnamefont
  {L.}~\bibnamefont {Sorba}}, \ and\ \bibinfo {author} {\bibfnamefont
  {F.}~\bibnamefont {Giazotto}},\ }\href {\doibase 10.1038/ncomms14984}
  {\bibfield  {journal} {\bibinfo  {journal} {Nat. Commun.}\ }\textbf {\bibinfo
  {volume} {8}},\ \bibinfo {pages} {14984} (\bibinfo {year}
  {2017})}\BibitemShut {NoStop}%
\bibitem [{\citenamefont {Cayao}\ \emph {et~al.}(2017)\citenamefont {Cayao},
  \citenamefont {San-Jose}, \citenamefont {Black-Schaffer}, \citenamefont
  {Aguado},\ and\ \citenamefont {Prada}}]{Cayao:PRB17}%
  \BibitemOpen
  \bibfield  {author} {\bibinfo {author} {\bibfnamefont {J.}~\bibnamefont
  {Cayao}}, \bibinfo {author} {\bibfnamefont {P.}~\bibnamefont {San-Jose}},
  \bibinfo {author} {\bibfnamefont {A.~M.}\ \bibnamefont {Black-Schaffer}},
  \bibinfo {author} {\bibfnamefont {R.}~\bibnamefont {Aguado}}, \ and\ \bibinfo
  {author} {\bibfnamefont {E.}~\bibnamefont {Prada}},\ }\href {\doibase
  10.1103/PhysRevB.96.205425} {\bibfield  {journal} {\bibinfo  {journal} {Phys.
  Rev. B}\ }\textbf {\bibinfo {volume} {96}},\ \bibinfo {pages} {205425}
  (\bibinfo {year} {2017})}\BibitemShut {NoStop}%
\bibitem [{\citenamefont {van Zanten}\ \emph {et~al.}(2020)\citenamefont {van
  Zanten}, \citenamefont {Sabonis}, \citenamefont {Suter}, \citenamefont
  {V{\"a}yrynen}, \citenamefont {Karzig}, \citenamefont {Pikulin},
  \citenamefont {O'Farrell}, \citenamefont {Razmadze}, \citenamefont
  {Petersson}, \citenamefont {Krogstrup},\ and\ \citenamefont
  {Marcus}}]{CharliePAT}%
  \BibitemOpen
  \bibfield  {author} {\bibinfo {author} {\bibfnamefont {D.~M.~T.}\
  \bibnamefont {van Zanten}}, \bibinfo {author} {\bibfnamefont
  {D.}~\bibnamefont {Sabonis}}, \bibinfo {author} {\bibfnamefont
  {J.}~\bibnamefont {Suter}}, \bibinfo {author} {\bibfnamefont {J.~I.}\
  \bibnamefont {V{\"a}yrynen}}, \bibinfo {author} {\bibfnamefont
  {T.}~\bibnamefont {Karzig}}, \bibinfo {author} {\bibfnamefont {D.~I.}\
  \bibnamefont {Pikulin}}, \bibinfo {author} {\bibfnamefont {E.~C.~T.}\
  \bibnamefont {O'Farrell}}, \bibinfo {author} {\bibfnamefont {D.}~\bibnamefont
  {Razmadze}}, \bibinfo {author} {\bibfnamefont {K.~D.}\ \bibnamefont
  {Petersson}}, \bibinfo {author} {\bibfnamefont {P.}~\bibnamefont
  {Krogstrup}}, \ and\ \bibinfo {author} {\bibfnamefont {C.~M.}\ \bibnamefont
  {Marcus}},\ }\href@noop {} {\bibfield  {journal} {\bibinfo  {journal} {Nature
  Physics}\ }\textbf {\bibinfo {volume} {16}},\ \bibinfo {pages} {663}
  (\bibinfo {year} {2020})}\BibitemShut {NoStop}%
\bibitem [{\citenamefont {\'Avila}\ \emph {et~al.}(2020)\citenamefont
  {\'Avila}, \citenamefont {Prada}, \citenamefont {San-Jose},\ and\
  \citenamefont {Aguado}}]{Avila-accompanying}%
  \BibitemOpen
  \bibfield  {author} {\bibinfo {author} {\bibfnamefont {J.}~\bibnamefont
  {\'Avila}}, \bibinfo {author} {\bibfnamefont {E.}~\bibnamefont {Prada}},
  \bibinfo {author} {\bibfnamefont {P.}~\bibnamefont {San-Jose}}, \ and\
  \bibinfo {author} {\bibfnamefont {R.}~\bibnamefont {Aguado}},\ }\href
  {\doibase 10.1103/PhysRevResearch.2.033493} {\bibfield  {journal} {\bibinfo
  {journal} {Phys. Rev. Research}\ }\textbf {\bibinfo {volume} {2}},\ \bibinfo
  {pages} {033493} (\bibinfo {year} {2020})}\BibitemShut {NoStop}%
\bibitem [{\citenamefont {Prada}\ \emph {et~al.}(2020)\citenamefont {Prada},
  \citenamefont {San-Jose}, \citenamefont {de~Moor}, \citenamefont {Geresdi},
  \citenamefont {Lee}, \citenamefont {Klinovaja}, \citenamefont {Loss},
  \citenamefont {Nyg{\aa}rd}, \citenamefont {Aguado},\ and\ \citenamefont
  {Kouwenhoven}}]{prada2019}%
  \BibitemOpen
  \bibfield  {author} {\bibinfo {author} {\bibfnamefont {E.}~\bibnamefont
  {Prada}}, \bibinfo {author} {\bibfnamefont {P.}~\bibnamefont {San-Jose}},
  \bibinfo {author} {\bibfnamefont {M.~W.~A.}\ \bibnamefont {de~Moor}},
  \bibinfo {author} {\bibfnamefont {A.}~\bibnamefont {Geresdi}}, \bibinfo
  {author} {\bibfnamefont {E.~J.~H.}\ \bibnamefont {Lee}}, \bibinfo {author}
  {\bibfnamefont {J.}~\bibnamefont {Klinovaja}}, \bibinfo {author}
  {\bibfnamefont {D.}~\bibnamefont {Loss}}, \bibinfo {author} {\bibfnamefont
  {J.}~\bibnamefont {Nyg{\aa}rd}}, \bibinfo {author} {\bibfnamefont
  {R.}~\bibnamefont {Aguado}}, \ and\ \bibinfo {author} {\bibfnamefont {L.~P.}\
  \bibnamefont {Kouwenhoven}},\ }\href@noop {} {\bibfield  {journal} {\bibinfo
  {journal} {Nature Reviews Physics}\ }\textbf {\bibinfo {volume} {2}},\
  \bibinfo {pages} {575} (\bibinfo {year} {2020})}\BibitemShut {NoStop}%
\bibitem [{\citenamefont {Das~Sarma}\ \emph {et~al.}(2012)\citenamefont
  {Das~Sarma}, \citenamefont {Sau},\ and\ \citenamefont
  {Stanescu}}]{Das-Sarma:PRB12}%
  \BibitemOpen
  \bibfield  {author} {\bibinfo {author} {\bibfnamefont {S.}~\bibnamefont
  {Das~Sarma}}, \bibinfo {author} {\bibfnamefont {J.~D.}\ \bibnamefont {Sau}},
  \ and\ \bibinfo {author} {\bibfnamefont {T.~D.}\ \bibnamefont {Stanescu}},\
  }\href {\doibase 10.1103/PhysRevB.86.220506} {\bibfield  {journal} {\bibinfo
  {journal} {Phys. Rev. B}\ }\textbf {\bibinfo {volume} {86}},\ \bibinfo
  {pages} {220506} (\bibinfo {year} {2012})}\BibitemShut {NoStop}%
\bibitem [{\citenamefont {Mishmash}\ \emph {et~al.}(2016)\citenamefont
  {Mishmash}, \citenamefont {Aasen}, \citenamefont {Higginbotham},\ and\
  \citenamefont {Alicea}}]{Mishmash:PRB16}%
  \BibitemOpen
  \bibfield  {author} {\bibinfo {author} {\bibfnamefont {R.~V.}\ \bibnamefont
  {Mishmash}}, \bibinfo {author} {\bibfnamefont {D.}~\bibnamefont {Aasen}},
  \bibinfo {author} {\bibfnamefont {A.~P.}\ \bibnamefont {Higginbotham}}, \
  and\ \bibinfo {author} {\bibfnamefont {J.}~\bibnamefont {Alicea}},\ }\href
  {\doibase 10.1103/PhysRevB.93.245404} {\bibfield  {journal} {\bibinfo
  {journal} {Phys. Rev. B}\ }\textbf {\bibinfo {volume} {93}},\ \bibinfo
  {pages} {245404} (\bibinfo {year} {2016})}\BibitemShut {NoStop}%
\bibitem [{\citenamefont {Lim}\ \emph {et~al.}(2012)\citenamefont {Lim},
  \citenamefont {Serra}, \citenamefont {L\'opez},\ and\ \citenamefont
  {Aguado}}]{PhysRevB.86.121103}%
  \BibitemOpen
  \bibfield  {author} {\bibinfo {author} {\bibfnamefont {J.~S.}\ \bibnamefont
  {Lim}}, \bibinfo {author} {\bibfnamefont {L.}~\bibnamefont {Serra}}, \bibinfo
  {author} {\bibfnamefont {R.}~\bibnamefont {L\'opez}}, \ and\ \bibinfo
  {author} {\bibfnamefont {R.}~\bibnamefont {Aguado}},\ }\href {\doibase
  10.1103/PhysRevB.86.121103} {\bibfield  {journal} {\bibinfo  {journal} {Phys.
  Rev. B}\ }\textbf {\bibinfo {volume} {86}},\ \bibinfo {pages} {121103}
  (\bibinfo {year} {2012})}\BibitemShut {NoStop}%
\bibitem [{\citenamefont {Prada}\ \emph {et~al.}(2012)\citenamefont {Prada},
  \citenamefont {San-Jose},\ and\ \citenamefont {Aguado}}]{Prada:PRB12}%
  \BibitemOpen
  \bibfield  {author} {\bibinfo {author} {\bibfnamefont {E.}~\bibnamefont
  {Prada}}, \bibinfo {author} {\bibfnamefont {P.}~\bibnamefont {San-Jose}}, \
  and\ \bibinfo {author} {\bibfnamefont {R.}~\bibnamefont {Aguado}},\ }\href
  {\doibase 10.1103/PhysRevB.86.180503} {\bibfield  {journal} {\bibinfo
  {journal} {Phys. Rev. B}\ }\textbf {\bibinfo {volume} {86}},\ \bibinfo
  {pages} {180503(R)} (\bibinfo {year} {2012})}\BibitemShut {NoStop}%
\bibitem [{\citenamefont {Rainis}\ \emph {et~al.}(2013)\citenamefont {Rainis},
  \citenamefont {Trifunovic}, \citenamefont {Klinovaja},\ and\ \citenamefont
  {Loss}}]{Rainis:PRB13}%
  \BibitemOpen
  \bibfield  {author} {\bibinfo {author} {\bibfnamefont {D.}~\bibnamefont
  {Rainis}}, \bibinfo {author} {\bibfnamefont {L.}~\bibnamefont {Trifunovic}},
  \bibinfo {author} {\bibfnamefont {J.}~\bibnamefont {Klinovaja}}, \ and\
  \bibinfo {author} {\bibfnamefont {D.}~\bibnamefont {Loss}},\ }\href {\doibase
  10.1103/PhysRevB.87.024515} {\bibfield  {journal} {\bibinfo  {journal} {Phys.
  Rev. B}\ }\textbf {\bibinfo {volume} {87}},\ \bibinfo {pages} {024515}
  (\bibinfo {year} {2013})}\BibitemShut {NoStop}%
\bibitem [{\citenamefont {San-Jose}\ \emph {et~al.}(2012)\citenamefont
  {San-Jose}, \citenamefont {Prada},\ and\ \citenamefont
  {Aguado}}]{PhysRevLett.108.257001}%
  \BibitemOpen
  \bibfield  {author} {\bibinfo {author} {\bibfnamefont {P.}~\bibnamefont
  {San-Jose}}, \bibinfo {author} {\bibfnamefont {E.}~\bibnamefont {Prada}}, \
  and\ \bibinfo {author} {\bibfnamefont {R.}~\bibnamefont {Aguado}},\ }\href
  {\doibase 10.1103/PhysRevLett.108.257001} {\bibfield  {journal} {\bibinfo
  {journal} {Phys. Rev. Lett.}\ }\textbf {\bibinfo {volume} {108}},\ \bibinfo
  {pages} {257001} (\bibinfo {year} {2012})}\BibitemShut {NoStop}%
\bibitem [{\citenamefont {Pikulin}\ and\ \citenamefont
  {Nazarov}(2012)}]{PhysRevB.86.140504}%
  \BibitemOpen
  \bibfield  {author} {\bibinfo {author} {\bibfnamefont {D.~I.}\ \bibnamefont
  {Pikulin}}\ and\ \bibinfo {author} {\bibfnamefont {Y.~V.}\ \bibnamefont
  {Nazarov}},\ }\href {\doibase 10.1103/PhysRevB.86.140504} {\bibfield
  {journal} {\bibinfo  {journal} {Phys. Rev. B}\ }\textbf {\bibinfo {volume}
  {86}},\ \bibinfo {pages} {140504} (\bibinfo {year} {2012})}\BibitemShut
  {NoStop}%
\bibitem [{\citenamefont {San-Jose}\ \emph {et~al.}(2013)\citenamefont
  {San-Jose}, \citenamefont {Cayao}, \citenamefont {Prada},\ and\ \citenamefont
  {Aguado}}]{San-Jose:NJP13}%
  \BibitemOpen
  \bibfield  {author} {\bibinfo {author} {\bibfnamefont {P.}~\bibnamefont
  {San-Jose}}, \bibinfo {author} {\bibfnamefont {J.}~\bibnamefont {Cayao}},
  \bibinfo {author} {\bibfnamefont {E.}~\bibnamefont {Prada}}, \ and\ \bibinfo
  {author} {\bibfnamefont {R.}~\bibnamefont {Aguado}},\ }\href
  {http://stacks.iop.org/1367-2630/15/i=7/a=075019} {\bibfield  {journal}
  {\bibinfo  {journal} {New J. Phys.}\ }\textbf {\bibinfo {volume} {15}},\
  \bibinfo {pages} {075019} (\bibinfo {year} {2013})}\BibitemShut {NoStop}%
\bibitem [{\citenamefont {Cayao}\ \emph {et~al.}(2018)\citenamefont {Cayao},
  \citenamefont {Black-Schaffer}, \citenamefont {Prada},\ and\ \citenamefont
  {Aguado}}]{Cayao:Belstein18}%
  \BibitemOpen
  \bibfield  {author} {\bibinfo {author} {\bibfnamefont {J.}~\bibnamefont
  {Cayao}}, \bibinfo {author} {\bibfnamefont {A.~M.}\ \bibnamefont
  {Black-Schaffer}}, \bibinfo {author} {\bibfnamefont {E.}~\bibnamefont
  {Prada}}, \ and\ \bibinfo {author} {\bibfnamefont {R.}~\bibnamefont
  {Aguado}},\ }\href {\doibase doi:10.3762/bjnano.9.127} {\bibfield  {journal}
  {\bibinfo  {journal} {Beilstein J. Nanotechnol.}\ }\textbf {\bibinfo {volume}
  {9}},\ \bibinfo {pages} {1339} (\bibinfo {year} {2018})}\BibitemShut
  {NoStop}%
\bibitem [{Note1()}]{Note1}%
  \BibitemOpen
  \bibinfo {note} {While strictly correct, the definition of
  $V_J^{bulk}(\varphi )$ given in Eq. \protect \textup {\hbox {\mathsurround
  \z@ \protect \normalfont (\ignorespaces \ref {VJbulk}\unskip \@@italiccorr
  )}} is non-optimal (since finding the correct lowest states of the NW
  spectrum as a function of $\varphi $ for all magnetic fields can be quite a
  cumbersome task). \protect \leavevmode {\protect \color {black}Instead, we
  just compute it by subtracting the low energy sector ($\protect \mathaccentV
  {hat}05E{\protect \mathcal {H}}$ in Eq. \protect \textup {\hbox
  {\mathsurround \z@ \protect \normalfont (\ignorespaces \ref
  {eq:lowenergy}\unskip \@@italiccorr )}} from the total spectrum sum $\DOTSB
  \sum@ \slimits@ \epsilon _p$)}.}\BibitemShut {Stop}%
\bibitem [{\citenamefont {San-Jose}()}]{MathQ}%
  \BibitemOpen
  \bibfield  {author} {\bibinfo {author} {\bibfnamefont {P.}~\bibnamefont
  {San-Jose}},\ }\href@noop {} {\enquote {\bibinfo {title} {\textrm{MathQ}, a
  \textrm{Mathematica} simulator for quantum systems},}\ }\bibinfo {note}
  {Http://www.icmm.csic.es/sanjose/MathQ/MathQ.html}\BibitemShut {NoStop}%
\bibitem [{\citenamefont {Albrecht}\ \emph {et~al.}(2016)\citenamefont
  {Albrecht}, \citenamefont {Higginbotham}, \citenamefont {Madsen},
  \citenamefont {Kuemmeth}, \citenamefont {Jespersen}, \citenamefont
  {Nyg{\aa}rd}, \citenamefont {Krogstrup},\ and\ \citenamefont
  {Marcus}}]{Albrecht:N16}%
  \BibitemOpen
  \bibfield  {author} {\bibinfo {author} {\bibfnamefont {S.~M.}\ \bibnamefont
  {Albrecht}}, \bibinfo {author} {\bibfnamefont {A.~P.}\ \bibnamefont
  {Higginbotham}}, \bibinfo {author} {\bibfnamefont {M.}~\bibnamefont
  {Madsen}}, \bibinfo {author} {\bibfnamefont {F.}~\bibnamefont {Kuemmeth}},
  \bibinfo {author} {\bibfnamefont {T.~S.}\ \bibnamefont {Jespersen}}, \bibinfo
  {author} {\bibfnamefont {J.}~\bibnamefont {Nyg{\aa}rd}}, \bibinfo {author}
  {\bibfnamefont {P.}~\bibnamefont {Krogstrup}}, \ and\ \bibinfo {author}
  {\bibfnamefont {C.~M.}\ \bibnamefont {Marcus}},\ }\href
  {http://dx.doi.org/10.1038/nature17162} {\bibfield  {journal} {\bibinfo
  {journal} {Nature}\ }\textbf {\bibinfo {volume} {531}},\ \bibinfo {pages}
  {206} (\bibinfo {year} {2016})}\BibitemShut {NoStop}%
\bibitem [{\citenamefont {Shen}\ \emph {et~al.}(2018)\citenamefont {Shen},
  \citenamefont {Heedt}, \citenamefont {Borsoi}, \citenamefont {van Heck},
  \citenamefont {Gazibegovic}, \citenamefont {Op~het Veld}, \citenamefont
  {Car}, \citenamefont {Logan}, \citenamefont {Pendharkar}, \citenamefont
  {Ramakers}, \citenamefont {Wang}, \citenamefont {Xu}, \citenamefont {Bouman},
  \citenamefont {Geresdi}, \citenamefont {Palmstr{\o}m}, \citenamefont
  {Bakkers},\ and\ \citenamefont {Kouwenhoven}}]{Shen2018}%
  \BibitemOpen
  \bibfield  {author} {\bibinfo {author} {\bibfnamefont {J.}~\bibnamefont
  {Shen}}, \bibinfo {author} {\bibfnamefont {S.}~\bibnamefont {Heedt}},
  \bibinfo {author} {\bibfnamefont {F.}~\bibnamefont {Borsoi}}, \bibinfo
  {author} {\bibfnamefont {B.}~\bibnamefont {van Heck}}, \bibinfo {author}
  {\bibfnamefont {S.}~\bibnamefont {Gazibegovic}}, \bibinfo {author}
  {\bibfnamefont {R.~L.~M.}\ \bibnamefont {Op~het Veld}}, \bibinfo {author}
  {\bibfnamefont {D.}~\bibnamefont {Car}}, \bibinfo {author} {\bibfnamefont
  {J.~A.}\ \bibnamefont {Logan}}, \bibinfo {author} {\bibfnamefont
  {M.}~\bibnamefont {Pendharkar}}, \bibinfo {author} {\bibfnamefont {S.~J.~J.}\
  \bibnamefont {Ramakers}}, \bibinfo {author} {\bibfnamefont {G.}~\bibnamefont
  {Wang}}, \bibinfo {author} {\bibfnamefont {D.}~\bibnamefont {Xu}}, \bibinfo
  {author} {\bibfnamefont {D.}~\bibnamefont {Bouman}}, \bibinfo {author}
  {\bibfnamefont {A.}~\bibnamefont {Geresdi}}, \bibinfo {author} {\bibfnamefont
  {C.~J.}\ \bibnamefont {Palmstr{\o}m}}, \bibinfo {author} {\bibfnamefont
  {E.~P. A.~M.}\ \bibnamefont {Bakkers}}, \ and\ \bibinfo {author}
  {\bibfnamefont {L.~P.}\ \bibnamefont {Kouwenhoven}},\ }\href@noop {}
  {\bibfield  {journal} {\bibinfo  {journal} {Nature Communications}\ }\textbf
  {\bibinfo {volume} {9}},\ \bibinfo {pages} {4801} (\bibinfo {year}
  {2018})}\BibitemShut {NoStop}%
\bibitem [{\citenamefont {Desp\'osito}\ and\ \citenamefont
  {Levy~Yeyati}(2001)}]{PhysRevB.64.140511}%
  \BibitemOpen
  \bibfield  {author} {\bibinfo {author} {\bibfnamefont {M.~A.}\ \bibnamefont
  {Desp\'osito}}\ and\ \bibinfo {author} {\bibfnamefont {A.}~\bibnamefont
  {Levy~Yeyati}},\ }\href {\doibase 10.1103/PhysRevB.64.140511} {\bibfield
  {journal} {\bibinfo  {journal} {Phys. Rev. B}\ }\textbf {\bibinfo {volume}
  {64}},\ \bibinfo {pages} {140511} (\bibinfo {year} {2001})}\BibitemShut
  {NoStop}%
\bibitem [{\citenamefont {Romero}\ \emph {et~al.}(2012)\citenamefont {Romero},
  \citenamefont {Lizuain}, \citenamefont {Shumeiko}, \citenamefont {Solano},\
  and\ \citenamefont {Bergeret}}]{PhysRevB.85.180506}%
  \BibitemOpen
  \bibfield  {author} {\bibinfo {author} {\bibfnamefont {G.}~\bibnamefont
  {Romero}}, \bibinfo {author} {\bibfnamefont {I.}~\bibnamefont {Lizuain}},
  \bibinfo {author} {\bibfnamefont {V.~S.}\ \bibnamefont {Shumeiko}}, \bibinfo
  {author} {\bibfnamefont {E.}~\bibnamefont {Solano}}, \ and\ \bibinfo {author}
  {\bibfnamefont {F.~S.}\ \bibnamefont {Bergeret}},\ }\href {\doibase
  10.1103/PhysRevB.85.180506} {\bibfield  {journal} {\bibinfo  {journal} {Phys.
  Rev. B}\ }\textbf {\bibinfo {volume} {85}},\ \bibinfo {pages} {180506}
  (\bibinfo {year} {2012})}\BibitemShut {NoStop}%
\bibitem [{\citenamefont {Bretheau}\ \emph {et~al.}(2014)\citenamefont
  {Bretheau}, \citenamefont {Girit}, \citenamefont {Houzet}, \citenamefont
  {Pothier}, \citenamefont {Esteve},\ and\ \citenamefont
  {Urbina}}]{PhysRevB.90.134506}%
  \BibitemOpen
  \bibfield  {author} {\bibinfo {author} {\bibfnamefont {L.}~\bibnamefont
  {Bretheau}}, \bibinfo {author} {\bibfnamefont {{\c C}.~{\"O}.}\ \bibnamefont
  {Girit}}, \bibinfo {author} {\bibfnamefont {M.}~\bibnamefont {Houzet}},
  \bibinfo {author} {\bibfnamefont {H.}~\bibnamefont {Pothier}}, \bibinfo
  {author} {\bibfnamefont {D.}~\bibnamefont {Esteve}}, \ and\ \bibinfo {author}
  {\bibfnamefont {C.}~\bibnamefont {Urbina}},\ }\href {\doibase
  10.1103/PhysRevB.90.134506} {\bibfield  {journal} {\bibinfo  {journal} {Phys.
  Rev. B}\ }\textbf {\bibinfo {volume} {90}},\ \bibinfo {pages} {134506}
  (\bibinfo {year} {2014})}\BibitemShut {NoStop}%
\bibitem [{\citenamefont {van Woerkom}\ \emph {et~al.}(2017)\citenamefont {van
  Woerkom}, \citenamefont {Proutski}, \citenamefont {van Heck}, \citenamefont
  {Bouman}, \citenamefont {V{\"a}yrynen}, \citenamefont {Glazman},
  \citenamefont {Krogstrup}, \citenamefont {Nyg{\aa}rd}, \citenamefont
  {Kouwenhoven},\ and\ \citenamefont {Geresdi}}]{vanWoerkom:NP17}%
  \BibitemOpen
  \bibfield  {author} {\bibinfo {author} {\bibfnamefont {D.~J.}\ \bibnamefont
  {van Woerkom}}, \bibinfo {author} {\bibfnamefont {A.}~\bibnamefont
  {Proutski}}, \bibinfo {author} {\bibfnamefont {B.}~\bibnamefont {van Heck}},
  \bibinfo {author} {\bibfnamefont {D.}~\bibnamefont {Bouman}}, \bibinfo
  {author} {\bibfnamefont {J.~I.}\ \bibnamefont {V{\"a}yrynen}}, \bibinfo
  {author} {\bibfnamefont {L.~I.}\ \bibnamefont {Glazman}}, \bibinfo {author}
  {\bibfnamefont {P.}~\bibnamefont {Krogstrup}}, \bibinfo {author}
  {\bibfnamefont {J.}~\bibnamefont {Nyg{\aa}rd}}, \bibinfo {author}
  {\bibfnamefont {L.~P.}\ \bibnamefont {Kouwenhoven}}, \ and\ \bibinfo {author}
  {\bibfnamefont {A.}~\bibnamefont {Geresdi}},\ }\href@noop {} {\bibfield
  {journal} {\bibinfo  {journal} {Nature Physics}\ }\textbf {\bibinfo {volume}
  {13}},\ \bibinfo {pages} {876} (\bibinfo {year} {2017})}\BibitemShut
  {NoStop}%
\bibitem [{\citenamefont {Hays}\ \emph {et~al.}(2018)\citenamefont {Hays},
  \citenamefont {de~Lange}, \citenamefont {Serniak}, \citenamefont {van
  Woerkom}, \citenamefont {Bouman}, \citenamefont {Krogstrup}, \citenamefont
  {Nyg\aa{}rd}, \citenamefont {Geresdi},\ and\ \citenamefont
  {Devoret}}]{PhysRevLett.121.047001}%
  \BibitemOpen
  \bibfield  {author} {\bibinfo {author} {\bibfnamefont {M.}~\bibnamefont
  {Hays}}, \bibinfo {author} {\bibfnamefont {G.}~\bibnamefont {de~Lange}},
  \bibinfo {author} {\bibfnamefont {K.}~\bibnamefont {Serniak}}, \bibinfo
  {author} {\bibfnamefont {D.~J.}\ \bibnamefont {van Woerkom}}, \bibinfo
  {author} {\bibfnamefont {D.}~\bibnamefont {Bouman}}, \bibinfo {author}
  {\bibfnamefont {P.}~\bibnamefont {Krogstrup}}, \bibinfo {author}
  {\bibfnamefont {J.}~\bibnamefont {Nyg\aa{}rd}}, \bibinfo {author}
  {\bibfnamefont {A.}~\bibnamefont {Geresdi}}, \ and\ \bibinfo {author}
  {\bibfnamefont {M.~H.}\ \bibnamefont {Devoret}},\ }\href {\doibase
  10.1103/PhysRevLett.121.047001} {\bibfield  {journal} {\bibinfo  {journal}
  {Phys. Rev. Lett.}\ }\textbf {\bibinfo {volume} {121}},\ \bibinfo {pages}
  {047001} (\bibinfo {year} {2018})}\BibitemShut {NoStop}%
\bibitem [{\citenamefont {Bretheau}\ \emph {et~al.}(2013)\citenamefont
  {Bretheau}, \citenamefont {Girit}, \citenamefont {Pothier}, \citenamefont
  {Esteve},\ and\ \citenamefont {Urbina}}]{Bretheau:N13}%
  \BibitemOpen
  \bibfield  {author} {\bibinfo {author} {\bibfnamefont {L.}~\bibnamefont
  {Bretheau}}, \bibinfo {author} {\bibfnamefont {C.~O.}\ \bibnamefont {Girit}},
  \bibinfo {author} {\bibfnamefont {H.}~\bibnamefont {Pothier}}, \bibinfo
  {author} {\bibfnamefont {D.}~\bibnamefont {Esteve}}, \ and\ \bibinfo {author}
  {\bibfnamefont {C.}~\bibnamefont {Urbina}},\ }\href
  {http://dx.doi.org/10.1038/nature12315} {\bibfield  {journal} {\bibinfo
  {journal} {Nature}\ }\textbf {\bibinfo {volume} {499}},\ \bibinfo {pages}
  {312} (\bibinfo {year} {2013})}\BibitemShut {NoStop}%
\bibitem [{\citenamefont {Vuik}\ \emph {et~al.}(2016)\citenamefont {Vuik},
  \citenamefont {Eeltink}, \citenamefont {Akhmerov},\ and\ \citenamefont
  {Wimmer}}]{Vuik:NJP16}%
  \BibitemOpen
  \bibfield  {author} {\bibinfo {author} {\bibfnamefont {A.}~\bibnamefont
  {Vuik}}, \bibinfo {author} {\bibfnamefont {D.}~\bibnamefont {Eeltink}},
  \bibinfo {author} {\bibfnamefont {A.~R.}\ \bibnamefont {Akhmerov}}, \ and\
  \bibinfo {author} {\bibfnamefont {M.}~\bibnamefont {Wimmer}},\ }\href
  {http://stacks.iop.org/1367-2630/18/i=3/a=033013} {\bibfield  {journal}
  {\bibinfo  {journal} {New J. Phys.}\ }\textbf {\bibinfo {volume} {18}},\
  \bibinfo {pages} {033013} (\bibinfo {year} {2016})}\BibitemShut {NoStop}%
\bibitem [{\citenamefont {Dom{\'\i}nguez}\ \emph {et~al.}(2017)\citenamefont
  {Dom{\'\i}nguez}, \citenamefont {Cayao}, \citenamefont {San-Jose},
  \citenamefont {Aguado}, \citenamefont {Yeyati},\ and\ \citenamefont
  {Prada}}]{Dominguez:NQM17}%
  \BibitemOpen
  \bibfield  {author} {\bibinfo {author} {\bibfnamefont {F.}~\bibnamefont
  {Dom{\'\i}nguez}}, \bibinfo {author} {\bibfnamefont {J.}~\bibnamefont
  {Cayao}}, \bibinfo {author} {\bibfnamefont {P.}~\bibnamefont {San-Jose}},
  \bibinfo {author} {\bibfnamefont {R.}~\bibnamefont {Aguado}}, \bibinfo
  {author} {\bibfnamefont {A.~L.}\ \bibnamefont {Yeyati}}, \ and\ \bibinfo
  {author} {\bibfnamefont {E.}~\bibnamefont {Prada}},\ }\href {\doibase
  10.1038/s41535-017-0012-0} {\bibfield  {journal} {\bibinfo  {journal} {npj
  Quantum Materials}\ }\textbf {\bibinfo {volume} {2}},\ \bibinfo {pages} {13}
  (\bibinfo {year} {2017})}\BibitemShut {NoStop}%
\bibitem [{\citenamefont {Antipov}\ \emph {et~al.}(2018)\citenamefont
  {Antipov}, \citenamefont {Bargerbos}, \citenamefont {Winkler}, \citenamefont
  {Bauer}, \citenamefont {Rossi},\ and\ \citenamefont {Lutchyn}}]{Antipov:18}%
  \BibitemOpen
  \bibfield  {author} {\bibinfo {author} {\bibfnamefont {A.~E.}\ \bibnamefont
  {Antipov}}, \bibinfo {author} {\bibfnamefont {A.}~\bibnamefont {Bargerbos}},
  \bibinfo {author} {\bibfnamefont {G.~W.}\ \bibnamefont {Winkler}}, \bibinfo
  {author} {\bibfnamefont {B.}~\bibnamefont {Bauer}}, \bibinfo {author}
  {\bibfnamefont {E.}~\bibnamefont {Rossi}}, \ and\ \bibinfo {author}
  {\bibfnamefont {R.~M.}\ \bibnamefont {Lutchyn}},\ }\href
  {https://arxiv.org/abs/1801.02616} {\  (\bibinfo {year} {2018})},\ \Eprint
  {http://arxiv.org/abs/1801.02616} {1801.02616} \BibitemShut {NoStop}%
\bibitem [{\citenamefont {Escribano}\ \emph {et~al.}(2018)\citenamefont
  {Escribano}, \citenamefont {Levy~Yeyati},\ and\ \citenamefont
  {Prada}}]{Escribano:BJN18}%
  \BibitemOpen
  \bibfield  {author} {\bibinfo {author} {\bibfnamefont {S.~D.}\ \bibnamefont
  {Escribano}}, \bibinfo {author} {\bibfnamefont {A.}~\bibnamefont
  {Levy~Yeyati}}, \ and\ \bibinfo {author} {\bibfnamefont {E.}~\bibnamefont
  {Prada}},\ }\href {\doibase 10.3762/bjnano.9.203} {\bibfield  {journal}
  {\bibinfo  {journal} {Beilstein J. Nanotechnol.}\ }\textbf {\bibinfo {volume}
  {9}},\ \bibinfo {pages} {2171} (\bibinfo {year} {2018})}\BibitemShut
  {NoStop}%
\bibitem [{\citenamefont {Winkler}\ \emph {et~al.}(2019)\citenamefont
  {Winkler}, \citenamefont {Antipov}, \citenamefont {van Heck}, \citenamefont
  {Soluyanov}, \citenamefont {Glazman}, \citenamefont {Wimmer},\ and\
  \citenamefont {Lutchyn}}]{PhysRevB.99.245408}%
  \BibitemOpen
  \bibfield  {author} {\bibinfo {author} {\bibfnamefont {G.~W.}\ \bibnamefont
  {Winkler}}, \bibinfo {author} {\bibfnamefont {A.~E.}\ \bibnamefont
  {Antipov}}, \bibinfo {author} {\bibfnamefont {B.}~\bibnamefont {van Heck}},
  \bibinfo {author} {\bibfnamefont {A.~A.}\ \bibnamefont {Soluyanov}}, \bibinfo
  {author} {\bibfnamefont {L.~I.}\ \bibnamefont {Glazman}}, \bibinfo {author}
  {\bibfnamefont {M.}~\bibnamefont {Wimmer}}, \ and\ \bibinfo {author}
  {\bibfnamefont {R.~M.}\ \bibnamefont {Lutchyn}},\ }\href {\doibase
  10.1103/PhysRevB.99.245408} {\bibfield  {journal} {\bibinfo  {journal} {Phys.
  Rev. B}\ }\textbf {\bibinfo {volume} {99}},\ \bibinfo {pages} {245408}
  (\bibinfo {year} {2019})}\BibitemShut {NoStop}%
\bibitem [{\citenamefont {Nijholt}\ and\ \citenamefont
  {Akhmerov}(2016)}]{PhysRevB.93.235434}%
  \BibitemOpen
  \bibfield  {author} {\bibinfo {author} {\bibfnamefont {B.}~\bibnamefont
  {Nijholt}}\ and\ \bibinfo {author} {\bibfnamefont {A.~R.}\ \bibnamefont
  {Akhmerov}},\ }\href {\doibase 10.1103/PhysRevB.93.235434} {\bibfield
  {journal} {\bibinfo  {journal} {Phys. Rev. B}\ }\textbf {\bibinfo {volume}
  {93}},\ \bibinfo {pages} {235434} (\bibinfo {year} {2016})}\BibitemShut
  {NoStop}%
\bibitem [{\citenamefont {Winkler}\ \emph {et~al.}(2017)\citenamefont
  {Winkler}, \citenamefont {Varjas}, \citenamefont {Skolasinski}, \citenamefont
  {Soluyanov}, \citenamefont {Troyer},\ and\ \citenamefont
  {Wimmer}}]{PhysRevLett.119.037701}%
  \BibitemOpen
  \bibfield  {author} {\bibinfo {author} {\bibfnamefont {G.~W.}\ \bibnamefont
  {Winkler}}, \bibinfo {author} {\bibfnamefont {D.}~\bibnamefont {Varjas}},
  \bibinfo {author} {\bibfnamefont {R.}~\bibnamefont {Skolasinski}}, \bibinfo
  {author} {\bibfnamefont {A.~A.}\ \bibnamefont {Soluyanov}}, \bibinfo {author}
  {\bibfnamefont {M.}~\bibnamefont {Troyer}}, \ and\ \bibinfo {author}
  {\bibfnamefont {M.}~\bibnamefont {Wimmer}},\ }\href {\doibase
  10.1103/PhysRevLett.119.037701} {\bibfield  {journal} {\bibinfo  {journal}
  {Phys. Rev. Lett.}\ }\textbf {\bibinfo {volume} {119}},\ \bibinfo {pages}
  {037701} (\bibinfo {year} {2017})}\BibitemShut {NoStop}%
\bibitem [{\citenamefont {Nowak}\ and\ \citenamefont
  {W\'ojcik}(2018)}]{PhysRevB.97.045419}%
  \BibitemOpen
  \bibfield  {author} {\bibinfo {author} {\bibfnamefont {M.~P.}\ \bibnamefont
  {Nowak}}\ and\ \bibinfo {author} {\bibfnamefont {P.}~\bibnamefont
  {W\'ojcik}},\ }\href {\doibase 10.1103/PhysRevB.97.045419} {\bibfield
  {journal} {\bibinfo  {journal} {Phys. Rev. B}\ }\textbf {\bibinfo {volume}
  {97}},\ \bibinfo {pages} {045419} (\bibinfo {year} {2018})}\BibitemShut
  {NoStop}%
\bibitem [{\citenamefont {W\'ojcik}\ and\ \citenamefont
  {Nowak}(2018)}]{PhysRevB.97.235445}%
  \BibitemOpen
  \bibfield  {author} {\bibinfo {author} {\bibfnamefont {P.}~\bibnamefont
  {W\'ojcik}}\ and\ \bibinfo {author} {\bibfnamefont {M.~P.}\ \bibnamefont
  {Nowak}},\ }\href {\doibase 10.1103/PhysRevB.97.235445} {\bibfield  {journal}
  {\bibinfo  {journal} {Phys. Rev. B}\ }\textbf {\bibinfo {volume} {97}},\
  \bibinfo {pages} {235445} (\bibinfo {year} {2018})}\BibitemShut {NoStop}%
\bibitem [{\citenamefont {Bargerbos}\ \emph {et~al.}(2019)\citenamefont
  {Bargerbos}, \citenamefont {Uilhoorn}, \citenamefont {Yang}, \citenamefont
  {Krogstrup}, \citenamefont {Kouwenhoven}, \citenamefont {de~Lange},
  \citenamefont {van Heck},\ and\ \citenamefont {Kou}}]{bargerbos2019}%
  \BibitemOpen
  \bibfield  {author} {\bibinfo {author} {\bibfnamefont {A.}~\bibnamefont
  {Bargerbos}}, \bibinfo {author} {\bibfnamefont {W.}~\bibnamefont {Uilhoorn}},
  \bibinfo {author} {\bibfnamefont {C.-K.}\ \bibnamefont {Yang}}, \bibinfo
  {author} {\bibfnamefont {P.}~\bibnamefont {Krogstrup}}, \bibinfo {author}
  {\bibfnamefont {L.~P.}\ \bibnamefont {Kouwenhoven}}, \bibinfo {author}
  {\bibfnamefont {G.}~\bibnamefont {de~Lange}}, \bibinfo {author}
  {\bibfnamefont {B.}~\bibnamefont {van Heck}}, \ and\ \bibinfo {author}
  {\bibfnamefont {A.}~\bibnamefont {Kou}},\ }\href@noop {} {\  (\bibinfo {year}
  {2019})},\ \Eprint {http://arxiv.org/abs/1911.10010} {arXiv:1911.10010}
  \BibitemShut {NoStop}%
\bibitem [{\citenamefont {Kringh{\o}j}\ \emph {et~al.}(2019)\citenamefont
  {Kringh{\o}j}, \citenamefont {van Heck}, \citenamefont {Larsen},
  \citenamefont {Erlandsson}, \citenamefont {Sabonis}, \citenamefont
  {Krogstrup}, \citenamefont {Casparis}, \citenamefont {Petersson},\ and\
  \citenamefont {Marcus}}]{kringhj2019}%
  \BibitemOpen
  \bibfield  {author} {\bibinfo {author} {\bibfnamefont {A.}~\bibnamefont
  {Kringh{\o}j}}, \bibinfo {author} {\bibfnamefont {B.}~\bibnamefont {van
  Heck}}, \bibinfo {author} {\bibfnamefont {T.~W.}\ \bibnamefont {Larsen}},
  \bibinfo {author} {\bibfnamefont {O.}~\bibnamefont {Erlandsson}}, \bibinfo
  {author} {\bibfnamefont {D.}~\bibnamefont {Sabonis}}, \bibinfo {author}
  {\bibfnamefont {P.}~\bibnamefont {Krogstrup}}, \bibinfo {author}
  {\bibfnamefont {L.}~\bibnamefont {Casparis}}, \bibinfo {author}
  {\bibfnamefont {K.~D.}\ \bibnamefont {Petersson}}, \ and\ \bibinfo {author}
  {\bibfnamefont {C.~M.}\ \bibnamefont {Marcus}},\ }\href@noop {} {\  (\bibinfo
  {year} {2019})},\ \Eprint {http://arxiv.org/abs/1911.10011}
  {arXiv:1911.10011} \BibitemShut {NoStop}%
\bibitem [{\citenamefont {Pita-Vidal}\ \emph {et~al.}(2019)\citenamefont
  {Pita-Vidal}, \citenamefont {Bargerbos}, \citenamefont {Yang}, \citenamefont
  {van Woerkom}, \citenamefont {Pfaff}, \citenamefont {Haider}, \citenamefont
  {Krogstrup}, \citenamefont {Kouwenhoven}, \citenamefont {de~Lange},\ and\
  \citenamefont {Kou}}]{pitavidal2019}%
  \BibitemOpen
  \bibfield  {author} {\bibinfo {author} {\bibfnamefont {M.}~\bibnamefont
  {Pita-Vidal}}, \bibinfo {author} {\bibfnamefont {A.}~\bibnamefont
  {Bargerbos}}, \bibinfo {author} {\bibfnamefont {C.-K.}\ \bibnamefont {Yang}},
  \bibinfo {author} {\bibfnamefont {D.~J.}\ \bibnamefont {van Woerkom}},
  \bibinfo {author} {\bibfnamefont {W.}~\bibnamefont {Pfaff}}, \bibinfo
  {author} {\bibfnamefont {N.}~\bibnamefont {Haider}}, \bibinfo {author}
  {\bibfnamefont {P.}~\bibnamefont {Krogstrup}}, \bibinfo {author}
  {\bibfnamefont {L.~P.}\ \bibnamefont {Kouwenhoven}}, \bibinfo {author}
  {\bibfnamefont {G.}~\bibnamefont {de~Lange}}, \ and\ \bibinfo {author}
  {\bibfnamefont {A.}~\bibnamefont {Kou}},\ }\href@noop {} {\  (\bibinfo {year}
  {2019})},\ \Eprint {http://arxiv.org/abs/1910.07978} {arXiv:1910.07978}
  \BibitemShut {NoStop}%
\bibitem [{\citenamefont {Vaitiek{\.e}nas}\ \emph {et~al.}(2020)\citenamefont
  {Vaitiek{\.e}nas}, \citenamefont {Winkler}, \citenamefont {van Heck},
  \citenamefont {Karzig}, \citenamefont {Deng}, \citenamefont {Flensberg},
  \citenamefont {Glazman}, \citenamefont {Nayak}, \citenamefont {Krogstrup},
  \citenamefont {Lutchyn} \emph {et~al.}}]{Vaitiekenas:S20}%
  \BibitemOpen
  \bibfield  {author} {\bibinfo {author} {\bibfnamefont {S.}~\bibnamefont
  {Vaitiek{\.e}nas}}, \bibinfo {author} {\bibfnamefont {G.}~\bibnamefont
  {Winkler}}, \bibinfo {author} {\bibfnamefont {B.}~\bibnamefont {van Heck}},
  \bibinfo {author} {\bibfnamefont {T.}~\bibnamefont {Karzig}}, \bibinfo
  {author} {\bibfnamefont {M.-T.}\ \bibnamefont {Deng}}, \bibinfo {author}
  {\bibfnamefont {K.}~\bibnamefont {Flensberg}}, \bibinfo {author}
  {\bibfnamefont {L.}~\bibnamefont {Glazman}}, \bibinfo {author} {\bibfnamefont
  {C.}~\bibnamefont {Nayak}}, \bibinfo {author} {\bibfnamefont
  {P.}~\bibnamefont {Krogstrup}}, \bibinfo {author} {\bibfnamefont
  {R.}~\bibnamefont {Lutchyn}},  \emph {et~al.},\ }\href@noop {} {\bibfield
  {journal} {\bibinfo  {journal} {Science}\ }\textbf {\bibinfo {volume} {367}}
  (\bibinfo {year} {2020})}\BibitemShut {NoStop}%
\bibitem [{\citenamefont {Pe\~naranda}\ \emph {et~al.}(2020)\citenamefont
  {Pe\~naranda}, \citenamefont {Aguado}, \citenamefont {San-Jose},\ and\
  \citenamefont {Prada}}]{Penaranda:19}%
  \BibitemOpen
  \bibfield  {author} {\bibinfo {author} {\bibfnamefont {F.}~\bibnamefont
  {Pe\~naranda}}, \bibinfo {author} {\bibfnamefont {R.}~\bibnamefont {Aguado}},
  \bibinfo {author} {\bibfnamefont {P.}~\bibnamefont {San-Jose}}, \ and\
  \bibinfo {author} {\bibfnamefont {E.}~\bibnamefont {Prada}},\ }\href
  {\doibase 10.1103/PhysRevResearch.2.023171} {\bibfield  {journal} {\bibinfo
  {journal} {Phys. Rev. Research}\ }\textbf {\bibinfo {volume} {2}},\ \bibinfo
  {pages} {023171} (\bibinfo {year} {2020})}\BibitemShut {NoStop}%
\end{thebibliography}%

\end{document}